\input harvmac
\input epsf
\input amssym.tex
\input xyv2
\input draft3.defs

\writedefs

\def\1{{\hat 1}}
\def\2{{\hat 2}}
\def\3{{\hat 3}}

\def\D{{\rm D}}
\def\F{{\rm F}}

\def\CM{{\cal M}}
\def\CV{{\cal V}}

\def\Dirac{{\displaystyle{\not} D }}

\def\bb{
\font\tenmsb=msbm10
\font\sevenmsb=msbm7
\font\fivemsb=msbm5
\textfont1=\tenmsb
\scriptfont1=\sevenmsb
\scriptscriptfont1=\fivemsb
}

\def\R{\hbox{$\bb R$}}

\def\Z{\hbox{$\bb Z$}}

\def\NZ{\Z_{\geq 0}}
\def\adm{m_{\rm ad}}



\lref\ZetaProd{
J.~R.~Quine, S.~H.~Heydari and R.~Y.~Song,
``Zeta regularized products,''
Trans.\ Amer.\ Math.\ Soc. {\bf 338}, 213 (1993).
}

\lref\SugishitaJCA{
  S.~Sugishita and S.~Terashima,
  ``Exact Results in Supersymmetric Field Theories on Manifolds with Boundaries,''
JHEP {\bf 1311}, 021 (2013).
[arXiv:1308.1973 [hep-th]].
}

\lref\HoriIKA{
  K.~Hori and M.~Romo,
  ``Exact Results In Two-Dimensional (2,2) Supersymmetric Gauge Theories With Boundary,''
[arXiv:1308.2438 [hep-th]].
}

\lref\DonagiCF{
  R.~Donagi and E.~Witten,
 ``Supersymmetric Yang-Mills theory and integrable systems,''
Nucl.\ Phys.\ B {\bf 460}, 299 (1996).
[hep-th/9510101].
}

\lref\NakajimaALE{
H.~Nakajima,
``Instantons on ALE spaces, quiver varieties, and Kac- Moody algebras,'' 
Duke.\ Math.\ {\bf 76}, 365416 (1994).
}

\lref\NakajimaRes{
H.~Nakajima, ``Resolutions of moduli spaces of ideal instantons on ${\bf R}^4$,'' in Topology, Geometry and Field Theory, World Scientific, 129136 (1994). 
}

\lref\NakajimaLec{
H.~Nakajima, ``Lectures on Hilbert schemes of points on surfaces,'' Univ.\ Lect.\ Ser.\ {\bf 18}, AMS, Providence, RI (1999).
}

\lref\LosevTP{
  A.~Losev, N.~Nekrasov and S.~L.~Shatashvili,
  ``Issues in topological gauge theory,''
Nucl.\ Phys.\ B {\bf 534}, 549 (1998).
[hep-th/9711108].
}

\lref\NekrasovTalk{
N.~Nekrasov,
``On the BPS/CFT correspondence,'' \hfill
\break
http://www.science.uva.nl/research/itf/strings/stringseminar2003-4.html.
}

\lref\NekrasovRJ{
  N.~Nekrasov and A.~Okounkov,
  ``Seiberg-Witten theory and random partitions,''
[hep-th/0306238].
}

\lref\LosevPY{
  A.~S.~Losev, A.~Marshakov and N.~A.~Nekrasov,
  ``Small instantons, little strings and free fermions,''
In *Shifman, M. (ed.) et al.: From fields to strings, vol. 1* 581-621.
[hep-th/0302191].
}
\lref\LossevBZ{
  A.~Lossev, N.~Nekrasov and S.~L.~Shatashvili,
  ``Testing Seiberg-Witten solution,''
In *Cargese 1997, Strings, branes and dualities* 359-372.
[hep-th/9801061].
}

\lref\MooreDJ{
  G.~W.~Moore, N.~Nekrasov and S.~Shatashvili,
  ``Integrating over Higgs branches,''
Commun.\ Math.\ Phys.\  {\bf 209}, 97 (2000).
[hep-th/9712241].
}

\lref\NekrasovQD{
  N.~A.~Nekrasov,
  ``Seiberg-Witten prepotential from instanton counting,''
Adv.\ Theor.\ Math.\ Phys.\  {\bf 7}, 831 (2004).
[hep-th/0206161].
}

\lref\GukovJK{
  S.~Gukov and E.~Witten,
  ``Gauge Theory, Ramification, And The Geometric Langlands Program,''
[hep-th/0612073].
}

\lref\GerasimovTD{
  A.~A.~Gerasimov and D.~R.~Lebedev,
  ``On topological field theory representation of higher analogs of classical special functions,''
JHEP {\bf 1109}, 076 (2011).
[arXiv:1011.0403 [hep-th]].
}

\lref\BraneBook{
P.~S.~Aspinwall, T.~Bridgeland, A.~Craw, M.~R.~Douglas, M.~Gross, A.~Kapustin, G.~W.~Moore, G.~Segal, B.~Szendr\"oi and P.~M.~H.~Wilson,
``Dirichlet Branes and Mirror Symmetry,''
Clay Mathematics Monographs.
}

\lref\GaddeDDA{
  A.~Gadde and S.~Gukov,
  ``2d Index and Surface operators,''
[arXiv:1305.0266 [hep-th]].
}

\lref\ItoEA{
  Y.~Ito, T.~Okuda and M.~Taki,
  ``Line operators on $\Bbb S^1\times \Bbb R^3$ and quantization of the Hitchin moduli space,''
JHEP {\bf 1204}, 010 (2012).
[arXiv:1111.4221 [hep-th]].
}
\lref\ItoKPA{
  Y.~Ito, K.~Maruyoshi and T.~Okuda,
  ``Scheme dependence of instanton counting in ALE spaces,''
JHEP {\bf 1305}, 045 (2013).
[arXiv:1303.5765 [hep-th]].
}

\lref\KapustinIW{
  A.~Kapustin, B.~Willett and I.~Yaakov,
  ``Exact results for supersymmetric abelian vortex loops in 2+1 dimensions,''
[arXiv:1211.2861 [hep-th]].
}

\lref\DrukkerSR{
  N.~Drukker, T.~Okuda and F.~Passerini,
  ``Exact results for vortex loop operators in 3d supersymmetric theories,''
[arXiv:1211.3409 [hep-th]].
}

\lref\DrukkerJM{
  N.~Drukker, J.~Gomis and D.~Young,
  ``Vortex Loop Operators, M2-branes and Holography,''
JHEP {\bf 0903}, 004 (2009).
[arXiv:0810.4344 [hep-th]].
}

\lref\GoddardQE{
  P.~Goddard, J.~Nuyts and D.~I.~Olive,
  ``Gauge Theories and Magnetic Charge,''
Nucl.\ Phys.\ B {\bf 125}, 1 (1977)..
}

\lref\GukovJK{
  S.~Gukov and E.~Witten,
  ``Gauge Theory, Ramification, And The Geometric Langlands Program,''
[hep-th/0612073].
}

\lref\tHooftHY{
  G.~'t Hooft,
  ``On the Phase Transition Towards Permanent Quark Confinement,''
Nucl.\ Phys.\ B {\bf 138}, 1 (1978).
}

\lref\KapustinPY{
  A.~Kapustin,
  ``Wilson-'t Hooft operators in four-dimensional gauge theories and S-duality,''
Phys.\ Rev.\ D {\bf 74}, 025005 (2006).
[hep-th/0501015].
}
\lref\GH{
P.~Griffiths and J.~Harris,
``Principles of Algebraic Geometry,''
Wiley Classics Library.
}

\lref\CG{
N.~Chriss and V.~Ginzburg,
``Representation theory and complex geometry,''
Modern Birkh\"auser Classics.
}
\lref\AtiyahSegal{
M.~Atiyah and G.~Segal, ``The index of elliptic operators: II.,'', Ann. of Math. (2) 87 1968 531-545.,}

\lref\Cylinder{
T.~Okuda,
in progress.
}

\lref\Donnely{
D.~Harold,
``Eta invariants for G-spaces,''
Indiana Univ. Math. J. {\bf 27}, 889 (1978).
}

\lref\CecottiME{
  S.~Cecotti and C.~Vafa,
  ``Topological antitopological fusion,''
Nucl.\ Phys.\ B {\bf 367}, 359 (1991)..
}

\lref\NekrasovUH{
  N.~A.~Nekrasov and S.~L.~Shatashvili,
  ``Supersymmetric vacua and Bethe ansatz,''
Nucl.\ Phys.\ Proc.\ Suppl.\  {\bf 192-193}, 91 (2009).
[arXiv:0901.4744 [hep-th]].
}
\lref\NekrasovUI{
  N.~A.~Nekrasov and S.~L.~Shatashvili,
  ``Quantum integrability and supersymmetric vacua,''
Prog.\ Theor.\ Phys.\ Suppl.\  {\bf 177}, 105 (2009).
[arXiv:0901.4748 [hep-th]].
}

\lref\AldayFS{
  L.~F.~Alday, D.~Gaiotto, S.~Gukov, Y.~Tachikawa and H.~Verlinde,
  ``Loop and surface operators in N=2 gauge theory and Liouville modular geometry,''
JHEP {\bf 1001}, 113 (2010).
[arXiv:0909.0945 [hep-th]].
}

\lref\GukovJK{
  S.~Gukov and E.~Witten,
  ``Gauge Theory, Ramification, And The Geometric Langlands Program,''
[hep-th/0612073].
}

\lref\DrukkerJP{
  N.~Drukker, D.~Gaiotto and J.~Gomis,
 ``The Virtue of Defects in 4D Gauge Theories and 2D CFTs,''
JHEP {\bf 1106}, 025 (2011).
[arXiv:1003.1112 [hep-th]].
}

\lref\FerrariSV{
  F.~Ferrari and A.~Bilal,
 ``The Strong coupling spectrum of the Seiberg-Witten theory,''
Nucl.\ Phys.\ B {\bf 469}, 387 (1996).
[hep-th/9602082].
}

\lref\GomisKV{
  J.~Gomis and B.~Le Floch,
``'t Hooft Operators in Gauge Theory from Toda CFT,''
JHEP {\bf 1111}, 114 (2011).
[arXiv:1008.4139 [hep-th]].
}

\lref\GomisPF{
  J.~Gomis, T.~Okuda and V.~Pestun,
``Exact Results for 't Hooft Loops in Gauge Theories on $\Bbb S^4$,''
JHEP {\bf 1205}, 141 (2012).
[arXiv:1105.2568 [hep-th]].
}

\lref\GaiottoFS{
  D.~Gaiotto,
  ``Surface Operators in N = 2 4d Gauge Theories,''
JHEP {\bf 1211}, 090 (2012).
[arXiv:0911.1316 [hep-th]].
}

\lref\WittenMH{
  E.~Witten and D.~I.~Olive,
  ``Supersymmetry Algebras That Include Topological Charges,''
Phys.\ Lett.\ B {\bf 78}, 97 (1978).
}

\lref\BonelliMMA{
  G.~Bonelli, A.~Sciarappa, A.~Tanzini and P.~Vasko,
  ``Vortex partition functions, wall crossing and equivariant Gromov-Witten invariants,''
[arXiv:1307.5997 [hep-th]].
}

\lref\SenSM{
  A.~Sen,
  ``Tachyon condensation on the brane anti-brane system,''
JHEP {\bf 9808}, 012 (1998).
[hep-th/9805170].
}

\lref\KapustinBI{
  A.~Kapustin and Y.~Li,
  ``D branes in Landau-Ginzburg models and algebraic geometry,''
JHEP {\bf 0312}, 005 (2003).
[hep-th/0210296].
}

\lref\HellermanBU{
  S.~Hellerman, S.~Kachru, A.~E.~Lawrence and J.~McGreevy,
  ``Linear sigma models for open strings,''
JHEP {\bf 0207}, 002 (2002).
[hep-th/0109069].
}

\lref\OkudaKE{
  T.~Okuda and V.~Pestun,
  ``On the instantons and the hypermultiplet mass of N=2* super Yang-Mills on $S^{4}$,''
JHEP {\bf 1203}, 017 (2012).
[arXiv:1004.1222 [hep-th]].
}

\lref\BeemMB{
  C.~Beem, T.~Dimofte and S.~Pasquetti,
  ``Holomorphic Blocks in Three Dimensions,''
[arXiv:1211.1986 [hep-th]].
}

\lref\JockersDK{
  H.~Jockers, V.~Kumar, J.~M.~Lapan, D.~R.~Morrison and M.~Romo,
  ``Two-Sphere Partition Functions and Gromov-Witten Invariants,''
[arXiv:1208.6244 [hep-th]].
}

\lref\PestunRZ{
  V.~Pestun,
``Localization of gauge theory on a four-sphere and supersymmetric Wilson loops,''
Commun.\ Math.\ Phys.\  {\bf 313}, 71 (2012).
[arXiv:0712.2824 [hep-th]].
}

\lref\DrukkerJP{
  N.~Drukker, D.~Gaiotto and J.~Gomis,
``The Virtue of Defects in 4D Gauge Theories and 2D CFTs,''
JHEP {\bf 1106}, 025 (2011).
[arXiv:1003.1112 [hep-th]].
}

\lref\GaiottoAK{
  D.~Gaiotto and E.~Witten,
  ``S-Duality of Boundary Conditions In N=4 Super Yang-Mills Theory,''
Adv.\ Theor.\ Math.\ Phys.\  {\bf 13} (2009).
[arXiv:0807.3720 [hep-th]].
}

\lref\GaiottoSA{
  D.~Gaiotto and E.~Witten,
  ``Supersymmetric Boundary Conditions in N=4 Super Yang-Mills Theory,''
J.\ Statist.\ Phys.\  {\bf 135}, 789 (2009).
[arXiv:0804.2902 [hep-th]].
}

\lref\DrukkerID{
  N.~Drukker, J.~Gomis, T.~Okuda and J.~Teschner,
``Gauge Theory Loop Operators and Liouville Theory,''
JHEP {\bf 1002}, 057 (2010).
[arXiv:0909.1105 [hep-th]].
}

\lref\AldayAQ{
  L.~F.~Alday, D.~Gaiotto and Y.~Tachikawa,
``Liouville Correlation Functions from Four-dimensional Gauge Theories,''
Lett.\ Math.\ Phys.\  {\bf 91}, 167 (2010).
[arXiv:0906.3219 [hep-th]].
}

\lref\WyllardHG{
  N.~Wyllard,
  ``A(N-1) conformal Toda field theory correlation functions from conformal N = 2 SU(N) quiver gauge theories,''
JHEP {\bf 0911}, 002 (2009).
[arXiv:0907.2189 [hep-th]].
}

\lref\GaiottoWE{
  D.~Gaiotto,
  ``N=2 dualities,''
JHEP {\bf 1208}, 034 (2012).
[arXiv:0904.2715 [hep-th]].
}

\lref\GaiottoHG{
  D.~Gaiotto, G.~W.~Moore and A.~Neitzke,
[arXiv:0907.3987 [hep-th]].
}

\lref\GaiottoTF{
  D.~Gaiotto, G.~W.~Moore and A.~Neitzke,
  ``Wall-Crossing in Coupled 2d-4d Systems,''
[arXiv:1103.2598 [hep-th]].
}

\lref\AldayFS{
  L.~F.~Alday, D.~Gaiotto, S.~Gukov, Y.~Tachikawa and H.~Verlinde,
``Loop and surface operators in N=2 gauge theory and Liouville modular geometry,''
JHEP {\bf 1001}, 113 (2010).
[arXiv:0909.0945 [hep-th]].
}

\lref\WittenZZ{
  E.~Witten,
``Mirror manifolds and topological field theory,''
In *Yau, S.T. (ed.): Mirror symmetry I* 121-160.
[hep-th/9112056].
}
\lref\WittenCD{
  E.~Witten,
  ``D-branes and K theory,''
JHEP {\bf 9812}, 019 (1998).
[hep-th/9810188].
}

\lref\WittenMK{
  E.~Witten,
  ``The N matrix model and gauged WZW models,''
Nucl.\ Phys.\ B {\bf 371}, 191 (1992).
}

\lref\BertramTX{
A.~Bertram, I.~Ciocan-Fontanine and B.~Kim,
``Two proofs of a conjecture of Hori and Vafa,''
Duke~Math.~J. {\bf 126}, 101 (2005).
[math/0304403].
}

\lref\MinasianMM{
  R.~Minasian and G.~W.~Moore,
  ``K theory and Ramond-Ramond charge,''
JHEP {\bf 9711}, 002 (1997).
[hep-th/9710230].
}

\lref\AspinwallJR{
  P.~S.~Aspinwall,
  ``D-branes on Calabi-Yau manifolds,''
[hep-th/0403166].
}

\lref\DouglasFJ{
  M.~R.~Douglas,
  ``Dirichlet branes, homological mirror symmetry, and stability,''
[math/0207021 [math-ag]].
}

\lref\HoriKT{
  K.~Hori and C.~Vafa,
  ``Mirror symmetry,''
[hep-th/0002222].
}

\lref\GerasimovTD{
  A.~A.~Gerasimov and D.~R.~Lebedev,
  ``On topological field theory representation of higher analogs of classical special functions,''
JHEP {\bf 1109}, 076 (2011).
[arXiv:1011.0403 [hep-th]].
}

\lref\LossevBZ{
  A.~Lossev, N.~Nekrasov and S.~L.~Shatashvili,
  ``Testing Seiberg-Witten solution,''
In *Cargese 1997, Strings, branes and dualities* 359-372.
[hep-th/9801061].
}

\lref\YoshidaAU{
  Y.~Yoshida,
  ``Localization of Vortex Partition Functions in $\mathcal{N}=(2,2) $ Super Yang-Mills theory,''
[arXiv:1101.0872 [hep-th]].
}

\lref\BonelliFQ{
  G.~Bonelli, A.~Tanzini and J.~Zhao,
  ``Vertices, Vortices and Interacting Surface Operators,''
JHEP {\bf 1206}, 178 (2012).
[arXiv:1102.0184 [hep-th]].
}

\lref\FujimoriAB{
  T.~Fujimori, T.~Kimura, M.~Nitta and K.~Ohashi,
  ``Vortex counting from field theory,''
JHEP {\bf 1206}, 028 (2012).
[arXiv:1204.1968 [hep-th]].
}

\lref\BeemMB{
  C.~Beem, T.~Dimofte and S.~Pasquetti,
  ``Holomorphic Blocks in Three Dimensions,''
[arXiv:1211.1986 [hep-th]].
}

\lref\TeschnerYF{
  J.~Teschner,
  ``On the Liouville three point function,''
Phys.\ Lett.\ B {\bf 363}, 65 (1995).
[hep-th/9507109].
}

\lref\MorrisonFR{
  D.~R.~Morrison and M.~R.~Plesser,
  ``Summing the instantons: Quantum cohomology and mirror symmetry in toric varieties,''
Nucl.\ Phys.\ B {\bf 440}, 279 (1995).
[hep-th/9412236].
}

\lref\GrisaruPX{
  M.~T.~Grisaru, A.~E.~M.~van de Ven and D.~Zanon,
 ``Four Loop beta Function for the N=1 and N=2 Supersymmetric Nonlinear Sigma Model in Two-Dimensions,''
Phys.\ Lett.\ B {\bf 173}, 423 (1986).
}

\lref\CandelasRM{
  P.~Candelas, X.~C.~De La Ossa, P.~S.~Green and L.~Parkes,
  ``A Pair of Calabi-Yau manifolds as an exactly soluble superconformal theory,''
Nucl.\ Phys.\ B {\bf 359}, 21 (1991)..
}

\lref\KrausNJ{
  P.~Kraus and F.~Larsen,
  ``Boundary string field theory of the D anti-D system,''
Phys.\ Rev.\ D {\bf 63}, 106004 (2001).
[hep-th/0012198].
}

\lref\TakayanagiRZ{
  T.~Takayanagi, S.~Terashima and T.~Uesugi,
  ``Brane - anti-brane action from boundary string field theory,''
JHEP {\bf 0103}, 019 (2001).
[hep-th/0012210].
}

\lref\HoriCK{
  K.~Hori, A.~Iqbal and C.~Vafa,
  ``D-branes and mirror symmetry,''
[hep-th/0005247].
}

\lref\DoreyPA{
  N.~Dorey, S.~Lee and T.~J.~Hollowood,
  ``Quantization of Integrable Systems and a 2d/4d Duality,''
JHEP {\bf 1110}, 077 (2011).
[arXiv:1103.5726 [hep-th]].
}

\lref\ShadchinYZ{
  S.~Shadchin,
  ``On F-term contribution to effective action,''
JHEP {\bf 0708}, 052 (2007).
[hep-th/0611278].
}

\lref\HerbstJQ{
  M.~Herbst, K.~Hori and D.~Page,
  ``Phases Of N=2 Theories In 1+1 Dimensions With Boundary,''
[arXiv:0803.2045 [hep-th]].
}

\lref\WuGE{
  T.~T.~Wu and C.~N.~Yang,
  ``Dirac Monopole Without Strings: Monopole Harmonics,''
Nucl.\ Phys.\ B {\bf 107}, 365 (1976)..
}

\lref\EguchiJX{
  T.~Eguchi, P.~B.~Gilkey and A.~J.~Hanson,
  ``Gravitation, Gauge Theories and Differential Geometry,''
Phys.\ Rept.\  {\bf 66}, 213 (1980)..
}

\lref\MaulikWI{
  D.~Maulik and A.~Okounkov,
  ``Quantum Groups and Quantum Cohomology,''
[arXiv:1211.1287 [math.AG]].
}

\lref\DoroudXW{
  N.~Doroud, J.~Gomis, B.~Le Floch and S.~Lee,
  ``Exact Results in D=2 Supersymmetric Gauge Theories,''
[arXiv:1206.2606 [hep-th]].
}

\lref\WittenXI{
  E.~Witten,
  ``The Verlinde algebra and the cohomology of the Grassmannian,''
In *Cambridge 1993, Geometry, topology, and physics* 357-422.
[hep-th/9312104].
}

\lref\HalversonEUA{
  J.~Halverson, V.~Kumar and D.~R.~Morrison,
  ``New Methods for Characterizing Phases of 2D Supersymmetric Gauge Theories,''
[arXiv:1305.3278 [hep-th]].
}

\lref\GomisWY{
  J.~Gomis and S.~Lee,
  ``Exact Kahler Potential from Gauge Theory and Mirror Symmetry,''
[arXiv:1210.6022 [hep-th]].
}

\lref\WessCP{
  J.~Wess and J.~Bagger,
  ``Supersymmetry and supergravity,''
Princeton, USA: Univ. Pr. (1992) 259 p.
}

\lref\WittenYC{
  E.~Witten,
  ``Phases of N=2 theories in two-dimensions,''
Nucl.\ Phys.\ B {\bf 403}, 159 (1993).
[hep-th/9301042].
}

\lref\NakajimaQAA{
H.~Nakajima,
  ``Quiver varieties and finite dimensional representations of quantum affine algebras,''
 J.\ Amer.\ Math.\ Soc.\ {\bf 14}, no.1\ 145-238 (2001).
[arXiv:math/9912158].
}

\lref\BeniniUI{
  F.~Benini and S.~Cremonesi,
  ``Partition functions of N=(2,2) gauge theories on $S^2$ and vortices,''
[arXiv:1206.2356 [hep-th]].
}

\lref\GaiottoSA{
  D.~Gaiotto, E.~Witten and ,
  ``Supersymmetric Boundary Conditions in N=4 Super Yang-Mills Theory,''
J.\ Statist.\ Phys.\  {\bf 135}, 789 (2009).
[arXiv:0804.2902 [hep-th]].
}

\lref\AbraHand{
M.~Abramowitz, I.~A.~Stegun, eds.,
``Handbook of Mathematical Function,''
Dover publications, New York, 1964.
}

\lref\WuDira{
T.~T.~Wu and C.~N.~Yang,
``Dirac Monopole Without Strings: Monopole Harmonics,''
Nucl.\ Phys.\ B {\bf 107}, 365 (1976).
}

\lref\WeinMono{
E.~J.~Weinberg,
``Monopole Vector Spherical Harmonics,''
Phys.\ Rev.\ D {\bf 49}, 1086 (1994),
hep-th/9308054.
}

\lref\BarrVect{
R.~G.~Barrera, G.~A.~Est\'evez and J.~Giraldo,
``Vector spherical harmonics and their application to magnetostatics,"
Eur.\ J.\ Phys.\ {\bf 6} 287 (1985).
}

\lref\GomiExac{
J.~Gomis, S.~Lee,
``Exact K\"ahler Potential from Gauge Theory and Mirror Symmetry,"
arXiv:1210.6022.
}

\lref\OoguriCK{
  H.~Ooguri, Y.~Oz and Z.~Yin,
  ``D-branes on Calabi-Yau spaces and their mirrors,''
Nucl.\ Phys.\ B {\bf 477}, 407 (1996).
[hep-th/9606112].
}

\lref\DoroExac{
N.~Doroud, J.~Gomis, B.~Le Floch and S.~Lee,
``Exact Results in D=2 Supersymmetric Gauge Theories,''
arXiv:1206.2606.
}

\lref\WarnerAY{
  N.~P.~Warner,
  ``Supersymmetry in boundary integrable models,''
Nucl.\ Phys.\ B {\bf 450}, 663 (1995).
[hep-th/9506064].
}

\writedefs

\newbox\tmpbox\setbox\tmpbox\hbox{\abstractfont}
\Title{\vbox{\baselineskip12pt \hbox{\hfill {\tt  UT-Komaba/13-8}}
}}
{
\vbox{
\vskip 1cm
\centerline{Exact results for boundaries and domain walls}
\vskip 0.5cm
\centerline{in 2d supersymmetric theories}
}
}

\centerline{Daigo Honda and Takuya Okuda}
\vskip 0.5cm
\centerline{\it University of Tokyo, Komaba,}
\centerline{\it Meguro-ku, Tokyo 153-8902, Japan}
\vskip 1.5cm

\noindent

We apply supersymmetric localization to ${\cal N}=(2,2)$ gauged linear sigma models on a hemisphere, with boundary conditions, {\it i.e.}, D-branes, preserving B-type supersymmetries.
We explain how to compute the hemisphere partition function for each object in the derived category of  equivariant coherent sheaves, and argue that it depends only on its K theory class.
The hemisphere partition function computes exactly the central charge of the D-brane, completing the well-known formula obtained by an anomaly inflow argument.
We also formulate supersymmetric domain walls as D-branes in the product of two theories.
In particular 4d line operators bound to a surface operator, corresponding via the AGT relation to certain defects in Toda CFT's, are constructed as domain walls.
Moreover we exhibit domain walls that realize the $sl(2)$ affine Hecke algebra.

\Date{}

\listtoc
\writetoc

\newsec{Introduction and summary}
\seclab\SecIntro

Two-dimensional ${\cal N}=(2,2)$ gauged linear sigma models \WittenYC\ are simple quantum field theories that exhibit very rich structures.
As such, they have a variety of applications.
When we put these theories on a surface with boundary, the boundary conditions describe D-branes.
A boundary condition in the product of two theories can be regarded as a domain wall that connects two regions where the two theories live.

In this paper we study boundaries and domain walls in ${\cal N}=(2,2)$ gauged linear sigma models using supersymmetric localization.
We focus on the hemisphere geometry, which has a single boundary component.
The resulting hemisphere partition function is roughly a half of the $\Bbb S^2$ partition function \refs{\BeniniUI,\DoroudXW} obtained by localization techniques similar to \PestunRZ.

There are two broad motivations for studying the hemisphere partition function.
The first is the study of D-branes in Calabi-Yau manifolds, with applications to mirror symmetry, Gromov-Witten invariants, D-brane stability, string phenomenology, etc.
In such contexts the two dimensional theory describes the worldsheet of a superstring, and one is especially interested in theories that flows to a non-linear sigma model with target space a compact Calabi-Yau.
Generically such a theory possesses no flavor symmetries.
The hemisphere partition function depends analytically on the complexified FI parameters, which we collectively denote as $t$ and use to parametrize the K\"ahler moduli space.
The second motivation, the main one for us, is to study the dynamics of the two-dimensional quantum field theory in its own right.
It is known that ${\cal N}=(2,2)$ theories are closely related to integrable models \refs{\NekrasovUH,\NekrasovUI}.
Such a theory also arises as the defining theory for a surface operator embedded in a four-dimensional theory \GaiottoFS.
It is natural to turn on twisted masses $m=(m_a)$, or equivariant parameters for flavor symmetries, in these contexts.
Boundaries are interesting ingredients in the physics of the theory, while domain walls ($\simeq$ line operators in two dimensions) provide a natural example of non-local disorder operators, and are akin to 't Hooft loops \refs{\tHooftHY,\KapustinPY,\GomisPF,\ItoEA}, vortex loops \refs{\DrukkerJM,\KapustinIW,\DrukkerSR},  surface operators \GukovJK, and domain walls \refs{\GaiottoSA,\GaiottoAK} in higher dimensions.

The type of boundary conditions ${\cal B}$ we study preserve B-type supersymmetries \refs{\WittenZZ}.
For abelian gauge theories general B-type boundary conditions were formulated in \HerbstJQ~and the references therein.
We extend these boundary conditions, in a straightforward way, to theories with non-abelian gauge groups and twisted masses.
We will argue that the hemisphere partition function $Z_{\rm hem}({\cal B};t;m)$ is the overlap 
$\langle \cal B|{\tt 1}\rangle$ of two states, where both the boundary state $\langle {\cal B}|$ and the state $|{\tt 1}\rangle$ created by a topological twist \CecottiME\ are zero-energy states in the Hilbert space for the Ramond-Ramond sector.

When the gauge theory flows to a non-linear sigma model with a smooth target space, there are refined and coarse classifications of B-branes:
\vskip 3mm
\centerline{
\hskip 38.5mm 
$\{$B-branes$\}$ $\simeq$ derived category of coherent sheaves}
\vskip 1mm
\centerline{
\hskip-20mm $\{$topological charges$\}$ $\simeq$ K theory}
\vskip 3mm
\noindent
The latter amounts to classifying B-branes up to dynamical creation and annihilation (tachyon condensation \SenSM) processes.
For details and precise treatments on these mathematical concepts, see for example \refs{\AspinwallJR,\BraneBook,\CG}.
In type II string theory compactified on a Calabi-Yau, such topological charges of branes determine the central charges \WittenMH\ of the extended supersymmetry algebra in non-compact dimensions.
This central charge is given precisely by the overlap $\langle \cal B|{\tt 1}\rangle$ \OoguriCK.
We will argue that the hemisphere partition function $Z_{\rm hem}({\cal B})$ indeed depends only on the K theory class of the brane.
The known formula for the central charge, which is valid in the large volume limit and was obtained by an anomaly inflow argument \MinasianMM, provides a useful check of our result and is completed by our exact formula.

More generally, our localization computation yields a pairing $\langle {\cal B}|{\tt f}\rangle$ between the boundary state $\langle {\cal B}|$ and a state obtained by the path integral with the insertion of an operator ${\tt f}$ annihilated by the supercharge used for localization.
With twisted masses for the flavor symmetry group $G_{\rm F}$ turned on, the sheaves and K theories are replaced by their $G_{\rm F}$-equivariant versions.
Related works that emphasize $G_{\rm F}$-equivariance include \refs{\GaddeDDA,\BonelliMMA}.
It was found by Nekrasov and Shatashvili \refs{\NekrasovUH,\NekrasovUI} that the relations in the equivariant quantum cohomology of certain models are precisely the Bethe ansatz equations of spin chains.
Our work is thus related to, and in fact most directly motivated by, the study of integrable structures in supersymmetric gauge theories.
Integrability suggests the presence of infinite dimensional quantum group symmetries, whose generators are expected to be realized as domain walls.
As mentioned domain walls are D-branes in product theories, and the quantum group symmetries are known to be realized geometrically as so-called convolution algebras in equivariant K theories and derived categories \CG.
In this work we take a modest step in this direction by realizing the $sl(2)$ affine Hecke algebra as the domain wall algebra.%
\foot{%
The connection between the domain wall and convolution algebras was explained to us by N.~Nekrasov and S.~Shatashvili, and had been discussed in the literature.
Realization of the affine Hecke algebra in two-dimensional field theory was also studied in \GukovJK.
}

Relatedly, the 2d ${\cal N}=(2,2)$ theories can also be embedded in a 4d ${\cal N}=2$ theory to define a surface operator \GaiottoFS.
Domain walls in the 2d theory can then be regarded as 4d line operators bound to the surface operator, and via the AGT correspondence \AldayAQ\ is related to certain defects in Toda conformal field theories \refs{\AldayFS}.
We use our results to identify the precise domain walls that correspond to the defects.

We also study Seiberg-like dualities.
In some dual pairs of theories, the hemisphere partition functions are found to be identical, while in the others they turn out to differ by a simple overall factor.
Such dualities also serve as nice checks of our results.

Besides investigations (see {\it e.g.} \refs{\DonagiCF}) directly relevant to the so-called class S theories \refs{\GaiottoHG,\GaiottoWE}, we note and emphasize that the relation between supersymmetric field theories and lower-dimensional models, regarding their integrable structures and symmetries, have been studied in different but related lines of development (see {\it e.g.} \refs{\LosevTP,\MooreDJ,\LossevBZ,\NekrasovQD,\LosevPY,\NekrasovRJ,\NakajimaALE,\NakajimaRes,\NakajimaLec}).
This was called BPS/CFT correspondence in \NekrasovTalk, and the AGT correspondence \AldayAQ\ can be considered a particular example.
Our interest in domain walls arose directly in this context.

The paper is organized as follows.
In Section \SecSUSYHemisphere\ we explain our set-up by specifying the geometry and the physical actions.
We analyze the symmetries of the set-up, and define the boundary conditions that preserve B-type supersymmetries.
In particular, we review two basic sets of boundary conditions for a chiral multiplet, which we call Neumann and Dirichlet conditions (for the entire multiplet).
These elementary boundary conditions are combined with the boundary interactions to provide more general boundary conditions.
In Section \SecLoc\ we perform localization and obtain the hemisphere partition function as an integral over scalar zero-modes.
We also provide its alternative expression as a linear combination of certain blocks given as infinite power series.
The geometric interpretation of the hemisphere partition function is explained in Section \SecGeom.
In particular, we explain how to compute the hemisphere partition function for a given object in the derived category.
We give examples of the hemisphere partition functions in Section \SecExamples.
We match the hemisphere partition functions with the large-volume formula for the central charges of D-branes in the quintic Calabi-Yau (and for more general complete intersection Calabi-Yau's in Appendix \AppCICY).
Section \SecDualities\ is devoted to the study of Seiberg-like dualities.
In Section \SecWall\ we study domain walls realized as D-branes in a product theory.
Such domain walls can be regarded as operators that act on a hemisphere partition function.
The action of certain walls are identified with monodromies of the partition function.
We also show that they realize certain defect operators of Toda theories in one case, and the $sl(2)$ affine Hecke algebra in another.
Appendices collect useful formulas and detailed computations.

{\it Note: We were informed by K.~Hori and M.~Romo of their overlapping project \HoriIKA.
We obtained our results independently, except calculations in Appendices \AppCICY\ and \AppFlop\ motivated by their results announced in several talks.
We also learned of a related ongoing work  \SugishitaJCA\ by S.~Sugishita and S.~Terashima.
The three groups coordinated the submission to the arXiv.
}

\newsec{${\cal N}=(2,2)$ theories on a hemisphere}
\seclab\SecSUSYHemisphere

In this section, we review the data for ${\cal N}=(2,2)$ theories and their symmetries.
We also explain the curved 2d geometries to consider, and review the definition of ${\cal N}=(2,2)$ theories on a two-sphere by specifying the physical Lagrangians \refs{\BeniniUI,\DoroudXW}, and modify the set-up by adding a boundary along the equator.
We also describe the boundary conditions, both for vector and chiral multiplets, with which we will perform localization.
We then review another ingredient, the boundary interactions that involve the Chan-Paton degrees of freedom \HerbstJQ.

\subsec{Bulk data for ${\cal N}=(2,2)$ theories}
\subseclab\BulkData

An ${\cal N}=(2,2)$ gauge theory in two dimensions can be thought of as a dimensional reduction of an ${\cal N}=1$ gauge theory in four dimensions, and in particular contains gauge and chiral multiplets.
Such a theory on the curved geometries we study is specified by the data
$$
(G,V_{\rm mat},t,W,m)\,.
$$
The gauge group $G$ is a compact Lie group, and  $V_{\rm mat}$ is the space carrying the matter representation $R_{\rm mat}$; for each irreducible representation $R_a$ in the decomposition
$$
R_{\rm mat}=\oplus R_a\,,
$$
we have a chiral multiplet whose scalar component we call $\phi_a$.
The symbol $t$ denotes a collection of complexified FI parameters.
If the gauge group is $U(N)$, it is given as $t=r-i\theta$, where $r$ is the FI parameter and $\theta$ is the theta angle.
The superpotential $W(\phi)$ is a gauge invariant holomorphic function of $\phi=(\phi_a)$ with R-charge $-2$, in our convention.
The complexified twisted masses $m=(m_a)$ are complex combinations of the real twisted masses
${\rm m}_a$ and the R-charges $q_a$:
$$
m_a=- {1\over 2} q_a - i \ell {\rm m}_a\,.
$$
Here $\ell$ is a length parameter of the geometry.
The vector R-symmetry group%
\foot{%
The axial R-symmetry, which may or may not be anomalous, is broken explicitly by couplings in the action defined on the curved geometries.
} 
$U(1)_{\rm R}$, more precisely its Lie algebra ${\frak u}(1)_{\rm R}$, acts on the fields $\phi_a$ according to the R-charges $q_a$.
If the superpotential is zero, 
$m_a$ are arbitrary complex parameters.
We can regard $m$ as taking values in the complexified Cartan subalgebra of the flavor symmetry group.
When $W$ is non-zero, they are constrained by the condition that for each term in the expansion of $W(\phi)$, $m_a$ for all the fields $\phi_a$ in the term sum to 1.
Correspondingly, the flavor symmetry group $G_{\rm F}$ is smaller than in the $W=0$ case.
A relation between $(m_a)$ and the reduced flavor symmetries will be given in \malpha.

\subsec{Conformal Killing spinors in 2d geometries with boundary}

Our aim is to compute the partition function of an ${\cal N}=(2,2)$ theory on a  hemisphere.
We will argue in Section \SecHilbert\ that the hemisphere partition function computes the overlap of the D-brane boundary state in the Ramond-Ramond sector and a closed string state corresponding to the identity operator.
For this purpose, it is useful to introduce a deformation parameter ($\ell/\tilde\ell$ below) that interpolates between a hemisphere with a round metric and a flat semi-infinite cylinder.
Let us study the conformal Killing spinors in these geometries.

\vskip 5mm
\noindent
{\bf Round hemisphere}
\vskip 3mm

We first consider the hemisphere with the round metric
\eqn\MetricRound{
ds^2=\ell^2 (d\vartheta^2+\sin^2\hskip -0.8mm \vartheta\, d\varphi^2)
}
in the region $0\leq\vartheta\leq \pi/2$, $0\leq\varphi\leq 2\pi$.
The corresponding vielbein are given by $e^{\hat 1}=\ell d\vartheta$, $e^{\hat 2}=\ell \sin\vartheta d\varphi$.
We denote by 
$$\gamma^{\hat 1}=\pmatrix{&1\cr 1&}
\,,
\quad 
\gamma^{\hat 2}=\pmatrix{&-i\cr i&}\,,
\quad
 \gamma^{\hat 3}=\gamma^3=\pmatrix{1&\cr &1}
$$ 
the usual Pauli matrices.
The conformal Killing spinor equations%
\foot{%
The non-zero component of the spin connection is
$\omega_{\hat 1\hat 2}=-\cos\vartheta d\varphi$,
and the covariant derivatives acting on a spinor are given by
$\nabla_\vartheta=\partial_\vartheta$, $\nabla_\varphi=\partial_\varphi-{i\over 2} \cos\vartheta \gamma^3$.
Note that $\tilde \epsilon=(1/2)\gamma^\mu\nabla_\mu\epsilon$.
}
$$
\nabla_\mu \epsilon=\gamma_\mu \tilde\epsilon
$$
have four independent solutions
\eqn\CKSRound{
\epsilon=
e^{-s{i\over 2}\vartheta \gamma_{\hat 2}} \pmatrix{e^{{i\over 2}\varphi} \cr 0}\,,
\quad
e^{-s{i\over 2}\vartheta \gamma_{\hat 2}} \pmatrix{0 \cr e^{-{i\over 2}\varphi}}\,,
}
with $s=\pm 1$.
The SUSY transformations on a round sphere were constructed in \refs{\BeniniUI,\DoroudXW}.
In our convention, these are obtained by taking $\tilde\ell=\ell$ in \DeformedVec\ and \DeformedChi.
The SUSY parameters $\epsilon$ and $\bar\epsilon$ that appear there are conformal Killing spinors, each having four independent solutions.
They parametrize the superconformal algebra on round $\Bbb S^2$, which contains eight fermionic charges.
The ${\cal N}=2$ SUSY algebra $SU(2|1)$ on $\Bbb S^2$, which does not contain dilatation and is compatible with masses, is generated by the spinors $\epsilon$ with $s=1$ and $\bar\epsilon$ with $s=-1$.
Thus $SU(2|1)$ contains four fermionic generators.
The boundary at $\vartheta=\pi/2$, however, breaks the isometry from $SU(2)$ to $U(1)$.
Thus we restrict to the subalgebra $SU(1|1)$ generated by two fermionic charges $\delta_\epsilon$ and $\delta_{\bar\epsilon}$ given by
\eqn\EqTwoSpinors{
\epsilon=
e^{-{i\over 2}\vartheta \gamma_{\hat 2}} \pmatrix{e^{{i\over 2}\varphi} \cr 0}\,,
\quad
\bar\epsilon=
e^{{i\over 2}\vartheta \gamma_{\hat 2}} \pmatrix{0 \cr e^{-{i\over 2}\varphi}}\,.
}
The isometry that appears in $\{\delta_\epsilon,\delta_{\bar\epsilon}\}$ shifts $\varphi$ by a constant and preserves the boundary.

Note that the spinors in \EqTwoSpinors\ are anti-periodic in $\varphi$.
Since bosons are periodic, fermions are all anti-periodic.
We will see in Section \SecBdryInt\ that there is a natural field redefinition that makes all the fields periodic in $\varphi$ along the boundary.

\vskip 5mm
\noindent
{\bf Deformed hemisphere}
\vskip 3mm

We will also consider the deformed metric \GomisWY
\eqn\MetricDeformed{\eqalign{
ds^2&\equiv h_{\mu\nu}dx^\mu dx^\nu = f^2(\vartheta)d\vartheta^2+ \ell^2
 \sin^2\vartheta d\varphi^2\,,
}}
where $f^2(\vartheta)=\ell^2 \cos^2\vartheta+\tilde \ell^2 \sin^2\vartheta$.
If we introduce the non-dynamical gauge field
\eqn\GaugeR{
V^{\rm R}={1\over 2} \left(1-{\ell \over f(\vartheta)}\right) d\varphi
}
for $U(1)_{\rm R}$, the spinors \EqTwoSpinors\ satisfy 
\eqn\CKSEq{
D_\mu\epsilon= {1\over 2f}\gamma_\mu\gamma_3\epsilon\,,
\qquad
D_\mu\bar\epsilon= -{1\over 2f}\gamma_\mu\gamma_3\bar\epsilon\,,
}
where the covariant derivatives act as $D_\mu \epsilon =(\nabla_\mu -i V^{\rm R}_\mu)\epsilon$, $D_\mu \bar\epsilon =(\nabla_\mu +i V^{\rm R}_\mu)\bar\epsilon$.
We assigned R-charges $+1$ and $-1$ to $\epsilon$ and $\bar\epsilon$ respectively.
These spinors generate the superalgebra $SU(1|1)$, which contains the isometry $U(1)$ that is compatible both with the deformed metric and the boundary $\vartheta=\pi/2$.
The corresponding fermionic transformations are listed in \DeformedVec\ and \DeformedChi.%
\foot{%
These formulas are essentially taken from \GomisWY\ except that we flip the sign of $q$.
}

\vskip 5mm
\noindent
{\bf Half-infinite cylinder}
\vskip 3mm

In the limit $\tilde\ell\rightarrow \infty$, the region near $\vartheta=\pi/2$ becomes a half-infinite cylinder; by replacing $\vartheta$ with $x=-\tilde\ell\cos\vartheta$, the deformed metric becomes
$$\eqalign{
ds^2
&=
dx^2 
+\ell^2 d\varphi^2
}$$
in the limit.
This geometry is flat, and the SUSY algebra gets enhanced.

\subsec{${\cal N}=(2,2)$ theories on a deformed hemisphere}

We now give the precise construction of an ${\cal N}=(2,2)$ theory on the deformed hemisphere for the data $(G,V_{\rm mat},t,W,m)$ defined in Section \BulkData.

The gauge multiplet for gauge group $G$ consists of the gauge field $A_\mu$, real scalars $\sigma_{1,2}$, gauginos $\lambda$, $\bar\lambda$, and the real auxiliary field ${\rm D}$.
Let us define
$$
\delta_Q\equiv \delta_\epsilon+ \delta_{\bar\epsilon}\,,
$$
where the SUSY transformations $\delta_\epsilon$ and $\delta_{\bar\epsilon}$ are given in \DeformedVec\ and \DeformedChi.
On a full deformed sphere the physical Lagrangian for a vector multiplet is \GomisWY
\eqn\LVecExact{
{\cal L}_{\rm vec}^{\rm exact}\equiv {1 \over g^2}\delta_Q\delta_{\bar\epsilon}\Tr \left(
{1\over 2}\bar\lambda \gamma^3 \lambda
-2i{\rm D}\sigma_2
+{i \over f(\vartheta)} \sigma_2^2
\right)\,.
}
See Appendix \SUSYTrans\ for our spinor conventions.
In general we can introduce a coupling $g$ for each simple or abelian factor in $G$.
Noting that $\delta_Q^2$ is a bosonic symmetry one can show that \LVecExact\ is  invariant under $\delta_Q$.
This Lagrangian can be written, up to total derivative terms, as
$$\eqalign{
\CL^{\rm bulk}_{\rm vec}&\equiv {1 \over 2g^2}\Tr\bigg[\left( F_{\1\2}+{\sigma_1 \over f}\right)^2 +D_\mu\sigma_1 D^\mu\sigma_1+D_\mu\sigma_2 D^\mu\sigma_2 -[\sigma_1,\sigma_2]^2+\D^2
\cr
&\qquad \qquad \qquad \qquad \quad
-{i \over 2}(D_\mu \bar\lambda \gamma^\mu \lambda -  \bar\lambda \gamma^\mu D_\mu\lambda) +i\bar\lambda[\sigma_1,\lambda]+\bar\lambda\gamma^3[\sigma_2, \lambda]\bigg]\,.
}$$
Since we are interested in manifolds with boundary it is important to keep the total derivative terms.
After some calculations, we obtain
$$
\int d^2x \sqrt{h}\,\CL^{\rm exact}_{\rm vec}=\int d^2x \sqrt{h}\,\CL^{\rm bulk}_{\rm vec} + \oint_{\vartheta={\pi\over 2}} d\varphi\, \CL^{\rm bdry}_{\rm vec}\,,
$$
where%
\foot{%
For general values of $\vartheta$, 
$\CL^{\rm exact}_{\rm vec}=
\CL^{\rm bulk}_{\rm vec} + (1/ g^2)D_\mu {\rm Tr}\big[-i\bar\epsilon\gamma^\mu\gamma^m\epsilon \CV_m \sigma_2 +(i/ 2)(\bar\lambda\gamma^3\epsilon)\bar\epsilon \gamma^\mu \lambda
+ \varepsilon^{\mu\nu} \sigma_1 D_\nu \sigma_2
+\bar\epsilon\gamma^\mu\epsilon \D\sigma_2
-\displaystyle (i/ 4)\bar\lambda \gamma^\mu\lambda
\big]$
and 
$\displaystyle \CL^{\rm exact}_{\rm chi}=
\CL^{\rm bulk}_{\rm chi} +D_\mu\big[i\varepsilon^{\mu\nu } \bar\epsilon\epsilon \bar\phi D_\nu \phi
+\bar\epsilon \gamma^3 \gamma^\mu \epsilon \bar\phi \sigma_1 \phi
+\bar\epsilon \gamma^\mu \epsilon \bar\phi \sigma_2 \phi
{-}\bar\epsilon \gamma^\mu \epsilon (q/ 2f) \bar\phi \phi
+i(\epsilon\bar\psi)\bar\epsilon\gamma^\mu \gamma^3 \psi
-(i/ 2)\bar\psi\gamma^\mu\psi
\big]\,.
$
}
$$\eqalign{
\CL^{\rm bdry}_{\rm vec}
&=
{1 \over g^2}\Tr
\bigg[
-{i \ell\over \tilde\ell} \sigma_2 D_1\sigma_2
+i\ell\left(F_{\1\2}+{1 \over \tilde\ell}\sigma_1\right)\sigma_2
+{i\ell \over 4 }(\bar\lambda_1 \lambda_2 -\bar\lambda_2 \lambda_1)
\bigg]\,.
}$$

A chiral multiplet consists of a complex scalar $\phi$, a fermion $\psi$, a complex auxiliary field ${\rm F}$, and their conjugate.
If the R-charge of $\phi$ is $q$, those of $\psi$ and ${\rm F}$ are $q+1$ and $q+2$ respectively.
The Lagrangian
\eqn\LChiExact{
\CL^{\rm exact}_{\rm chi} \equiv 
\delta_Q \delta_{\bar\epsilon} \left( -\bar\psi \gamma^3 \psi
+2\bar\phi\left(\sigma_2 -i{q +1\over 2 f} \right)\phi \right)\,,
}
has the structure
$$
\int d^2x \sqrt{h}\,\CL^{\rm exact}_{\rm chi}=\int d^2x \sqrt{h}\,\CL^{\rm bulk}_{\rm chi} + \oint_{\vartheta={\pi\over 2}} d\varphi\, \CL^{\rm bdry}_{\rm chi}\,,
$$
with
\eqn\LChiBulk{\eqalign{
\CL^{\rm bulk}_{\rm chi}& \equiv
\bigg[
D_\mu \bar\phi D^\mu\phi+\bar\phi\left(\sigma_1^2+\sigma_2^2 -i{q +1 \over f}\sigma_2-{q^2 \over 4 f^2} -{q \over 4}{\cal R}\right)\phi
+\bar\F \F+i\bar\phi {\rm D}\phi
\cr
&\quad
+{i \over 2}(D_\mu\bar\psi \gamma^\mu \psi
-\bar\psi \gamma^\mu D_\mu \psi)
+\bar\psi \big( i\sigma_1-\left(\sigma_2 -{i q \over 2 f}\right)\gamma^3 \big)\psi
+i\bar\psi\lambda\phi -i\bar\phi\bar\lambda\psi\bigg]\,,
}}
and
$$
\CL^{\rm bdry}_{\rm chi}
=
\ell \left[\bar\phi\, \sigma_1 \phi +i\bar\psi\Big (1+{\gamma_\1\over 2} \Big) \psi \right]\,,
$$
where ${\cal R}$ is the scalar curvature.
The twisted mass ${\rm m}$ can be introduced by the replacement $\sigma_2 \rightarrow \sigma_2 +{\rm m}$.
In general the action involves an arbitrary number of chiral fields $\phi_a$ with R-charge $q_a$ and twisted mass ${\rm m}_a$.

If the gauge group $G$ contains an abelian factor we should also include the topological term.
For $G=U(N)$ this is $-i (\theta / 2\pi)\int {\rm Tr}\,F$, which on the hemisphere is a Wilson loop.
It should be supersymmetrized into
\eqn\Stheta{
S_\theta \equiv -{\theta \over 2\pi}\oint _{\vartheta={\pi \over 2}}{\rm Tr}  \left(iA_\varphi -\ell \sigma_2\right)d\varphi\,.
}
This is further supplemented by the Fayet-Iliopoulos (FI) term
\eqn\SFI{
S_{\rm FI}\equiv -i {r\over 2\pi} \int d^2x \sqrt h\, {\rm Tr} \left({\rm D}- {\sigma_2 \over f}\right)\,.
}
Both $S_\theta$ and $S_{\rm FI}$ are invariant under $\delta_Q$ by themselves.

Finally, if the superpotential $W(\phi)$ is non-zero we also have%
\eqn\LW{
{\cal L}_{ W}
= -{ i\over 2}\Big ({\rm F}^i \partial_i { W}-{1\over 2}\psi^i\psi^j \partial_i\partial_j{ W}\Big)-{ i\over 2}
\Big(\bar {\rm F}_i \bar\partial^i \bar{ W} - {1\over 2} \bar\psi_i\bar\psi_j \bar\partial^i \bar\partial^j\bar { W}
\Big)\,.
}
Here  $\phi^i$ collectively denote the components of $\phi=(\phi_a)$.
Noting that $W$ is gauge invariant with R-charge $-2$, one can show that its variation is a total derivative
\eqn\Warner{
\delta_{ Q} {\cal L}_{{W}}= {1\over 2} D_\mu\left(
 \epsilon  \gamma^\mu  \psi^i   \partial_i { W}
+
\bar\epsilon \gamma^\mu\bar\psi_i\bar\partial^i\bar { W}
\right)\,,
}
known as the Warner term \WarnerAY.
This needs to be cancelled by the SUSY variation of the boundary interaction that we will discuss in Section \SecBdryInt.

We define our supersymmetric theory by the functional integral of
$$
 \exp(-S_{\rm phys})\times ({\rm boundary\ interaction})
$$
with the total physical action
\eqn\SPhys{
S_{\rm phys}\equiv
\int d^2x \sqrt{h}\left(\CL^{\rm bulk}_{\rm vec} + \CL^{\rm bulk}_{\rm chi} +\CL_{{ W} }\right)
+S_\theta + S_{\rm FI}\,.
}
For the theory to be supersymmetric, the total integrand has to be invariant under supersymmetry transformations.
We focus on the supercharge ${ Q}$ of our choice.
For the vector multiplet we need to impose such boundary conditions that annihilate $\delta_{Q} \int \sqrt h\,{\cal L}^{\rm bulk}_{\rm vec}= - \delta_{Q} \oint d\varphi\, {\cal L}^{\rm bdry}_{\rm vec}$.
Similarly $ \delta_{Q} \oint d\varphi\, {\cal L}^{\rm bdry}_{\rm chi}$ must vanish under the boundary conditions for chiral multiplets.
In Section \SecBoundary\ we will see that the boundary conditions introduced in \HerbstJQ\ do the job.
We will also see there, following \HerbstJQ, that the Warner term \Warner\ can be cancelled by a suitable boundary interaction.

\subsec{Basic boundary conditions for vector and chiral multiplets}
\subseclab\SecBoundary

Let us introduce several basic boundary conditions that are compatible with the supercharge $Q$.
These are straightforward generalizations of the boundary conditions found in \HerbstJQ\ for abelian gauge groups.

\vskip 5mm
\noindent
{\bf Vector multiplets}
\vskip 3mm

The boundary condition for a vector multiplet we consider in this paper%
\foot{%
The boundary condition \EqBCPreserve\ preserves the full gauge symmetry $G$ along the boundary.
It should also be possible to formulate a boundary condition that preserves a subgroup $H$, as in \GaiottoSA.
} 
consists of the following set of boundary conditions on the component fields at $\vartheta=\pi/2$:
\eqn\EqBCPreserve{\eqalign{
&\sigma^{}_1=0\,,\quad D_1 \sigma_2=0\,, \quad A_1=0\,,\quad F_{12}=0\,,
\cr
&\bar\epsilon \lambda=\epsilon\bar\lambda=0\,,
\quad 
D_1(\bar\epsilon\gamma_3\lambda) = D_1(\epsilon\gamma_3\bar\lambda) =0\,,
\cr
& 
\hskip 18mm
D_{\hat 1}({\rm D} -i D_{\hat 1} \sigma_1)=0\,.
}}
The term ${\cal L}^{\rm bdry}_{\rm vec}$ vanishes with this condition imposed.
In particular we have $ \delta_{Q} \oint d\varphi\, {\cal L}^{\rm bdry}_{\rm vec}=0$, as needed for preserving ${Q}$.

\vskip 5mm
\noindent
{\bf Chiral multiplets}
\vskip 3mm

For a chiral multiplet, we study two sets of boundary conditions for the component fields at $\vartheta=\pi/2$.
The {\it Neumann boundary condition} for a chiral multiplet is given by
\eqn\EqNeumann{\eqalign{
&
\hskip 20mm
D_1\phi=D_1\bar\phi=0\,, 
\cr
& \bar\epsilon \gamma_3 \psi=\epsilon\gamma_3 \bar \psi=0\,,  
\quad
D_1(\bar\epsilon \psi)=D_1(\epsilon \bar \psi)=0\,,
\cr
&
\hskip 25mm
 {\rm F}=0\,. 
}}
Chiral multiplets with this boundary condition describe the target space directions tangent to a submanifold wrapped by the D-brane.
In particular, for space-filling D-branes all the chiral multiplets obey the Neumann boundary condition.
The {\it Dirichlet boundary condition} for a chiral multiplet is given by%
\foot{%
After the field redefinition \EqRedef, the last line simply reads $D_1 ({\rm F}^{\rm new}+iD_{\1}\phi^{\rm new})=0$.
}
\eqn\EqDirichlet{\eqalign{
 &
\hskip25mm
\phi=\bar\phi=0\,,
 \cr
 &\bar\epsilon \psi=\epsilon \bar \psi=0\,,
\quad
D_1(\bar\epsilon\gamma_3 \psi)=D_1(\bar\epsilon\gamma_3\bar\psi)=0\,,
\cr
&
\hskip15mm
D_1 (e^{-i\varphi}{\rm F}+iD_{\1}\phi)=0\,.
}}
The complex scalar field $\phi$ parametrizes a direction normal to a submanifold.
In either case the boundary condition implies that ${\cal L}^{\rm bdry}_{\rm chi}=0$, ensuring that $\delta_{Q} \oint d\varphi\, {\cal L}^{\rm bdry}_{\rm chi}=0$.

We will see in Section \SecSheaves,  generalizing an argument in the abelian case studied by \HerbstJQ, that any lower dimensional D-brane can be described as a bound state of space-filling D-branes carrying Chan-Paton fluxes.

\subsec{Boundary interactions}
\subseclab\SecBdryInt

Following \HerbstJQ, we now introduce supersymmetric boundary interactions that will play an important role.
First we introduce the Chan-Paton vector space
$$
{\cal V}= {\cal V}^{\rm e} \oplus {\cal V}^{\rm o}\,.
$$
This is $\Bbb Z_2$-graded, and accordingly ${\rm End}({\cal V})$ can be given the structure of a superalgebra.
The space of fields is also a superalgebra, and (by implicitly taking the tensor product of superalgebras), we can make fermions anti-commute with odd linear operators acting on ${\cal V}$.
The boundary interaction  will be constructed using a conjugate pair of odd operators ${\cal Q}(\phi)$ and $\bar{\cal Q}(\bar\phi)$, called a {\it tachyon profile}.
These are respectively polynomials of $\phi$ and $\bar\phi$, and must satisfy the conditions we describe below.

Gauge group $G$, flavor group $G_{\rm F}$, and the vector R-symmetry group $U(1)_{\rm R}$ act on the space ${\cal V}$.
In other words, there is a representation, or equivalently a homomorphism%
\foot{%
More precisely, we allow $\rho$ to be a projective representation.
See Sections \BulkData\ and \BCcomplexes.
We denote the induced representation of the Lie algebra by $\rho_*$.
}
$$
\rho: G\times G_{\rm F}\times U(1)_{\rm R} \rightarrow {\rm End}({\cal V})\,.
$$
We demand that the tachyon profile is invariant under $G$ and $G_{\rm F}$:
\eqn\GaugeInvQ{
\rho(g){\cal Q}(g^{-1}\cdot \phi) \rho(g)^{-1}={\cal Q}(\phi) \,, 
\qquad
\rho(g)\bar{\cal Q}(\bar\phi\cdot g) \rho(g)^{-1}={\cal Q}(\bar \phi) 
}
for $g\in G\times G_{\rm F}$.
For the R-symmetry, let us denote the generator by $R$.
It acts on a chiral multiplet $\phi_a$, in the notation of Section \BulkData, as
\eqn\GenR{
R\cdot \phi_a=q_a \phi_a\,,
}
where $q_a$ is the R-charge.
We require that the tachyon profile satisfies the conditions
\eqn\RChargeQ{\eqalign{
&\rho(e^{i\alpha R}) {\cal Q}(e^{-i\alpha R}\cdot \phi) \rho(e^{-i\alpha R})
= e^{i\alpha}{\cal Q}(\phi)
 \,,
\cr
&
\rho(e^{i\alpha R}) \bar{\cal Q}(\bar \phi\cdot e^{i\alpha  R}) \rho(e^{-i\alpha R})
= e^{-i\alpha}\bar {\cal Q}(\bar \phi)
\,.
}}

We can now define the {\it boundary interaction} \refs{\HellermanBU,\HerbstJQ},  an ${\rm End}({\cal V})$-valued 1-form along the boundary circle at $\vartheta=\pi/2$:
\eqn\EqBoundaryInt{\eqalign{
{\cal A}_{\hat \varphi}&=
\rho_*( A_{\hat \varphi} + i \sigma_2 )
+ {\rho_{*}(R)\over 2\ell}
+i \rho_{*}({\rm m})
\cr
&\qquad
+ {i\over 2}\{{\cal Q},\bar {\cal Q}\}
+{1\over2}(\psi_1-\psi_2)^i
 \partial_i {\cal Q}+{1\over 2} (\bar\psi_1-\bar\psi_2)_i \partial^i \bar {\cal Q}\,.
}}
Here the representation $\rho_*$  of the Lie algebra of $G\times G_{\rm F}\times U(1)_{\rm R}$ is induced from $\rho$.
In the path integral we include
\eqn\BoundaryIntExp{
{\rm Str}_{\cal V}\hskip -1mm \left[ P \exp\left(i \oint d\varphi{\cal A}_{ \varphi}\right)\right]
\,.
}
As in \refs{\KapustinBI,\HerbstJQ}, one can show with some calculations that the $Q$ variation of the boundary interaction ${\cal A}_{\hat\varphi}$ cancels the Warner term $\delta_{ Q} {\cal L}_{{ W}}$ in \Warner,
$$\eqalign{
&\,\quad \delta_{ Q}
{\rm Str}_{\cal V}\hskip -1mm \left[P e^{i \oint d\varphi{\cal A}_{ \varphi}}e^{-\int d^2x \sqrt{h} {\cal L}_{{ W}}}
\right]
\cr
&=
{\rm Str}_{\cal V}\hskip -1mm \left[P e^{i \oint d\varphi{\cal A}_{ \varphi}}e^{-\int d^2x \sqrt{h} {\cal L}_{{ W}}}
\left(
i\oint d\varphi  \,\delta_{ Q} {\cal A}_{\varphi}
-\int d^2x \sqrt{h}\,\delta_{Q} {\cal L}_{{W}}\right)
\right]
\cr
&=0\,,
}$$
 if ${\cal Q}$ and $\bar{\cal Q}$ satisfy
\eqn\MatFac{
{\cal Q}^2={ W}\cdot {\bf 1}_{\cal V}\,,
\qquad
\bar {\cal Q}^2=\bar {W}\cdot {\bf 1}_{\cal V}\,.
}
When the conditions \MatFac\ are satisfied, we say that 
the tachyon profile ${\cal Q}$ 
is a {\it matrix factorization} of the superpotential $W$.
The boundary interaction \EqBoundaryInt\ allows us to construct interesting supersymmetric theories on a hemisphere.

In order to compare \EqBoundaryInt\ with \HerbstJQ, it is useful to introduce a version of vector R-symmetry group (in general distinct from the original) and perform a field redefinition.
This will also be important to understand the target space interpretation in Section  \SecHilbert.

Consider first the case $W=0$.
Because an R-symmetry mixed with flavor symmetries%
\foot{%
Mixing with gauge symmetries plays no role, so we exclude the possibility from discussion.
}
 is also an R-symmetry, we can define a new R-symmetry by
$$
R_{\rm deg}= R- q_a F^a\,,
$$
where $F^a$ are the flavor generators (for $W=0$) such that
$$
F^a\cdot \phi_b=\delta^a_b \phi_b\,.
$$
The R-charges for the new R-symmetry for all $\phi_a$ vanish, and those of the superpartners $\psi_a$ and  ${\rm F}_a$ are $+1$ and $+2$, respectively.
The first condition in \RChargeQ\ applied to $R_{\rm deg}$ implies that the tachyon profile ${\cal Q}$ increases the eigenvalue of $R_{\rm deg}$ by one: $[\rho_{*}(R_{\rm deg}),{\cal Q}]={\cal Q}$.
We require that the eigenvalues of $R_{\rm deg}$ in  ${\cal V}$ are all integers.
Then we can decompose ${\cal V}$ into the eigenspaces ${\cal V}^i$ of $R_{\rm deg}$ with eigenvalue $i$.
Since $W=0$, ${\cal Q}$ defines a differential of the cochain complex
$$
\ldots 
\longrightarrow
{\cal V}^i
\longrightarrow
{\cal V}^{i+1}
\longrightarrow
\ldots
$$

Whether $W$ is zero or not, we will require that there is an R-symmetry generator $R_{\rm deg}$ that has only even (odd) integer eigenvalues in ${\cal V}^{\rm e}$ (respectively ${\cal V}^{\rm o}$), and even integer eigenvalues ${\rm d}_a$ on $\phi_a$.
Any such generator is related to the previous R-symmetry generator $R$ as
\eqn\RdegGeneral{
R_{\rm deg}= R- q_\alpha F^\alpha\,,
}
where $F^\alpha$ are the Cartan generators of the flavor group $G_{\rm F}$ preserved by $W$, and $q_\alpha$ take real values.
As we will see in Section \BCcomplexes, there is a natural choice of $R_{\rm deg}$ when the gauge theory flows to a non-linear sigma model.
Using ${\rm d}_a$, we can parametrize the complexified twisted masses by the Cartan of $G_{\rm F}$ as
$ m_a=-(1/ 2) {\rm d}_a + m_\alpha (F^\alpha)_a$, where%
\foot{%
The symbols $(q_\alpha, F^\alpha, m_\alpha)$, labeled by the directions $\alpha$ in the Cartan of $G_{\rm F}$,  should be distinguished from $(q_a,F^a,m_a)$ labeled by $a$ parametrizing irreducible matter representations.
The term $-(1/2){\rm d}_a$ in $m_a$ is analogous to a shift in the 4d mass on $\Bbb S^4$ noticed in \OkudaKE.
}
\eqn\malpha{
m_\alpha=-{1\over 2} q_\alpha -i\ell {\rm m}_\alpha\,.
}
When the superpotential $W$ breaks all flavor symmetries, $m_a$ are simply R-charges rescaled, $m_a=-{\rm d}_a/2$.

Let us consider the simultaneous redefinition
\eqn\EqRedef{
\Phi(\vartheta,\varphi)
\rightarrow 
\Phi^{\rm new}(\vartheta,\varphi)
=
e^{-{i\over 2}R_{\rm deg}\varphi}
\cdot \Phi(\vartheta,\varphi)
}
of all the bosonic and fermionic fields $\Phi$ in the theory.
Since we demanded that $R_{\rm deg}$ has even integers as eigenvalues on the scalars $\phi_a$, bosonic fields remain periodic while fermions become periodic from anti-periodic.

In the new description, which is valid in the neighborhood of the boundary, the background gauge field \GaugeR\ for (the original) $U(1)_{\rm R}$ is shifted as
\eqn\EqRBackground{
V^{\rm R} \rightarrow V^{\rm R,new}=V^{\rm R}-{1\over 2}d\varphi=-{\ell\over 2 f(\vartheta)}d\varphi\,.
}
In addition, the field redefinition induces an extra background gauge field for the flavor symmetry:
\eqn\FlavorBack{
V^{\rm F}=   {1\over 2} q_\alpha F^\alpha d\varphi\,.
}
The full covariant derivative
$$
D_\mu= \nabla_\mu -i A_\mu -i V_\mu  R
$$
becomes
$$
D^{\rm new}_\mu =\nabla_\mu 
-i A_\mu -i V^{\rm R,new}_\mu R -i V^{\rm F}_\mu\,.
$$
If we apply the redefinition to SUSY parameters, they become at $\vartheta=\pi/2$
\eqn\SpinorsFlat{
\epsilon_{\rm flat}= {1\over\sqrt 2}
\pmatrix{1 \cr 1}\,,
\quad
\bar\epsilon_{\rm flat}= {1\over\sqrt 2}
\pmatrix{1 \cr 1}\,.
}
Each spinor gives rise to a linear combination of left- and right-moving, barred or unbarred, supercharges.
Thus they correspond to the B-type supersymmetries \WittenZZ.

The field redefinition \EqRedef\ removes from ${\cal A}_{\hat\varphi}$ the R-symmetry background and induces a flavor background \FlavorBack, with ${\cal Q}$ and ${\bar {\cal Q}}$ redefined in a natural way:
\eqn\BoundaryIntNew{
{\cal A}^{\rm new}_{\hat \varphi}=
\rho_*( A_{\hat \varphi} + i \sigma_2 )
+\rho_{*}(V^{\rm F}_{\hat \varphi} + i {\rm m})
+{i\over 2}\{{\cal Q}^{\rm new},\bar {\cal Q}^{\rm new}\} +\ldots\,.
}
This expression agrees with the interaction found in \HerbstJQ\ when the flavor part is taken into account.

Let us summarize Sections \SecBoundary\ and \SecBdryInt.
Given a theory specified by the bulk data $(G,V_{\rm mat},t,W,m)$, we can define a boundary condition ${\cal B}$, or a {\it D-brane}, by the data
$$
{\cal B}=(
{\bf Neu}, {\bf Dir}, {\cal V}, {\cal Q})\,.
$$
The vector multiplet obey the boundary condition \EqBCPreserve.
The symbols ${\bf Neu}$ and ${\bf Dir}$ denote that set of chiral multiplets that obey the Neumann and the Dirichlet boundary conditions \EqNeumann\ and \EqDirichlet, respectively.
We will often assume that ${\bf Dir}=\emptyset$ and simply write ${\cal B}=({\cal V},{\cal Q})$.
The Chan-Paton space ${\cal V}={\cal V}^{\rm e}\oplus {\cal V}^{\rm o}$ is $\Bbb Z_2$-graded and carries a representation of $G\times G_{\rm F}\times U(1)_{\rm R}$.
It must admit a new R-symmetry generator $R_{\rm deg}$ that is a mixture of the original R-symmetry (encoded in $m$) and flavor symmetries, and has integer eigenvalues on ${\cal V}$ that descend to the $\Bbb Z_2$-grading.
The tachyon profile ${\cal Q}$ is a matrix factorization of $W$, {\it i.e.}, an odd linear operator on ${\cal V}$ that squares to  $W \cdot {\bf 1}_{\cal V}$.

\newsec{Localization on a hemisphere}
\seclab\SecLoc
\subsec{Localization action and locus}
\subseclab\SecLocAc

In a supersymmetric quantum field theory, we know a priori that the path integral receives contributions from the field configurations that are annihilated by the supercharges.%
\foot{%
One of the early references that discusses this explicitly is \WittenMK.
}
Moreover, if the locus of such invariant configurations is finite dimensional, the path integral can be exactly performed by evaluating the one-loop determinant in the normal directions.
This statement holds for any action that preserves supersymmetry as long as its behavior for large values of fields is reasonable.

Though the one-loop determinant depends on the choice of the action,
there is still redundancy; if the action is modified by adding an exact term, the one-loop determinant does not change by the standard argument.
In the following, we will use \LVecExact\ and \LChiExact\ to define the localization action
\eqn\Sloc{
S_{\rm loc}\equiv \int d^2x\sqrt h ({\cal L}^{\rm exact}_{\rm vec}+{\cal L}^{\rm exact}_{\rm chi})\,.
}
Namely, we will consider the path integral
$$
Z_{\rm hem}\equiv \int [DA_\mu\ldots D\phi\ldots]
\,{\rm Str}_{\cal V} \hskip -1mm \left[
P 
 \exp\left(i \oint d\varphi{\cal A}_{ \varphi}\right)
\right]
\exp\left(-S_{\rm phys} -{\rm t} S_{\rm loc}\right)\,,
$$
where the boundary interaction ${\cal A}_\varphi$ and the physical action $S_{\rm phys}$ are defined in \EqBoundaryInt\ and \SPhys, respectively.
Since $S_{\rm loc}$ is $Q$-exact, the path integral is independent of ${\rm t}$.
We evaluate the path integral in the limit ${\rm t}\rightarrow +\infty$; the one-loop determinant can be obtained from the quadratic part of $S_{\rm loc}$.

For a generic assignment of R-charges, the localization locus for the theory on a (deformed) two-sphere was determined in \refs{\BeniniUI,\DoroudXW,\GomisWY}.
On the hemisphere with the symmetry-preserving boundary condition \EqBCPreserve, we have a further simplification that the flux $B$ vanishes.
Then the only non-vanishing field in the locus is
\eqn\LocusHemisphere{
\sigma_2={\rm const}\,.
}
In this locus, the physical action $S_{\rm phys}$ contributes to the path integral
\eqn\EqVectorClassical{
e^{-i\ell t{\rm Tr}\,\sigma_2}\,,
}
which comes from 
$S_{\theta}$ in \Stheta\ and $S_{\rm FI}$ in \SFI.
Here we have set $t=r - i\theta$.
As part of the classical contribution, we also need to evaluate the supertrace \BoundaryIntExp.
It is most cleanly evaluated using the expression \BoundaryIntNew\ after the field redefinition \EqRedef.
In the localization locus \LocusHemisphere, the supertrace becomes
\eqn\EqBdryCont{
{\rm Str}_{\cal V} 
\left[
e^{- 2\pi \ell\rho_*( \sigma_2)}
e^{- 2\pi i\rho_{*}( - {1\over 2}q_\alpha F^\alpha -i\ell{\rm m})}
\right]
=
{\rm Str}_{\cal V} 
\left[
e^{- 2\pi i\rho_{*}(\sigma+m_\alpha F^\alpha)}
\right]
\,.
}
where we defined
$\sigma=-i\ell \sigma_2$.
In most of the paper we will simply write \EqBdryCont\ as
$$
{\rm Str}_{\cal V} 
\left[
e^{- 2\pi i(\sigma+ m)}
\right]\,.
$$

\subsec{One-loop determinants}
\subseclab\SecOneLoop

In this section we compute the one-loop determinant for the saddle point configuration \LocusHemisphere.
Because the computations are easier for chiral multiplets than for vector multiplets, we first treat the former.
For simplicity we work with the round metric \MetricRound\ and suppress $\ell$ during computations.

Let us consider a chiral multiplet in a representation $R$ of the gauge group.
Around the localization locus \LocusHemisphere, the chiral multiplet part of the localization action \Sloc\ reads, to the quadratic order,
$$\eqalign{
S_{\rm chi}^{(2)}&=\int d^2x \sqrt h\Big[\bar\phi\Big({\rm M}^2  -i(q+1)\sigma_2 -{q^2 +2q \over 4 } \Big)\phi +\bar\F \F 
-\bar\psi \gamma^3\Big( i \gamma^3\gamma^\mu D_\mu +\sigma_2 -{i q \over 2 } \Big)\psi\Big]\,,
}$$
where
$$
{\rm M}^2\equiv-D^\mu D_\mu +\sigma_2^2\,.
$$
The Gaussian integral over ${\rm F}$ and $\bar{\rm F}$ does not depend on any parameter and will be ignored.
As we show in Appendix \EigenProb, the Dirac operator in the particular combination $\gamma^3\gamma^\mu D_\mu$ is self-adjoint on the hemisphere---the naive one $i\gamma^\mu D_\mu$ is not---when the relevant boundary conditions are imposed on the spinors.

Let us denote the weights of $R$ by $w$.
To avoid clutter we assume that each weight $w$ has multiplicity $1$; it is trivial to drop the assumption.
Each field can be expanded in an orthonormal basis consisting of weight vectors $e_w$  such that $\sigma_2\cdot e_w= w(\sigma_2)e_w$.
We write $\bar e^w\equiv (e_w)^\dagger$.
Using the scalar spherical harmonics $Y_{jm}$ and the spinor harmonics $\chi^\pm_{jm}(\vartheta,\varphi)$ reviewed in Appendix \AppMonoHarm, we expand
\eqn\ExpansionChiral{\eqalign{
&
\phi=\sum_w\sum^\infty_{j=0} \mathop{\sum{}'}^j_{m=-j} \phi^w_{jm}Y_{jm}(\vartheta,\varphi)e_w\,,
\qquad\ \,\,
\bar\phi=
\sum_w
\sum^\infty_{j=0} \mathop{\sum{}'}^j_{m=-j} (\phi^w_{jm})^* Y_{jm}(\vartheta,\varphi)^*\bar e^w
\,,
\cr
&
\psi=\sum_w\sum_{s=\pm} \sum^\infty_{j={1 \over 2}}\mathop{\sum{}'}^j_{m=-j}\psi^{ws}_{jm} \chi^s_{jm}(\vartheta,\varphi) e_w\,,
\
\bar \psi=\sum_w\sum_{s=\pm} \sum^\infty_{j={1 \over 2}} \mathop{\sum{}'}^j_{m=-j}\bar \psi^{s}_{wjm} \chi^s_{jm}(\vartheta,\varphi) \bar e^w\,.
}}
The symbol $\Sigma'$ indicates that the sum is restricted to such $m$ that
$$
j - m = \left\{\matrix{
{\rm even} & {\rm for}& \phi \ {\rm and} \ \bar\phi\,,
\cr
{\rm odd} & {\rm for}& s=+ \ {\rm in}\ \psi \ {\rm and} \ \bar\psi\,,
\cr
{\rm even} & {\rm for}& s=- \ {\rm in}\ \psi \ {\rm and} \ \bar\psi\,.
}\right.
$$ 
for the Neumann-type boundary conditions \EqNeumann, and
$$
j-m = \left\{\matrix{
{\rm odd} & {\rm for}& \phi \ {\rm and} \ \bar\phi\,,
\cr
{\rm even} & {\rm for}& s=+ \ {\rm in}\ \psi \ {\rm and} \ \bar\psi\,,
\cr
{\rm odd} & {\rm for}& s=- \ {\rm in}\ \psi \ {\rm and} \ \bar\psi\,.
}\right.
$$ 
for the Dirichlet-type boundary conditions \EqDirichlet.
Using the mode expansions, the eigenvalues, and the orthogonality relations reviewed in Appendix \AppMonoHarm,
we obtain
\eqn\ActionModeChi{\eqalign{
S^{(2)}_{\rm chi}&=
{1\over 2}
\sum_w\sum^\infty_{j=0}\mathop{\sum{}'}^j_{m=-j}
(\phi^{w}_{jm})^*
\left[\left(j+{1\over 2}  \right)^2
+\left(w\cdot \sigma_2  - i {q +1 \over 2}\right)^2
\right]
\phi^{w}_{jm}
\cr
&\quad +
{1 \over 2} \sum_w \sum_{j=1/2}^\infty \mathop{\sum{}'}_{m=-j}^j \sum_{s=\pm}
{
(-1)^{m+{1\over 2}}s\,
\bar\psi^{-s}_{wj,-m}
}
\left[s \, i\left(j+{1\over 2}\right) +w\cdot \sigma_2  -i{q\over 2}\right]\psi^{ws}_{jm}\,.
}}
From this we can calculate the one-loop determinant.
\eqn\OneLoopChiDerived{\eqalign{
Z_{\rm 1\mathchar`-loop}^{\rm chi}
&=
\prod_{w}
 {\displaystyle
\prod_{j=1/2}^\infty 
\left[\left(j+{1 \over 2}\right)^2 +\left(w\cdot\sigma_2 - i{q \over 2}\right)^2\right]^{j+1/2}
\over \displaystyle
\prod_{j=0}^\infty \left[\left(j+{1 \over 2}\right)^2 +\left(w\cdot\sigma_2  -i {q+1 \over 2}\right)^2\right]^{(j+1\ {\rm or}\ j)}
}
\cr
&=
\prod_{w}
\left\{
\matrix{
\displaystyle
1\bigg/\prod_{j=0}^\infty \left[j-i\left(w\cdot\sigma_2  - i {q\over 2}\right)\right] & {\rm (Neumann) }\,,
\cr
\displaystyle
\prod_{j=0}^\infty \left(j+1+i\left(w\cdot\sigma_2  - i{ q\over 2}\right)\right) & {\rm (Dirichlet)}\,.
}\right.
}}
The twisted mass ${\rm m}$ can be introduced by replacing $w\cdot \sigma_2 \rightarrow w\cdot\sigma_2 +{\rm m}$.
The infinite products can be regularized by the gamma function $\Gamma(1+z)=e^{-\gamma z} \prod_{k=1}^\infty e^{z/k} (1+z/k)^{-1}$, where $\gamma$ is the Euler constant.
Even if we use the gamma function so that we get the required zeros and poles, there are ambiguities in the overall $z$-dependent normalizations.
For reasons we explain in Sections \SecResults\ and \SecKoszul, we choose the relative factor between the Neumann and the Dirichlet cases such that%
\foot{%
Determining the overall factor requires a more careful treatment to be discussed elsewhere.
}
\eqn\OneLoopChiral{
Z_{\rm 1\mathchar`-loop}^{\rm chi}(\sigma; m)
=
\left\{
\matrix{
Z_{\rm 1\mathchar`-loop}^{\rm chi,\, Neu} \equiv &
\displaystyle
 \prod_{w\in R}\Gamma(w\cdot\sigma +  m) & {\rm (Neumann) }\,,
\cr
Z_{\rm 1\mathchar`-loop}^{\rm chi,\, Dir} \equiv &
\displaystyle
{
-2\pi i\, e^{\pi i(w\cdot\sigma+ m)}
\over
 \prod_{w\in R}\Gamma(1-w\cdot\sigma- m) 
}& {\rm (Dirichlet)}\,,
}\right.
}
where
the product is over all the weights in the representation $R$, and
$$
\sigma \equiv -i \ell \sigma_2\,,
\qquad
 m \equiv - {q\over 2} -i \ell {\rm  m} \,.
$$
We have recovered $\ell$ for the definition of $\sigma$.

The infinite products require UV regularization and result in the running of the effective FI parameters.
As in \DoroudXW, we take into account the effect of renormalization by replacing the UV complexified FI parameter $t$ with its renormalized value $t_{\rm ren}$.
For each abelian factor in the gauge group $G$, this gives
\eqn\renormFI{
t \rightarrow t_{\rm ren}=t -\sum_a Q_a \ln(\ell M_{\rm UV})\,,
}
where $Q_a$ are the charges of the chiral multiplets, and $M_{\rm UV}$ is the UV cut-off.%
\foot{%
By the same mechanism, effective FI parameters are generated for flavor symmetries \BeniniUI.
The partition function is then multiplied by the factor $e^{-m \ln(\ell M_{\rm UV})}$ for each twisted mass $m$.
}
In the Calabi-Yau case $\sum_a Q_a=0$, we have $t_{\rm ren}=t$.

We turn to the vector multiplet for the gauge group $G$.
In the $R_\xi$ gauge, the localization action $S_{\rm loc}$ augmented by the ghost action \DoroudXW, around the locus \LocusHemisphere, reads
\eqn\ActionQuadraticVec{\eqalign{
S^{(2)}_{\rm vec}
=\int  d^2x \sqrt h \Tr \bigg[& A^\mu \left({\rm M}^2 + 1 \right)A_\mu
+2\tilde \sigma_1\varepsilon^{\mu\nu}\nabla_{\mu} A_\nu
+\tilde\sigma_1\left({\rm M}^2 + 1 \right)\tilde\sigma_1 
\cr
&\qquad\qquad
+ \tilde \sigma_2 {\rm M}^2 \tilde \sigma_2 +\D^2
+\bar\lambda\gamma^3\left(i \gamma^3 \gamma^\mu D_{\mu}
+\sigma_2\right)\lambda
+ \overline c\, {\rm M}^2 c\bigg]
}}
up to the quadratic order, where $\tilde\sigma_r$ are  the fluctuations of the fields $\sigma_r$, and
$$ 
{\rm M}^2:=-D^\mu D_\mu + \sigma_2^2.
$$
The Gaussian integral over ${\rm D}$ is trivial and will be neglected.

On the vector multiplet we impose the boundary condition \EqBCPreserve.
Let us denote the basis of ${\frak g}_{\Bbb C}$ by $H_i$ ($i=1,\ldots,{\rm rk}\,G$) and $E_\alpha$, where $H_i$ span the Cartan subalgebra, and $\alpha$ are the roots of $G$: $[H_i,E_\alpha]=\alpha(H_i)E_\alpha$, $E_\alpha^\dagger= E_{-\alpha}$.
We choose a decomposition 
of the root system into the positive and the negative roots.
For $r=1,2$, we expand
$$
\tilde \sigma_r = \sum_{\alpha>0} \sum^\infty_{j=0} \mathop{\sum{}'}^j_{m=-j}\tilde \sigma^\alpha_{rjm}Y_{jm}(\vartheta,\varphi) E_\alpha + h.c. +\ldots
$$
The ellipses indicate terms in the Cartan subalgebra, whose contributions are independent of physical parameters and will be dropped.
Ghosts $(c, \bar c)$ are expanded in a way similar to $(\phi,\bar\phi)$ with coefficients $(c^\alpha_{jm},\bar c_{\alpha jm})$, respectively.
The expansions of the gauginos $(\lambda,\bar\lambda)$ are similar to those of $(\psi,\bar\psi)$, and have respectively the coefficients $(\lambda^{s\alpha}_{jm},\bar \lambda^s_{\alpha jm})$.
For the gauge field,
 $$
A_\mu=\sum_{\alpha>0} \sum^2_{\lambda=1}\sum^\infty_{j=1} \mathop{\sum{}'}^j_{m=-j}A^{\alpha\lambda}_{jm} (C^\lambda_{jm})_\mu E_\alpha
+ h.c. +\ldots\,,
$$
where $ (C^\lambda_{jm})_\mu$ are the vector spherical harmonics reviewed in Appendix \AppMonoHarm.
The sums $\sum'_m$ are restricted to those $m$ which satisfy
$$
j - m = \left\{\matrix{
{\rm even(odd)} & {\rm for} & \lambda = 1(2) \ {\rm in} \ A_\mu \,,
\cr
{\rm odd} & {\rm for}& \tilde \sigma_1 , c , \bar c \,,
\cr
{\rm even} & {\rm for}& \tilde \sigma_2 \,,
\cr
{\rm even(odd)} & {\rm for}& s=+(-) \ {\rm in}\ \lambda \ {\rm and} \ \bar\lambda \,.
}\right.
$$

The eigenvalues of the kinetic operators as well as the pairings of the eigenmodes can be found by using the properties of the spherical harmonics reviewed in Appendix \AppMonoHarm.
Let us split the quadratic action \ActionQuadraticVec\ into the bosonic and the fermionic parts.
The bosonic part $S^{(2){\rm b}}_{\rm vec}$ reads 
\eqn\ActionModeVecBos{\eqalign{
S^{(2){\rm b}}_{\rm vec}&=\sum_{\alpha>0}
\Bigg(
\sum^2_{\lambda=1}\sum^\infty_{j=1}\mathop{\sum{}'}^j_{m=-j}
(A^{\alpha\lambda}_{jm} )^*
\left[j(j+1)+(\alpha\cdot \sigma_2)^2\right]A^{\alpha\lambda}_{jm}
\cr
& \qquad\qquad
- \sum^\infty_{j=1} \mathop{\sum{}'}^j_{m=-j} \left[(\tilde\sigma^\alpha_{1jm})^*\sqrt{j(j+1)}A^{\alpha 2}_{jm} 
+c.c.\right]
\cr
& \quad\qquad\qquad
+ \sum^2_{r=1}\sum^\infty_{j=0}\mathop{\sum{}'}^j_{m=-j}
 (\tilde \sigma^{\alpha}_{rjm})^*
\left[j(j+1)+(\alpha\cdot\sigma_2)^2+2-r\right]\tilde\sigma^\alpha_{rjm}\Bigg)\,.
}}
The gaugino part is similar to the fermionic part in the chiral multiplet action \ActionModeChi.
The ghost part is 
$$
\sum_\alpha \sum^\infty_{j=0}\mathop{\sum{}'}^j_{m=-j}
\bar c_{-\alpha,j,-m}\left[ j(j+1) + (\alpha\cdot \sigma_2)^2 \right] c^\alpha_{jm}
\,.
$$

Let us now calculate the one-loop determinant $Z_{\rm 1\mathchar`-loop}^{\rm vec}$ for the vector multiplet.
The combined contribution from $A^{\alpha 2}_{jm}$ and $\tilde\sigma_1$ to $Z_{\rm 1\mathchar`-loop}^{\rm vec}$ is
$$\eqalign{
&
\prod_{\alpha>0 } \prod^\infty_{j=1}
\left|
\matrix{
j(j+1) + (\alpha \cdot \sigma_2) ^2 & \sqrt{j(j+1)}
\cr
\sqrt{j(j+1)} & j(j+1)+(\alpha \cdot \sigma_2)^2+1
}\right|^{-j}
\cr
=&
\prod_{\alpha>0} \prod^\infty_{j=1} 
\left[j^2+ (\alpha\cdot\sigma_2)^2 \right]^{-j} \left[(j+1)^2+ (\alpha \cdot \sigma_2)^2 \right]^{-j}
\,.
}$$
The contributions from the other modes can be computed straightforwardly.
Combining everything together, we have
$$
Z_{\rm 1\mathchar`-loop}^{\rm vec} \sim \prod_{\alpha>0} \prod^\infty_{j=0} \left[ j^2 + (\alpha \cdot \sigma_2)^2 \right]\,.
$$
Recall the notation $\sigma=-i\ell \sigma_2$.
After regularization, we obtain%
\foot{%
An analogous factor appears in an integral representation of a vortex partition function \GerasimovTD.}
\eqn\OneLoopVector{
Z_{\rm 1\mathchar`-loop}^{\rm vec} =
\prod_{\alpha>0} \alpha \cdot  \sigma \sin( \pi \alpha \cdot  \sigma)\,.
}

\subsec{Results for the hemisphere partition function}
\subseclab\SecResults

We now write down the partition function of the ${\cal N}=(2,2)$ theory  $(G,V_{\rm mat},t,W,m)$ on a hemisphere with boundary condition ${\cal B}=({\bf Neu}, {\bf Dir}, {\cal V}, {\cal Q})$.
Putting together the calculations in Sections \SecLocAc\ and \SecOneLoop, we obtain the partition function%
\foot{%
We divided each sine by $-\pi$, so that the hemisphere partition functions behave better under dualities discusses in Section \SecDualities.
}
\eqn\ResultPreservedGeneral{
Z_{\rm hem}({\cal B}; {t_{\rm ren}} ;m)
= 
{1\over |W(G)|
}
\int_{ \sigma\in i {\frak t}}
{
d^{{\rm rk}(G)}\sigma
\over
(2\pi i)^{{\rm rk}(G)} 
}
{\rm Str}_{\cal V} 
[
e^{ - 2\pi i ( \sigma + m)}
]
e^{ t_{\rm ren}\cdot  \sigma}
Z_{\rm 1\mathchar`-loop}({\cal B};\sigma; m)\,,
}
where the one-loop determinant is
\eqn\ResultPreservedGeneralOneLoop{\eqalign{
Z_{\rm 1\mathchar`-loop}({\cal B};\sigma; m)&=
\Big(
\prod_{\alpha>0} \alpha \cdot  \sigma 
{
\sin( \pi \alpha \cdot  \sigma)
\over
-\pi
}
\Big)
\prod_{a\in \bf{Neu}} 
\prod_{w\in R_{a}}
\Gamma(w\cdot  \sigma+ m_a)
\cr
&
\qquad\qquad \times
\prod_{a\in \bf{Dir}} 
\prod_{w\in R_{a}}
{-2\pi i  e^{\pi i(w\cdot\sigma+ m_a)} \over \Gamma(1- w\cdot  \sigma-  m_a)}
\,,
}}
Here $W(G)$ is the Weyl group, ${\frak t}={\frak t}(G)$ is the Cartan subalgebra, and ${\rm rk}$ denotes the rank.
Recall also that $t_{\rm ren}\cdot \sigma$ with $t_{\rm ren}=r_{\rm ren} - i\theta$ denotes the renormalized FI and the topological couplings \renormFI\ for the abelian factors in the gauge group $G$.%
\foot{%
If $G=U(N)$, $t_{\rm ren}\cdot\sigma=t_{\rm ren} {\rm Tr}\,\sigma$.
}
The complexified twisted masses $m=(m_a)$ are defined as the combinations $m_a=- {1\over 2} q_a - i \ell {\rm m}_a$ of the R-charges $q_a$ and the real twisted masses ${\rm m}_a$.
In the rest of the paper, we will refer to $m_a$ simply as twisted masses.

In the special case $G=U(1)$, the partition function becomes
\eqn\ResultPreservedUOne{
Z_{\rm hem}= 
\int {d\sigma \over 2\pi i} e^{t_{\rm ren} \sigma} {\rm Str}_{\cal V} [e^{- 2\pi i ( \sigma +  m)}]
\prod_{a\in \bf{Neu}} \Gamma( Q_a  \sigma+  m_a)
\prod_{a\in \bf{Dir}}{-2\pi i\, e^{\pi i(Q_a\sigma +  m_a)} \over \Gamma(1- Q_a  \sigma- m_a)}\,,
}
where $Q_a$ is the $U(1)$ charge for the $a$-th chiral multiplet.

Depending on the representations in which the chiral fields transform, it may be necessary to deform the contour in the asymptotic region so that the integral is convergent.
For $r$ deep inside the K\"ahler cone of a geometric phase, the integral \ResultPreservedGeneral\ can be evaluated explicitly by the residue theorem.

In particular for theories whose axial R-symmetry is non-anomalous in flat space,%
\foot{%
This is equivalent to the condition $\sum_a \sum_{w\in R_a} w=0$, which makes the asymptotic behavior of the integrand to be determined by $e^{t\cdot\sigma}$.
}
we can write down a general formula for $Z_{\rm hem}$ using multi-dimensional residues, as in the case of the $\Bbb S^2$ partition function \HalversonEUA.
Let $H_i$, $i=1,\ldots {\rm rk}(G)$, be the simple coroots, which we treat as a basis of ${\frak t}_{\Bbb C}$.
Let us expand 
\eqn\abelianize{
\sigma=\sum_j \sigma^{j} H_j\,,
\qquad
w\cdot \sigma=\sum_j w_j \sigma^j\,,
\qquad
t\cdot\sigma=\sum_j t_j \sigma^j
}
and write $\vec\sigma=(\sigma^{j})$, $\vec w=(w_j)$, $\vec t=\vec r-i\vec \theta=(t_j=r_j-i\theta_j)$.
When $G$ is non-abelian, $t_j$ in \abelianize\ are not all independent.
Let $I$ be a subset of $\{(a,w)|a\in{\bf Neu}, w\in R_a\}$ with $|I|={\rm rk}(G)$ such that the weights $w$ that appear are linearly independent.
Denote by ${\frak I}$ the set of such subsets $I$.
Each $I$ is associated with gamma function factors $\Gamma(w\cdot \sigma+ m_a)$, $(a,w)\in I$.
We denote by $P_I$ the set of the points $p$ with $\sigma(p)\in {\frak t}_{\Bbb C}$ satisfying
\eqn\PoleCone{
(w\cdot \sigma(p)+m_a )_{(a,w)\in I} \in \Bbb Z_{\leq 0}^{{\rm rk}(G)}\,.
}
Following \HalversonEUA, define
\eqn\CI{
C(I):=\Bigg\{\vec r=\sum_{(a,w)\in I} r_{aw}\vec w \, \Bigg| \, r_{aw}>0 {\rm \ for\ all}\ (a,w)\in I \Bigg\}\,.
}
The hemisphere partition function \ResultPreservedGeneralOneLoop\ is then given as
\eqn\ResultResidue{\eqalign{
Z_{\rm hem}({\cal B})
&
= {1\over |W(G)|} \mathop{\sum_{I\in{\frak I}:}}_{\vec r\in C(I)}
\sum_{p\in P_I} \mathop{\rm Res}_{\sigma=\sigma(p)}
\Big(
{\rm Str}_{\cal V} 
[
e^{ - 2\pi i ( \sigma + m)}
]
e^{ t_{\rm ren}\cdot  \sigma}
Z_{\rm 1\mathchar`-loop}({\cal B};\sigma; m)
\Big)
\,.
}}
The definition of Res, the multi-dimensional residue \GH,  will be apparent from the next paragraph.

An elementary way to understand the formula \ResultResidue\ goes as follows.
For given FI parameters $\vec r$, \ResultPreservedGeneral\ can be evaluated in principle by successive integrations over $\sigma^{1}$, $\sigma^{2}$, etc.
There are many gamma function factors of which we pick poles, and the combinatorics in such a calculation becomes quite complicated.
The combinatorics for the total contribution from the set of factors specified by $I$, however, is not affected by the presence of other factors, and is in fact captured by a simple change of integration variables.
Namely we take $\{w\cdot\sigma+m_a|(a,w)\in I\}$ as new variables to be integrated over along the imaginary axis and compute the residues of the chosen factors.
Unless $r_{aw}>0$ for all $(a,w)\in I$, the contribution vanishes.

Although we do not do this explicitly, it should be possible to obtain the infinite sum expression \ResultResidue\ by localization with a different $Q$-exact action \refs{\BeniniUI,\DoroudXW}.
In such a computation, the saddle point configurations correspond to the discrete Higgs vacua, namely the solutions to the D-term and F-term equations satisfying $(w\cdot \sigma+m_a)\phi_a=0$ for all $a$.
The label $I$ specifies the chiral fields that take non-zero vevs.
Indeed the decomposition $\vec r=\sum_{(a,w)\in I} r_{aw} \vec w$ implies that the D-term equations%
\foot{%
The D-term equations read $D^I\propto \mu^I=0$, where $\mu^I$ are given in \MomentMap.
} 
can be solved by setting $\phi_a^w = (r_{aw}/2\pi)^{1/2}$ for $(a,w)\in I$ with other $\phi_a^w=0$.
The value of $\sigma$ is fixed by the condition $w\cdot \sigma+m_a=0$ for $(a,w)\in I$, corresponding to the tip of the cone determined by \PoleCone.
Each infinite sum specified by $I$ is a power series in the exponentiated FI-parameters, and defines an analog of the 3d holomorphic block \BeemMB.

The results above were obtained by explicit localization calculations on a hemisphere with the round metric \MetricRound.
We now argue that they should also be valid for the deformed metric \MetricDeformed\ by interpreting the one-loop determinants 
 \OneLoopChiral\ and \OneLoopVector\ using the equivariant index theorem as in \refs{\PestunRZ,\GomisPF,\BeniniUI}.
With an appropriate choice of localization action $S_{\rm loc}=\delta_{ Q}\Bbb V$, the one-loop determinant should be given from the equivariant index by converting a sum into a product according to
$$
{\rm ind}\,\Bbb D =\sum_j c_j e^{\lambda_j} \rightarrow Z_{\rm 1\mathchar`-loop}=\prod_j \lambda_j^{-c_j/2}\,,
$$
where ${\Bbb D}$ is a differential operator in $\Bbb V$, $j$ parametrize the eigenmodes of the bosonic symmetry generator $\delta_{Q}^2$, $c_j=\pm 1$, and $\lambda_j$ are the eigenvalues of $\delta_{Q}^2$.
When the geometry has no boundary, the index ${\rm ind}\,\Bbb D$ is given as a sum of contributions from the fixed points of $\delta_{Q}^2$.
In the presence of boundary, at least with suitable boundary conditions such as those in \Donnely, the equivariant index is a sum of fixed point contributions and the boundary contributions.
Thus the one-loop determinant $Z_{\rm 1\mathchar`-loop}$ should also factorize into such local contributions.

For a chiral multiplet, it was shown in \BeniniUI\ that the combined contribution from the north and the south poles ($\vartheta=0$ and $\pi$ respectively) of the round two-sphere is
$$
\prod_w {\Gamma(w\cdot\sigma +  m) \over \Gamma(1-w\cdot\sigma- m)}
\sim
Z_{\rm 1\mathchar`-loop}^{{\rm chi}, {\Bbb S^2}}
\sim
Z_{\rm 1\mathchar`-loop}^{\rm chi,\, Neu} Z_{\rm 1\mathchar`-loop}^{\rm chi,\, Dir}\,,
$$
where by $\sim$ we mean the match of zeros and poles.
It was also shown in \GomisWY\ that the full sphere one-loop determinant is independent of the metric deformation \MetricDeformed.
As in the four-dimensional case \refs{\PestunRZ,\GomisPF}, we interpret the square-root $(Z_{\rm 1\mathchar`-loop}^{{\rm chi}, {\Bbb S^2}})^{1/2}\sim (Z_{\rm 1\mathchar`-loop}^{\rm chi,\, Neu} Z_{\rm 1\mathchar`-loop}^{\rm chi,\, Dir})^{1/2}$ as the local contribution from each of the north and the south poles.%
\foot{%
In \BeniniUI, $Z_{\rm 1\mathchar`-loop}^{\rm chi,\, Neu}$ and $Z_{\rm 1\mathchar`-loop}^{\rm chi,\, Dir}$ were assigned to distinct poles.
}
Then \OneLoopChiral\ implies,  in the case of the round sphere, that the single-boundary  contribution to the one-loop determinant is
\eqn\OneLoopBdryNeu{
(\sin[\pi(w\cdot\sigma+ m)])^{-1/2}
}
for the Neumann boundary condition, and
\eqn\OneLoopBdryDir{
(\sin[\pi(w\cdot\sigma+ m)])^{1/2} 
}
for the Dirichlet boundary condition (up to ambiguities in the overall factors).
On the other hand,  the local approximate form of $\Bbb D$ and the action of $\delta_{Q}^2$ near the boundary is essentially independent of deformation.
Thus we expect that the single-boundary  contribution to the one-loop determinant is given by the same formulas \OneLoopBdryNeu\ and \OneLoopBdryDir, even after deformation.%
\foot{%
As a check, one can compute the one-loop determinant on $\Bbb S^1\times ({\rm interval})$ by mode expansion and confirm that it is the product of two boundary contributions, for any pair of boundary conditions on the two boundaries.
}
Then, the formula \OneLoopChiral\ for the one-loop determinant on a hemisphere should also be valid for the deformed metric \MetricDeformed.
We can apply the same logic to the vector multiplet, recalling that the full sphere one-loop determinant is $\prod_{\alpha>0} (\alpha \cdot  \sigma)^2 $ \refs{\BeniniUI,\DoroudXW}.
It follows that the single-boundary contribution to one-loop determinant is
$$
\prod_{\alpha>0} \sin( \pi \alpha \cdot  \sigma)\,.
$$

The local contributions to the one-loop determinant from the poles and the boundary are determined by $\delta_Q^2$, and cannot be affected by the deformation parameter $\tilde\ell$.
The classical contributions computed in \SecLocAc\ are also independent of $\tilde\ell$.
These arguments suggest that the expression of the hemisphere partition function \ResultPreservedGeneral\ should also be valid for the deformed metric \MetricDeformed.

\subsec{Hilbert space interpretation}
\subseclab\SecHilbert

We argued above that the partition function on the deformed sphere is independent of the parameter $\tilde\ell$.
In the limit that $\tilde\ell\rightarrow \infty$, the geometry near the boundary $\vartheta=\pi/2$ becomes flat, and the non-dynamical gauge field $V^{\rm R,new}$ in \EqRBackground\ for $U(1)_{\rm R}$ vanishes in the frame where all the fields are periodic.

The boundary condition ${\cal B}$ on a hemisphere $0\leq \vartheta\leq\pi/2$ defines the boundary state $\langle {\cal B}|$ in the Hilbert space of the theory on a spatial circle.
Since all the fields are periodic in the frame with $V^{\rm R,new}(\tilde \ell\rightarrow \infty)=0$,  $\langle {\cal B}|$ is in the {\it Ramond-Ramond sector}.
The hemisphere partition function \ResultPreservedGeneral\ is the overlap $\langle \cal B|{\tt 1}\rangle$ between $\langle {\cal B}|$ and a state $|{\tt 1}\rangle$ created by the path integral on the hemisphere with no operator insertion.
Let ${\tt f}(\sigma)$ be a gauge invariant polynomial of $\sigma$.
The result \ResultPreservedGeneral\ can be generalized to include a twisted chiral operator ${\tt f}({\sigma_1-i\sigma_2})$:
\eqn\BfOverlap{\eqalign{
\langle {\cal B}|{\tt f}\rangle
&=
\int_{\cal B} {\cal D}A\ldots e^{-S_{\rm phys}}
{\rm Str}_{\cal V}\hskip -1mm \left[ P \exp\left(i \oint d\varphi{\cal A}_{ \varphi}\right)\right]
{\tt f}({\sigma_1-i\sigma_2})
\cr
&=
{1\over |W(G)|
}
\int_{ \sigma\in i {\frak t}}
{
d^{{\rm rk}\,G}\sigma
\over
(2\pi i)^{{\rm rk}\,G}
}
{\rm Str}_{\cal V} 
[
e^{ - 2\pi i ( \sigma +  m)}
]
e^{ t_{\rm ren}\cdot  \sigma}
Z_{\rm 1\mathchar`-loop}({\cal B};\sigma; m)
{\tt f}(  \sigma)\,,
}}
where $\int_{\cal B}$ indicates functional integration with the boundary condition ${\cal B}$.
The Ramond-Ramond state $|{\tt f}\rangle$ is created by the path integral, defined using the physical action \SPhys, with the insertion of ${\tt f}({\sigma_1-i\sigma_2})$ at $\vartheta=0$.
The argument in \GomisWY\ suggests that it is closely related to the state defined by the path integral of the A-twisted theory \CecottiME.%
\foot{%
The argument was used to justify the proposal that the $\Bbb S^2$ partition function is related to the K\"ahler potential on the K\"ahler moduli space \JockersDK. 
}
We will identify the boundary state $\langle {\cal B}|$ with its projection to the BPS subspace.

The partition function on the full sphere $0\leq \vartheta\leq \pi$, as computed in \refs{\BeniniUI,\DoroudXW}, is the overlap $Z_{\Bbb S^2}=\langle {\tt 1}|{\tt 1}\rangle$.
By generalizing to include ${\cal O}_1\equiv {\tt f}({\sigma_1-i\sigma_2})$ at $\vartheta=0$, and  $\overline{\cal O}_2\equiv {\tt g}({-\sigma_1-i\sigma_2})$ at $\vartheta=\pi$, we obtain
\eqn\Spheregf{\eqalign{
\langle {\tt g}|{\tt f}\rangle
&
=
\langle \overline{\cal O}_2(\vartheta=\pi) {\cal O}_1(\vartheta=0)\rangle
=
\int {\cal D}A\ldots e^{-S_{\rm phys}}
{\tt g}({-\sigma_1-i\sigma_2})
{\tt f}({\sigma_1-i\sigma_2})
\cr
&=
{c\over |W(G)|}
\sum_{B\in \Lambda_{\rm cochar}}
\int_{ \sigma\in i {\frak t}} 
{
d^{{\rm rk}(G)}\sigma
\over
(2\pi i)^{{\rm rk}(G)}
}
e^{ t_{\rm ren}\cdot ( \sigma -B/2)}
e^{{\bar t_{\rm ren}}\cdot ( \sigma +B/2)}
{(-1)^{w_0\cdot B}}
{\tt g}\Big( \sigma+{B \over 2}\Big)
\cr
&\ 
\times
{\tt f}\Big(  \sigma-{B \over 2}\Big)
\prod_{\alpha>0}
\Big[ 
{(\alpha\cdot B)^2\over 4}
-
(\alpha\cdot\sigma)^2
\Big]
\prod_a
\prod_{w\in R}
{
\Gamma(w\cdot ( \sigma-B/2)+ m_a)
\over
\Gamma(1-w\cdot(  \sigma+B/2)- m_a)
}
\,.
}}
We have included a normalization constant $c$ and used a weight $w_0$ to parametrize the ambiguity in the normalization of the flux sectors labeled by GNO charges \GoddardQE\ $B\in \Lambda_{\rm cochar}(G)$.%
\foot{%
The lattice $\Lambda_{\rm cochar}(G)$ consists of the elements of the Cartan subalgebra which have integer pairings with the weights that appear in all the representations of the group $G$ (rather than  ${\frak g}$).
}

The path integral on the other half of the sphere ($\pi/2\leq \vartheta\leq \pi$) gives 
\eqn\fBOverlap{\eqalign{
\langle {\tt g}| {\cal B}\rangle
&=
\int_{\cal B} {\cal D}A\ldots e^{-S_{\rm phys}}
{\rm Str}_{\cal V}\hskip -1mm \left[ P \exp\left(
+i \oint d\varphi\tilde{\cal A}_{ \varphi}\right)\right]
{\tt g}({-\sigma_1 - i\sigma_2})
\cr
&=
{1\over |W(G)|}
\int_{ \sigma\in i {\frak t}}
{
d^{{\rm rk}\,G}\sigma
\over
(2\pi i)^{{\rm rk}\,G}
}
{\rm Str}_{\cal V} 
[
e^{  2\pi i ( \sigma +  m)}
]
e^{ {\bar t_{\rm ren}}\cdot  \sigma}
Z_{\rm 1\mathchar`-loop}({\cal B};\sigma; m)
{\tt g}(  \sigma)\,,
}}
where
$$
\tilde {\cal A}_{\hat\varphi}=
\rho_*( A_{\hat \varphi} + i \sigma_2 ) +{\rho_*(R)\over 2\ell}
+i \rho_{*}({\rm m})
- {i\over 2}\{{\cal Q},\bar {\cal Q}\}
+{i\over2}
\left(
(\psi_1-\psi_2)^i
 \partial_i {\cal Q}+ (\bar\psi_1-\bar\psi_2)_i \partial^i \bar {\cal Q}
\right)
\,.
$$
It is also natural to consider the partition function on a cylinder with boundary conditions ${\cal B}_{1,2}$ along the two boundaries
\eqn\CylOverlap{\eqalign{
\langle {\cal B}_1|{\cal B}_2\rangle
&=
\int_{{\cal B}_1,{\cal B}_2}
\hskip-5mm
 {\cal D}A\ldots e^{-S_{\rm phys}}
{\rm Str}_{{\cal V}_1}\hskip -1mm \Big[ P \exp\Big(
i \oint d\varphi{\cal A}^{+}_{ \varphi}\Big)\Big]
{\rm Str}_{{\cal V}_2}\hskip -1mm \Big[ P \exp\Big(
i \oint d\varphi {\cal A}^{-}_{ \varphi}\Big)\Big]\,,
}}
with
$$
{\cal A}^\pm_\varphi=
\rho_*( A_{\hat \varphi} + i \sigma_2 )
+ \rho_{*}(V^{\rm F}_{\hat\varphi}+i{\rm m})
\pm {i\over 2}\{{\cal Q},\bar {\cal Q}\}
+{1\over2}e^{{\pi i \over 4}(1\mp 1)} 
\left(
(\psi_1-\psi_2)^i
 \partial_i {\cal Q}+ (\bar\psi_1-\bar\psi_2)_i \partial^i \bar {\cal Q}
\right)
\,.
$$
This is a supersymmetric index of the theory on a spatial interval.
Since it is independent of the width, this quantity can be computed by a supersymmetric quantum mechanics or classical formulas involving characteristic classes, as we will see in Section \SecSheaves.
In particular there is no ambiguity in this quantity.

\vskip -17mm

\bigskip
\centerline{
\epsfxsize 1.15 in \epsfbox{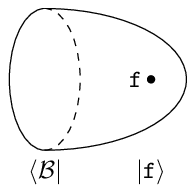}
\hskip 5mm
\epsfxsize 1.6 in \epsfbox{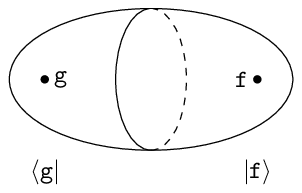}
\hskip 8mm
\epsfxsize 1.15 in \epsfbox{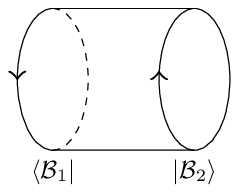}
}
\centerline{\sl \hskip 0cm (a) \hskip 4cm (b)  \hskip 4cm (c)}
  \centerline{\sl {\bf Figure 1} (a) Hemisphere with an operator insertion.}
\centerline{\sl
 (b) Twisted chiral/twisted anti-chiral 2-point function.  (c) Cylinder partition function.
}
\bigskip
\noindent

The Hilbert space interpretation implies that the $\Bbb S^2$ partition function (or its generalization \Spheregf) is determined by the hemisphere partition functions (or their generalizations) and the cylinder partition function \CylOverlap.
Namely, by choosing boundary states $|{\cal B}_a\rangle$ that form a basis of the BPS Hilbert space, we set
$$
\chi_{ab}=\langle {\cal B}_a|{\cal B}_b\rangle
$$
and denote the inverse matrix by $\chi^{ab}$.
Then
$$
\langle {\tt g}|{\tt f}\rangle= \langle {\tt g}|{\cal B}_a\rangle \chi^{ab}\langle {\cal B}_b|{\tt f}\rangle\,.
$$
In some examples with twisted masses, we will introduce another basis $\{|{\bf v}\rangle\}$ that is orthonormal.
In that case we can write $\langle {\tt g}|{\tt f}\rangle=\sum_{\bf v} \langle {\tt g}|{\bf v}\rangle \langle {\bf v}|{\tt f}\rangle$.
In Section \SecTstarGr\ we will demonstrate such factorizations, and see how they allow us to fix the parameters $c$ and $w_0$ that parametrize the ambiguities in the $\Bbb S^2$ partition function of the $T^*{\rm Gr}(N,N_{\rm F})$ model studied there.

\newsec{Hemisphere partition functions and geometry}
\seclab\SecGeom

\subsec{Target space interpretation of the gauge theory}
\subseclab\SecTarget

In this paper we are concerned with the geometric phases in which the theory reduces to a non-linear sigma model with a smooth target space.
We consider two cases.

\vskip 5mm
\noindent
{\bf Case 1: $W=0$, target space $X$}

This is the setup where the gauge theory has no superpotential, and flows in the IR to a non-linear sigma model with target space $X$, which takes the form of a K\"ahler quotient
$$
X=\mu^{-1}(0)/G\,.
$$
The moment map $\mu =(\mu^I)_{I=1}^{\dim G}
:  V_{\rm mat}\rightarrow {{\frak g}^*}$ is given by
\eqn\MomentMap{
\mu^{I}\equiv
\left\{
\matrix{
\bar\phi T^{I}\phi & {\rm for} \ I\ {\rm non\mathchar`-abelian}\,,
\cr
\displaystyle
\bar\phi T^{I}\phi
-{r_I\over 2\pi} &  {\rm for}\ I\ {\rm abelian}\,,
}
\right.
}
where $T^I$ are the generators of $G$ which we split into abelian and non-abelian simple factors.
The complex structure of $X$ can also be specified by viewing it as a holomorphic quotient:
\eqn\HoloQuot{
X=(V_{\rm mat}\backslash {\rm deleted\ set})/G_{\Bbb C}\,.
}
Here  $G_{\Bbb C}$ is the complexification of $G$, and the deleted set consists of those points whose $G_{\Bbb C}$-orbits do not intersect with $\mu^{-1}(0)$.
If the gauge group $G$ is abelian, $X$ is a toric variety.

\vskip 5mm
\noindent
{\bf Case 2: $W=P\cdot {\rm G}(x)$, target space} $M$

In the second situation we consider, the theory has a superpotential of the form
$$
W=P\cdot {\rm G}(x)= P_\alpha {\rm G}^\alpha(x)\,,
$$
where we split the chiral fields $\phi$ into two groups as $\phi=(x,P_\alpha)$.
Assuming that the space
$$
M= {\mu^{-1}(0)} \cap {\rm G}^{-1}(0)/ G 
$$
is smooth, the F-term equations ${\partial\over \partial \phi^i}W(\phi)=0$ reduce to
$$
P_\alpha=0\,,\qquad {\rm G}^\alpha(x)=0\,.
$$
Thus $M$ is the target space of the low-energy theory, and is a submanifold of $X=\mu^{-1}(0)|_{P=0}/G$.
If we focus on the complex structure, $M$ is given as
\eqn\HoloQuotM{
M= (V_{\rm mat}\backslash {\rm deleted\ set})\cap {\rm G}^{-1}(0)\cap \{P_\alpha=0\}/ G_{\Bbb C}\,.
}

\vskip 5mm
 
Let us now consider the target space interpretation of the boundary interaction ${\cal A}$.
For simplicity we turn off the twisted masses, work in the flat limit ($\tilde \ell\rightarrow \infty$ with finite $x=-\tilde\ell\cos\vartheta$), and assume that the gauge group is $G=U(N)$, for which
the D-term equations take the form 
\eqn\DtermEq{
\bar\phi T^{I}\phi-{r\over 2\pi}\delta^{{I}0}=0
}
with $T^{{I}=0}=(1/N){\bf 1}$  
corresponding to the abelian part.
We take the FI parameter to be large and positive $r\gg 0$.
In the IR limit $g^2\rightarrow \infty$, the gauge theory flows to the non-linear sigma model with the target space $X$ in Case 1 and $M$ in Case 2.
We assume that the target space is smooth.
The equations of motion that follow from \LChiBulk\ imply that in the present limit \HerbstJQ,
$$
A_\mu=M^{-1}_{IJ}\left(
i  \bar\phi T^I (
 \mathop\partial^\leftarrow -\mathop\partial^\rightarrow
 )_\mu \phi
+\bar\psi T^I\gamma_\mu\psi\right)T^J\,,
$$
$$
\sigma_1=-i M^{-1}_{IJ} (\bar\psi T^I \psi) T^J\,,
\qquad
\sigma_2=
M^{-1}_{IJ}\left(i{1{+}q\over f} \bar\phi T^{I}\phi +\bar\psi\gamma_3  T^{I}\psi\right)T^{J}\,,
$$
where the derivatives $\displaystyle\mathop{\partial}^\leftarrow $ and $\displaystyle\mathop{\partial}^\rightarrow$ act on $\bar\phi$ and $\phi$ respectively, and
$M^{-1}_{IJ}$ is the inverse of the matrix
$M^{IJ}=\bar\phi\{T^{I},T^{J}\}\phi$.
Under $\phi(x) \rightarrow g(x)\phi(x)$, we get the correct transformation $d -i A \rightarrow g(x)(d-iA)g^{-1}(x)$, etc.
Let $R$ be a representation of $G$.
As noted in the context of an abelian gauge theory in \HerbstJQ, the expression $\displaystyle M^{-1}_{IJ}\left(i\bar\phi T^{I}(\mathop\partial^\leftarrow -\mathop\partial^\rightarrow)_\mu\phi\right)$, contracted with the generators $T^{J}$ acting on a vector space $V$, is the pull-back of a connection on the natural holomorphic vector bundle constructed from $V$.
This bundle is defined as
\eqn\NaturalBundleReal{
(({\rm solutions\ of\ the\ D\mathchar`-term\ and\  F\mathchar`-term\ equations})
\times V)/G\,.
}
Thus the Chan-Paton space ${\cal V}$ descends to a collection of holomorphic vector bundles.

We can also see that how the theta angle $\theta$ and the FI-parameter $r$ are related to the B-field and the K\"ahler form of the target space, respectively.
Since the theta term involves only the abelian part $I=0$, the discussion is essentially the same as in the abelian case.  (See for example \HoriCK.)
First note that the matrix $M^{IJ}$ is block-diagonal; the entries with $({I}=0,{J}\neq0)$ or $({I}\neq0,{J}=0)$ vanish because of the D-term equations \DtermEq.
Thus the $U(1)$ part of the gauge field is given, in the current approximation, by
$$
{\rm Tr}\,A
={2\pi i\over r} (d\bar\phi\cdot\phi-\bar\phi\cdot d\phi)\,.
$$
The $\theta$-term \Stheta\ gives a factor 
$
\exp 
(
 -{2\theta\over r } \int d\phi \wedge d\bar\phi
)
$
in the path integral.
This should be identified with  the B-field coupling $\exp(2\pi i \int B)$.
Thus
$$
B={i\theta \over \pi r} d\phi\wedge d\bar\phi\,,
$$
where $\phi$ and $\bar\phi$ are constrained by the D-term equations \DtermEq.
On the other hand the K\"ahler form of the target space is given, in the large volume limit, by
$$
\omega= {i\over 2\pi}d\phi\wedge d\bar\phi\,.
$$
In order to understand the natural combinations of parameters, let us temporarily consider the A-model where $\phi$ is holomorphic on the world-sheet and the kinetic term in \LChiBulk\ gives a factor $\exp(-2\pi \int \omega)$ for a world-sheet instanton.
By combining it with the B-field and the boundary interaction for bundle,
we get
\eqn\ABomega{
{\rm Tr}\,
P
\exp
\big(
 i\oint_{\partial\Sigma} 
\iota^*  A_{\rm target}
\big)
\exp
\big(
2\pi i \int_\Sigma \iota^* (B+i\omega)
\big)
}
where $A_{\rm target}$ is a connection on  the bundle and $\iota^*$ is the pullback by the embedding $\iota:\Sigma\hookrightarrow X{\rm \ or}\ M$.

\subsec{Hemisphere partition function, derived category of coherent sheaves, and K theory}
\subseclab\SecSheaves

In \ResultPreservedGeneral\ we derived an expression of the hemisphere partition function for arbitrary  boundary data ${\cal B}=({\bf Neu},{\bf Dir}, {\cal V}, {\cal Q})$.
We assumed that the whole gauge multiplet satisfies the symmetry preserving boundary condition \EqBCPreserve.
The collections of chiral multiplets satisfying the Neumann condition \EqNeumann\ and  the Dirichlet boundary condition \EqDirichlet\ are denoted by ${\bf Neu}$ and ${\bf Dir}$, respectively.
The Chan-Paton vector space ${\cal V}$ is a representation of $G\times G_{\rm F}\times U(1)_{\rm R}$, and its $\Bbb Z_2$-grading is given by the $U(1)_{\rm R}$ charge (weight) modulo 2.
The tachyon profile ${\cal Q}$ is an odd linear transformation on ${\cal V}$.

Suppose that an ${\cal N}=(2,2)$ non-linear sigma model has as target space a non-singular algebraic variety.
In this paper we are interested in an ${\cal N}=(2,2)$ gauge theory that flows at low energy to such a non-linear sigma model.
As in Section \SecTarget, we denote the target space as $X$ if it is the quotient of a linear space minus a deleted set, and as $M$ if it is the zero-locus of some section on such $X$.
Two high-energy boundary conditions that give rise to the same  boundary condition (D-brane) at low energy should be considered as the same.
It is believed that the low-energy branes that preserve B-type supersymmetry form a category equivalent to what is known as the (bounded) derived category of coherent sheaves, which we denote by $D(X) $ or $D(M)$.
We argue that the hemisphere partition function gives a well-defined  map
\eqn\HPFmap{
Z_{\rm hem}: D(X\ {\rm or}\ M) \rightarrow \{{\rm functions\ of\ }(t,m)\}\,.
}
Let us discuss what this means and how to show it.

Physically, a coherent sheaf is a D-brane whose world-volume does not necessarily wrap the whole target space.
An object of the derived category is a complex of coherent sheaves, up to an equivalence relation called quasi-isomorphism.
An important point is that 
{\it any object in the derived category of (non-equivariant) coherent sheaves on a reasonable space $X$ or $M$ is quasi-isomorphic to a complex of holomorphic vector bundles.}%
\foot{%
Any equivariant coherent sheaf has a {\it locally free resolution}, {\it i.e.}, a representative of the        quasi-isomorphism class by a complex of equivariant holomorphic vector bundles. (Proposition 5.1.28 of \CG).
Though we personally do not know that every object in the derived category has the property, this seems likely and will be assumed.
}
Thus an arbitrary D-brane, even one with lower dimensions, can be represented as a bound state of space-filling branes.

Indeed there is an operation to bind D-branes.
Given two complexes ${\cal E}, {\cal F}$ defined respectively as
$$
\ldots  \mathop{\longrightarrow}^{d^{i-1}_{\cal E}} {\cal E}^i \mathop{\longrightarrow}^{d^i_{\cal E}} {\cal E}^{i+1} \mathop{\longrightarrow}^{d^{i+1}_{\cal E}}\ldots\,,
\qquad d^{i+1}_{\cal E}d^{i}_{\cal E}=0\,,
$$
$$
\ldots  \mathop{\longrightarrow}^{d^{i-1}_{\cal F}} {\cal F}^i \mathop{\longrightarrow}^{d^i_{\cal F}} {\cal F}^{i+1} \mathop{\longrightarrow}^{d^{i+1}_{\cal F}}\ldots\,,
\qquad d^{i+1}_{\cal F}d^{i}_{\cal F}=0\,,
$$
and a collection $f$ of homomorphisms $f^i: {\cal E}^i\rightarrow {\cal F}^i$ such that $f^{i+1}\cdot d_{\cal E}^i=d^i_{\cal F}\cdot f^i$,%
\foot{%
Such a collection of homomorphisms is called a cochain map.
}
 the mapping cone of $f$, denoted as $C(f)$, is the complex whose $i$-th term is $C(f)^i={\cal E}^{i+1}\oplus {\cal F}^i$
with differential $d^i_{C(f)}(x,y)=(-d^{i+1}_{\cal E}(x),f^{i+1}(x)+d^i_{\cal F}(y))$.
The brane $C(f)$ is the bound state of ${\cal E}$ and the anti-brane of ${\cal F}$.
It is known that $f:{\cal E}\rightarrow {\cal F}$ is a quasi-isomorphism if and only if $C(f)$ is exact.

Thus in order to show that \HPFmap\ is well-defined, we need to i) define a map%
\foot{%
In Case 2, {\it i.e.}, for target space $M\subset X$, our construction, given in Section \BCcomplexes, of $Z_{\rm hem}$ for an object of $D(M)$ involves resolving the pushforward of the object to $X$ by a complex of vector bundles.
Thus the relevant bundles in \ComplexToBC\ are those on $X$, not $M$.
}
\eqn\ComplexToBC{
{\rm complex\ of\ holomorphic\ vector\ bundles}
\longmapsto
{\rm boundary\ condition\ } {\cal B}
}
and then ii) show that an exact complex of vector bundles has a vanishing hemisphere partition function.
Part i) will be done in Section \BCcomplexes.
Part ii) will be discussed in Section \BCcomplexes\ and Appendix \AppExact.
Since vector bundles are carried by space-filling branes, we can assume that all chiral multiplets obey the Neumann boundary condition in \ResultPreservedGeneral.

The Grothendieck group of the derived category, which is isomorphic to the K theory of the target space, is an additive group generated by $[{\cal E}]$ for any complex ${\cal E}$ of holomorphic vector bundles, with the relation
$$
[C(f)]=[{\cal E}]-[{\cal F}]
$$
for any $f:{\cal E}\rightarrow {\cal F}$.
The relation is clearly respected by $Z_{\rm hem}$.
Thus $Z_{\rm hem}$ depends only on the K theory class.

\subsec{From complexes of vector bundles to boundary conditions}
\subseclab\BCcomplexes

The aim here is to define the map \ComplexToBC\ that yields a boundary condition for a given complex of holomorphic vector bundles.
We will treat separately Cases 1 and 2.

\vskip 5mm

\noindent
{\bf Case 1}

 When the target space is a quotient space $X$ of the form \HoloQuot, we have a natural $G_{\rm F}$-equivariant holomorphic vector bundle for each representation of $(G\times G_{\rm F})_{\Bbb C}$ as in \NaturalBundleReal;
if $V$ is the representation space, focusing on the holomorphic structure, the bundle is given as%
\foot{%
If $G=U(N)$, $V_{\rm mat}=\{(Q^i{}_f)\}={\bf N}^{\oplus N_{\rm F}}$, deleted set $=\{Q:{\rm rk}(Q)<N\}$, the anti-fundamental representation $\bar{\bf N}$ gives the tautological bundle over the Grassmannian ${\rm Gr}(N,N_{\rm F})$.
}
\eqn\NaturalBundle{
\left(
(V_{\rm mat}\backslash {\rm deleted\ set})\times {\cal V}
\right)/G_{\Bbb C}\,.
}
We will assume that any object in $D(X)$ can be represented as a complex of holomorphic vector bundles constructed in this way.

Given a complex ${\cal E}$ of vector bundles of the form \NaturalBundle, one can  construct the corresponding boundary condition ${\cal B}$ using a straightforward generalization of a procedure in~\HerbstJQ.
Suppose that the $i$-th term ${\cal E}^i$ in the complex arises from the representation $V^i$ of $(G\times G_{\rm F})_{\Bbb C}$.
Then we simply take as the Chan-Paton space ${\cal V}={\cal V}^{\rm e}\oplus {\cal V}^{\rm o}$ with ${\cal V}^{\rm e}=\oplus_{i:{\rm even}}V^i$,  ${\cal V}^{\rm o}=\oplus_{i:{\rm odd}}V^i$.
Since the chiral fields serve as target space coordinates, it is natural to choose an R-symmetry $R_{\rm deg}$, introduced in Section \SecBdryInt, so that $R_{\rm deg}\cdot\phi^a=0$.
We let $R_{\rm deg}$ have eigenvalue $i\in \Bbb Z$ on $V^i$.
The differential%
\foot{%
It is a differential in the sense of homological algebra, and is an algebraic operation.
}
 $d_{\cal E}=(d_{\cal E}^i)$ naturally pulls back to the tachyon profile ${\cal Q}$ that squares to zero.
Thus we obtain the map 
\eqn\MapEToBForX{
{\cal E} \longmapsto {\cal B}=({\cal V},{\cal Q})\,.
}
In the case that $G$ is abelian and $G_{\rm F}$ is trivial, many examples of this construction were studied in \HerbstJQ.
Non-abelian and equivariant examples will be given in Section \SecExamples.

In order to show that the map \HPFmap\ is well-defined, we need to show that the hemisphere partition function for an exact complex vanishes.
The proof that \HPFmap\ is well-defined amounts to showing that the supertrace in the integrand cancels all the poles that could potentially contribute in \ResultResidue.
This is explained in Appendix \AppExact, by using the resolved conifold as an example.

\vskip 5mm
\noindent
{\bf Case 2}

The construction of the map \ComplexToBC\ for target space $M$ in \HoloQuotM\ is also a generalization of the procedure in the abelian, non-equivariant setting introduced in \HerbstJQ.%
\foot{%
Though this construction was referred to as the ``compact'' case in \HerbstJQ, we adapt it to any manifold $M$, such as $T^*{\rm Gr}(N,N_{\rm F})$, obtained as the zero-locus $s^{-1}(0)$ of a section $s$.
}
This is a little more involved than in Case 1.

Recall that the chiral fields $x$ parametrize the ambient space $X$.
The superpotential is
$$
W=P\cdot {\rm G} = P_\alpha {\rm G}^\alpha(x)\,,
$$
where ${\rm G}=({\rm G}^\alpha)$ represents a section $s$ of a vector bundle $E$ and the field $P$ takes values in the dual $E^*$ by the construction in \NaturalBundle.
Given an object ${\cal E}$ of the derived category $D(M)$, 
we first push it forward by the inclusion $i:M\rightarrow X$.
The resulting object of $D(X)$ is quasi-isomorphic to a complex $\hat {\cal E}$ of vector bundles over $X$
\eqn\ComplexAmbient{
\ldots
\mathop{\longrightarrow}^{ d}
\hat{\cal E}^j 
\mathop{\longrightarrow}^{ d}
\hat{\cal E}^{j+1} 
\mathop{\longrightarrow}^{ d}
\ldots\,.
}
In the present case, we define the new R-symmetry $R_{\rm deg}$ in Section \SecBdryInt\ so that 
$$
R_{\rm deg}\cdot x=0\,,\qquad
R_{\rm deg}\cdot P_\alpha=-2P_\alpha\,.
$$
As in Case 1, $\hat{\cal E}$ and $d$ naturally lifts to a Chan-Paton space ${\cal V}$ and an odd operator ${\cal Q}_{(0)}$ on ${\cal V}$, which squares to zero: ${\cal Q}_{(0)}^2=0$.
Since we have a superpotential ${\cal W}$, we need a matrix factorization as the boundary interaction in order to cancel the Warner term \Warner\ and preserve supersymmetry.
This can be constructed by the ansatz
\eqn\QAnsatz{
{\cal Q}={\cal Q}_{(0)} 
+ \sum_\alpha P_\alpha {\cal Q}_{(1)}^\alpha
+ {1\over 2!}\sum_{\alpha,\beta} P_\alpha P_\beta {\cal Q}_{(1)}^{\alpha\beta}
+ \ldots
}
The equation ${\cal Q}^2= W\cdot {\bf 1}$ can be used recursively to find
 ${\cal Q}^{\alpha_1\ldots \alpha_k}_{(k)}$.
The existence of a solution to the equation was shown in \HerbstJQ.
Thus the boundary interaction is purely determined by the geometric consideration, except a subtlety that we now discuss.

In Case 2 we need to shift the assignment, to ${\cal V}$, of overall charges for the abelian part of $G\times G_{\rm F}$.
The shift is from the charges specified by the representations $V^i$.
We now argue for the necessity of the shift by generalizing an argument in \HerbstJQ.
First note that if we know the overall charge assignment for one D-brane on $M$, then the relative charge assignment for other D-branes is automatically determined.
Thus we focus on the simplest D-brane, the space-filling brane carrying no gauge flux.
This corresponds to the trivial line bundle over $M$, or in other words to the structure sheaf ${\cal O}_M$.
Its pushforward $i_* {\cal O}_M$ to the ambient space $X$ is known to be quasi-isomorphic to the so-called Koszul complex
$$
\wedge^{\rm r} E^* 
\longrightarrow
\ldots
\longrightarrow
\wedge^2 E^* 
\longrightarrow
E^*
\longrightarrow
{\cal O}_X
 \,,
$$
where ${\rm r}={\rm rk}\,E$ and the last term has degree zero.
The differential is the contraction by the section $s$ that defines $M$.
The natural way to implement the Koszul complex in the gauge theory is to quantize free fermions living along the boundary \refs{\TakayanagiRZ,\KrausNJ}.
After quantization we obtain fermionic oscillators $\eta_\alpha$, $\bar\eta^\alpha$ satisfying the anti-commutation relations $\{\eta_\alpha,\bar\eta^\beta\}=\delta_\alpha^\beta$.
Let $|0\rangle$ be the Clifford vacuum: $\eta_\alpha |0\rangle=0$.
Then the Koszul complex is realized by
$$
\Bbb C\bar\eta^{1}\ldots \bar\eta^{\rm r}|0\rangle
\longrightarrow
\ldots
\longrightarrow
\bigoplus_\alpha \Bbb C \bar\eta^\alpha |0\rangle
\longrightarrow
 \Bbb C|0\rangle
$$
with the differentials given by ${\cal Q}_{(0)}=\eta_\alpha {\rm G}^\alpha(x)$.
The recursive procedure above terminates in one step, and simply gives
\eqn\QForOM{
{\cal Q}=\eta_\alpha {\rm G}^\alpha(x) + \bar\eta^\alpha P_\alpha\,.
}
This is manifestly a matrix factorization: ${\cal Q}^2=W\cdot {\bf 1}$.

The question is which amount of abelian charges we should assign to $|0\rangle$.
Suppose that the bundle $E$ arises from representation $\rho_E$ of $G\times G_{\rm F}$.
The trivial line bundle ${\cal O}_X$, and hence the space $\Bbb C |0\rangle$, corresponds to the trivial representation in the construction \NaturalBundle.
Physically, however, the canonical choice is to assign one-dimensional projective%
\foot{%
As in the worldsheet theory of a superstring these are representations of a covering of $G\times G_{\rm F}$, and may be interpreted as charge fractionalization introduced by hand.
} representations to $|0\rangle$ and $\bar\eta^{1}\ldots \bar\eta^{\rm r}|0\rangle$ symmetrically:
\eqn\VacuaReps{
\Bbb C |0\rangle \leftrightarrow (\det \rho_E)^{1/2}
\,,
\qquad\quad
\Bbb C \bar\eta^{1}\ldots \bar\eta^{\rm r}|0\rangle \leftrightarrow (\det \rho_E)^{-1/2}\,.
}

This suggests the map
\eqn\MapEToB{
{\cal E}\in D(M) \mapsto {\cal B}=({\cal V},{\cal Q})
}
defined as follows.
For the complex \ComplexAmbient\ quasi-isomorphic to $i_* {\cal E}$, suppose that the vector bundle $\hat{\cal E}^i$ arises  via \NaturalBundle\ from a representation $\rho_i$ of $G\times G_{\rm F}$.
Then we take
\eqn\ChanPatonM{
{\cal V}=\bigoplus_i {\cal V}^i\,,
}
as the Chan-Paton space, where ${\cal V}^i$ is the representation space of 
\eqn\RepVihat{
\rho_i\otimes (\det \rho_E)^{1/2}\,.
}
The tachyon profile ${\cal Q}$ is determined by the procedure explained around in \QAnsatz.

The validity of \MapEToB\ will be checked by comparing the hemisphere partition function with the large volume formula of the D-brane central charge in Section \SecQuintic, as well as by showing that the resulting hemisphere partition functions for the structure sheaf in certain target spaces are invariant under various dualities.

\newsec{Examples}
\seclab\SecExamples

\subsec{D0-brane on $\Bbb C^n$}
\subseclab\SecKoszul

Let us consider the theory of $n$ free chiral multiplets $\phi^i$, $i=1,\ldots, n$, with target space $X=\Bbb C^n$.
The flavor symmetry $G_{\rm F}=U(n)$ allows us to consider equivariant sheaves.
In particular, the skyscraper sheaf at the origin, {\it i.e.}, the D0-brane can be resolved by the Koszul complex
\eqn\EqKoszul{
\Lambda^{n,0}
\mathop{\longrightarrow}
\Lambda^{n-1,0}
\mathop{\longrightarrow}
\ldots
\mathop{\longrightarrow}
\Lambda^{0,0}={\cal O}
\,,
}
where $\Lambda^{p,q}$ is the vector bundle of $(p,q)$-forms, and the differential is the contraction by $\phi^i \partial_i$.
The map \MapEToBForX\ can be described by fermionic oscillators obeying
$\{\eta_i,\bar\eta^j\}=\delta_i^j$ with $ i,j=1,\ldots, n$,
and the Clifford vacuum $|0\rangle$ such that $\eta_i |0\rangle =0$ for any $i$.
The tachyon profile
$$
{\cal Q}(\phi)=\phi^i \eta_i\,,
\qquad
\bar {\cal Q}(\bar\phi)=\bar\phi_i \bar\eta^i
$$
gives a realization of the differential.
The boundary contribution \EqBdryCont\  is
$ \prod_j (1-e^{2\pi i  m_j})$.
The one-loop determinant should be computed for the Neumann conditions for all $\phi^i$ since the D0-brane is constructed as a bound state of space-filling branes.
It is simply
$
\prod_j \Gamma( m_j)
$.
The hemisphere partition function of the model is therefore
\eqn\ZhemKoszul{
Z_{\rm hem}({\rm D0\mathchar`-brane})=\prod_j \Gamma( m_j) (1-e^{2\pi i  m_j})
=\prod_j {-2 \pi i e^{\pi i m_j}\over \Gamma(1- m_j)}
\,.
}
This gives the hemisphere partition function for the full Dirichlet condition.%
\foot{%
The zeros due to the gamma functions in the denominator of \ZhemKoszul\ coincide with the zeros in \OneLoopChiDerived\ for the full Dirichlet condition.
The relative normalization in \OneLoopChiral\ between the Neumann and the Dirichlet conditions was chosen to agree with \ZhemKoszul.
}

\subsec{Quintic Calabi-Yau}
\subseclab\SecQuintic

Let us consider a $G=U(1)$ theory with chiral fields $(P,\phi_1,\ldots,\phi_5)$ with charges $(-5,1,1,1,1,1)$.
We assign R-charges $(q_P,q_1,\ldots,q_5)=(-2,0,\ldots,0)$ respectively.
If we include the superpotential $W=P {\rm G}(\phi)$, where ${\rm G}$ is a degree-five polynomial, the theory with $r\gg 0$ flows to the non-linear sigma model with target space the quintic $M$, which is the hypersurface in $\Bbb P^4$ given by ${\rm G}(\phi)=0$.
Let us consider the line bundle ${\cal O}_M(n)$ obtained by pulling ${\cal O}_{\Bbb P^4}(n)$ back to $M$.
We can apply the map \MapEToB\ to construct the boundary condition ${\cal B}=({\cal V},{\cal Q})$.
The Chan-Paton space ${\cal V}$ is the fermionic Fock space spanned by $|0\rangle$ and $\bar\eta|0\rangle$ with $\{\eta,\bar\eta\}=1$, and the tachyon profile is given by
$$
{\cal Q}={\rm G}(\phi)\eta+ P\bar\eta\,.
$$
Following \RepVihat\ we assign gauge charge $n+5/2$ to $|0\rangle$.
Thus
\eqn\QuinticHemPF{
Z_{\rm hem}[{\cal O}_M(n)]
=\int_{i\Bbb R} {d\sigma\over 2\pi i}e^{-2\pi i  n\sigma}(e^{-5\pi i\sigma}-e^{5\pi i\sigma}) e^{t\sigma} \Gamma(\sigma)^5\Gamma(1-5\sigma) \,.
}
As mentioned after \ResultPreservedUOne,
convergence requires a deformation of the contour for large $|\sigma|$.
Specifically, we choose the contour to approach straight lines tilted to the left by angle $\delta>0$ from the imaginary axis, and demand that $r\delta > \theta+2\pi n$.
Deep in the geometric phase where $r\gg 0$, we can choose $\delta$ to be small.
We also demand that the contour crosses the real axis with positive ${\rm Re}\,\sigma$.%
\foot{%
One can also realize such a contour as a Lagrangian brane by a boundary condition~\HoriIKA.
}
The integral can then be evaluated by the Cauchy theorem, and is expressed as a power series in  $e^{-t}$, together with cubic polynomial terms in $t$:
\eqn\ZhemQuintic{
Z_{\rm hem}[{\cal O}_M(n)]
=
{
-{20 \over 3}
\pi^4 
}
\Big({ t\over 2\pi i}-n\Big)
\Big(
2
\Big({ t\over 2\pi i}-n\Big)^2
+5
\Big)
{-400\pi i} \zeta(3)
+{\cal O}(e^{-t})\,.
}
We can compare this with the large volume formula for the central charge (see, {\it e.g.}, \refs{\MinasianMM,\DouglasFJ,\AspinwallJR})%
\foot{%
In our convention, ${\rm ch}\,E ={\rm Tr}\exp\left(F/ 2\pi\right)$, $B+i\omega=-( t/ 2\pi i){\bf e}$, and  $F+2\pi B$ is the gauge invariant combination.
See \ABomega.
}
\eqn\ChargeLargeVolume{
\int_M {\rm ch}({\cal O}_M(n))e^{B+i\omega} \sqrt{\hat A(TM)}\,.
}
Our conventions for $B$ and $\omega$ can be found in Section \SecTarget.
Let ${\bf e}$ be the generator of $H^2(M,\Bbb Z)$ such that $\int_M {\bf e}^3=5$.
If we make the identification
$$
B+i\omega = {i t\over  2\pi}{\bf e} +{\cal O}(e^{-t})
$$ 
in the large volume limit $t\rightarrow +\infty$, \ChargeLargeVolume\ becomes
$$
 \int_M e^{n{\bf e}} e^{i t{\bf e}/2\pi} \Big(1+{5\over 6}{\bf e}^2\Big)^{1/2}
=
-{5\over 12}
\Big({ t\over 2\pi i}-n\Big)
\Big(
2\Big({ t\over 2\pi i}-n\Big)^2+5
\Big)\,,
$$
which agrees with the hemisphere partition function \ZhemQuintic\ up to an overall numerical factor, as well as constant and exponentially suppressed terms.
This is the most direct demonstration that our hemisphere partition function computes the central charge of the D-brane, or more precisely the overlap of the D-brane boundary state in the Ramond-Ramond sector and the identity closed string state.
We see that the hemisphere partition function also captures the constant term proportional to $\zeta(3)$; it is expected to arise at the four-loop order in the non-linear sigma model \refs{\GrisaruPX,\CandelasRM}.

In Appendix \AppCICY, we generalize the results here and exhibit the agreement between the hemisphere partition function and the large volume formula \ChargeLargeVolume\ for branes in an arbitrary complete intersection Calabi-Yau in a product of projective spaces.

One can also show that $Z_{\rm hem}$ satisfies a differential equation
$$
\left(\partial_t^4- 5^5 e^{-t}\prod_{j=1}^4(\partial_t-j/5)\right)
Z_{\rm hem}[{\cal O}_M(n)]=0\,.
$$
This is the well-known Picard-Fuchs equation obeyed by the periods of the mirror quintic.

\subsec{Projective spaces and Grassmannians}
\subseclab\SecProjGrass

Let us consider the theory with gauge group $G=U(1)$, $N_{\rm F}$ fundamental chiral multiplets $Q_f$ ($f=1,\ldots, N_{\rm F}$), and without a superpotential.
We denote the complexified twisted masses by $-m_f$.
For $r\gg 0$ and $m_f=0$, the classical space of vacua is the complex projective space $X=\Bbb P^{N_{\rm F}-1}$.
This is the simplest example of Case 1 discussed in Section \SecSheaves; the space $V_{\rm mat}=\Bbb C^{N_{\rm F}}$ of matter fields carries charge $+1$ under $G=U(1)$ and the anti-fundamental representation $\bar{\bf N}_{\rm F}$ of the flavor group $G_{\rm F}=U(N_{\rm F})$.

The D-brane carrying $n$ units of the gauge flux is the line bundle ${\cal O}(n)$.
The derived category of coherent sheaves $D(X)$, as well as the K theory $K(X)$ and their $G_{\rm F}$-equivariant versions, is known to be generated by the Beilinson basis, ${\cal O}(n)$ with $0\leq n\leq N_{\rm F}-1$.
The hemisphere partition function of ${\cal O}(n)$ is given by
$$
Z_{\rm hem}({\cal O}(n))
=\int_{-i\infty}^{i\infty} {d\sigma\over 2\pi i}e^{-2\pi i  n\sigma} e^{t_{\rm ren}\sigma}\prod_{f=1}^{N_{\rm F}} \Gamma(\sigma-m_f)\,.
$$
If $r\gg0$, for convergence we tilt the contour in the asymptotic region toward the negative real direction as ${\rm Im}\,\sigma\rightarrow \pm\infty$.
If ${\rm Re}\, m_f<0$ we simply close the contour along the imaginary axis to the left and compute the integral by picking up the poles at $\sigma=m_f-k$, $k\in \Bbb Z_{\geq 0}$.
For other values of $m_f$ we define the integral by analytic continuation, or equivalently by choosing the contour in the intermediate region so that we pick the same poles.
$$
Z_{\rm hem}({\cal O}(n))
=
\sum_{v=1}^{N_{\rm F}} e^{m_v(t_{\rm ren}-2\pi i n)}
\sum_{k=0}^\infty
e^{-k t_{\rm ren}}
{(-1)^k \over k!}
\prod_{f\neq v}\Gamma(m_{vf}-k)\,,
$$
where $m_{vf}=m_v-m_f$.

Next we consider the theory with gauge group $G=U(N)$, $N_{\rm F}$ fundamental chiral multiplets $Q^i_f$ ($i=1,\ldots,N$ and $f=1,\ldots, N_{\rm F}$), and with no superpotential.
Again the complexified twisted masses will be denoted by $-m_f$.
For $r\gg 0$ and $N\leq N_{\rm F}$
the target space of the low-energy theory is the Grassmannian $X={\rm Gr}(N,N_{\rm F})$ of $N$-dimensional subspaces in $\Bbb C^{N_{\rm F}}$.
The flavor group $G_{\rm F}=U(N_{\rm F})$ acts on $X$ naturally.
Let $V$ be a vector space in some representation of $G\times G_{\rm F}$.
For the corresponding holomorphic vector bundle $E$ given by \NaturalBundle, the hemisphere partition function is given by
$$
Z_{\rm hem}({\cal O}(E))=
{1\over N!} \int_{i\Bbb R^N} {d^N\sigma\over (2\pi i)^N}
{\rm Tr}_V\big[ e^{-2\pi i (\sigma+m)}\big] 
e^{t_{\rm ren} {\rm Tr}\,\sigma}
\prod_{i<j}\sigma_{ij}{\sin\pi\sigma_{ji}\over \pi}
\prod_{f=1}^{N_{\rm F}}\prod_{j=1}^N \Gamma(\sigma_j-m_f)\,.
$$
We take the traces by viewing $\sigma$ as a diagonal matrix, and  abbreviate symbols as $\sigma_{ij}=\sigma_i-\sigma_j$, $m_{fg}=m_f-m_g$.
Let us assume that $r\gg0$.
The integral can be computed by the residue theorem.
We will frequently use the notation
\eqn\Defv{
{\bf v}=\{f_1<f_2<\ldots<f_N\}\subseteq \{1,\ldots,N_{\rm F}\}
}
to label the sequences of poles.
These should correspond to the classical Higgs vacua that are the saddle points in a different localization scheme \refs{\BeniniUI,\DoroudXW}.
We also denote the complement sets as
$$
{\bf v}^\vee= \{1,\ldots,N_{\rm F}\} \backslash{\bf v}\,.
$$
Let us define $m^{\bf v}=(m^{\bf v}_j)$ by
\eqn\Defmv{
m^{\bf v}_j=m_{f_j}\,.
}
Picking up the poles at
\eqn\UNPoles{
\sigma_j=
m^{\bf v}_{\vec k} \equiv m^{\bf v}_j-k_j\,,
\qquad
k_j\in\Bbb Z_{\geq 0}\,,
}
and using the vortex partition function defined in \DefVortex, we obtain
\eqn\EqPFGrassmann{ \eqalign{
Z_{\rm hem}({\cal O}(E))
=
\sum_{{\bf v}}
{\rm Tr}_V\big( e^{-2\pi i (m^{\bf v}+m)}\big)
e^{t_{\rm ren} {\rm Tr}\, m^{\bf v}}
\Big(
\prod_{f\in{\bf v}} \prod_{g\in {\bf v}^\vee} \Gamma( m_{fg})
\Big)
Z_{\rm vortex}^{{\bf v}}(t_{\rm ren};  m)
\,.
}}

\subsec{Cotangent bundles of Grassmannians $T^* {\rm Gr}(N,N_{\rm F})$}
\subseclab\SecTstarGr
Let us consider the theory with gauge group $G=U(N)$, $N_{\rm F}$ fundamentals $Q^i{}_f$ and anti-fundamentals $\tilde Q^f{}_i$ and one adjoint $\Phi^i{}_j$  ($i,j=1,\ldots,N$ and $f=1,\ldots,N_{\rm F}$).
We include the superpotential
$$
W={\rm Tr}\,\tilde Q \Phi Q\,.
$$
For $r\gg0$, the theory flows to the non-linear sigma model with target space the cotangent bundle of the Grassmannian  $M=T^* {\rm Gr}(N,N_{\rm F})$, with $\Phi$ playing the role of $P$ in Section \SecTarget.
We denote the twisted masses of $(Q_f,\tilde Q^f,\Phi)$ by $(-m_f,1+m_f-m_{\rm ad},m_{\rm ad})$ respectively.

We illustrate the Hilbert space interpretation in Section \SecHilbert\ using this model.
We choose $w_0$ in the formula \Spheregf\ for the two-point function $\langle {\tt g}|{\tt f}\rangle$ so that $w_0\cdot B=(N-1)\sum B_j$.
The integral \Spheregf\ can be evaluated as in \BeniniUI.
It becomes
\eqn\gfTStar{\eqalign{
\langle {\tt g}|{\tt f}\rangle
&=
c\sum_{\bf v}
e^{(t+\bar t){\rm Tr}\, m^{\bf v}}
\prod_{f\in {\bf v}}
\prod_{g\in {\bf v}^\vee}
{
\Gamma( m_{fg})\Gamma(1- m_{fg} - m_{\rm ad}) 
\over
\Gamma(1- m_{fg})
\Gamma( m_{fg} + m_{\rm ad})
}
\cr
&\qquad
\times
Z^{\bf v}_{\rm vortex}(\bar t;  m;{\tt g})
Z^{\bf v}_{\rm vortex}(t; m;{\tt f})
\,,
}}
where ${\bf v}$ and $m^{\bf v}$ were defined in Section \SecProjGrass, and $Z_{\rm vortex}^{{\bf v}}(t;m;{\tt f})$ is a generalization of the vortex partition function \DefVortex
$$\eqalign{
&\qquad
Z_{\rm vortex}^{{\bf v}}(t; m {;{\tt f}})
\cr
&=
\sum_{{\vec  k}\in {\Bbb Z}_{\geq 0}^N}
e^{-|\vec k| t}
{ {\tt f}(m^{\bf v}_{\vec k} })
\prod_{i}
\bigg(
\prod_{j} { ( m_{f_i f_j}+\adm- k_i)_{k_j} \over ( m_{f_i f_j} - k_i)_{k_j} }
\prod_{f \in {\bf v}^\vee } { ( m_{f_i f} +\adm - k_i)_{k_i} \over (  m_{f_i f} - k_i)_{k_i} }
\bigg)\,.
}$$

By defining
\eqn\vfTStarGr{\eqalign{
\langle{\bf v}|{\tt f}\rangle
&=
c^{1\over 2}
e^{t {\rm Tr}\, m^{\bf v}}
\bigg[
\prod_{f\in {\bf v}}\prod_{g\in {\bf v}^\vee}
{
\Gamma( m_{fg})\Gamma(1- m_{fg} - m_{\rm ad}) 
\over
\Gamma( m_{fg} + m_{\rm ad})
\Gamma(1- m_{fg})
}
\bigg]^{1\over 2}
Z^{\bf v}_{\rm vortex}(
t ;  m;{\tt f})
}}
and
$$\eqalign{
\langle {\tt g}|{\bf v}\rangle
&=
c^{1\over 2}
e^{\bar t {\rm Tr}\, m^{\bf v}}
\bigg[
\prod_{f\in {\bf v}}\prod_{g\in {\bf v}^\vee}
{
\Gamma( m_{fg})\Gamma(1- m_{fg} - m_{\rm ad}) 
\over
\Gamma( m_{fg} + m_{\rm ad})
\Gamma(1- m_{fg})
}
\bigg]^{1\over 2}
Z^{\bf v}_{\rm vortex}(
\bar t;  m;{\tt g})
}$$
we can write
$
\langle {\tt g}|{\tt f}\rangle=\sum_{\bf v} \langle {\tt g}|{\bf v}\rangle \langle {\bf v}|{\tt f}\rangle
$.

In order to justify our choice of $w_0$ and relate $c$ to the normalization of hemisphere partition functions, let us compute the hemisphere partition function $Z_{\rm hem}({\cal O}_M)=\langle {\cal B}[{\cal O}_M]|{\tt 1}\rangle$ and more generally $\langle {\cal B}[{\cal O}_M]|{\tt f}\rangle$ for the structure sheaf ${\cal O}_M$.
We can use the matrix factorization \QForOM.
In the present notation we introduce oscillators $(\eta^i{}_j, \bar\eta^i{}_j)$ satisfying $\{\eta^i{}_j,\bar\eta^k{}_l\}=\delta^i_l\delta^k_j$, and let $|0\rangle$ be the Clifford vacuum: $\eta^i{}_j|0\rangle=0$.
Then 
$$
{\cal Q}=
Q  \tilde Q \eta+\Phi\bar\eta
$$
with the indices contracted.
Assigning the abelian charges symmetrically between $|0\rangle$ and $\prod_{i,j}\bar\eta^i{}_j|0\rangle$ as in \VacuaReps, we find the contribution
$
\prod_{i,j=1}^N 
2i\sin\pi( \sigma_{ij}+ m_{\rm ad})
$
from the boundary interaction.
We will see in Section \SecTStarDuality\ that for a geometrically expected duality to hold, we need to multiply the hemisphere partition function \ResultPreservedGeneral\ by an extra $N$-dependent overall factor, {\it e.g.}, ${ (2\pi i)^{-N^2}}$.
We thus go ahead and include it.
Then\foot{%
Compared with \ResultPreservedGeneralOneLoop, we see that the boundary interaction has an effect of changing the boundary condition for $\Phi$ from Neumann to Dirichlet.
}
\eqn\HemPFTGrOM{\eqalign{
Z_{\rm hem}({\cal O}_M)
&=
\int^{i\infty}_{-i\infty} {d^N  \sigma \over (2\pi i)^N N!}
e^{t{\rm Tr}\,\sigma} \prod_{i<j}\sigma_{ij} {\sin\pi \sigma_{ji}\over \pi}
\prod_{i,j=1}^N 
{\sin\pi( \sigma_{ij}+ m_{\rm ad}) \over \pi}
\cr
&\qquad\times
\prod_{i,j=1}^N\Gamma(\sigma_{ij}+ m_{\rm ad})
\prod_{j=1}^N \prod_{f=1}^{N_{\rm F}} \Gamma(\sigma_j - m_f)\Gamma({1}-\sigma_j + m_f - m_{\rm ad})\,.
}}
By applying \ResultResidue\
we find 
\eqn\ZhemOMTStar{
Z_{\rm hem}({\cal O}_M)
=
\sum_{\bf v}
e^{t{\rm Tr}\,m^{\bf v}}
\bigg[
\prod_{f\in{\bf v}}\prod_{g\in{\bf v}^\vee} \Gamma(m_{fg})\Gamma(1- m_{fg} - m_{\rm ad})
\bigg]
Z_{\rm vortex}^{\bf v}(t;m)\,.
}
Note that the same argument $t$ as in \gfTStar\ appears in the vortex partition function here; this is only possible for our choice of $w_0$.
We can compute $\langle {\cal B}[{\cal O}_M]|{\tt f}\rangle$ similarly.
Comparing with \vfTStarGr, we find that 
$
\langle {\cal B}[{\cal O}_M]|{\tt f}\rangle =\sum_{\bf v}\langle {\cal B}[{\cal O}_M]|{\bf v}\rangle\langle{\bf v}|{\tt f}\rangle$,
where 
\eqn\BOMvTStar{
\langle {\cal B}[{\cal O}_M]|{\bf v}\rangle=
c^{-1/2}
\Big[\prod_{f\in {\bf v}}\prod_{g\in {\bf v}^\vee} {\pi \over \sin\pi  m_{fg}}
{\pi \over \sin\pi( m_{fg}+ m_{\rm ad})}\Big]^{1/2}
\,.
}
A parallel consideration shows that $\langle {\cal B}[{\cal O}_M]|{\bf v}\rangle=\langle {\bf v}|{\cal B}[{\cal O}_M]\rangle$, giving an expression for the cylinder partition function.
It is expected to coincide with the equivariant index of the Dirac operator on $M$.
Indeed $\langle {\cal B}[{\cal O}_M]|{\cal B}[{\cal O}_M]\rangle$ determined by \BOMvTStar\ 
agrees with%
\foot{%
It is possible to show by localization that the equivariant Dirac index given by \ZcylAB, or more generally by \ZcylABTwo, is indeed the corresponding partition function on the cylinder.
} 
\eqn\ZcylAB{
{\rm ind}(\displaystyle{\not} D)
=
\sum_{p:\, {\rm fixed \, points}}
{
1
\over
\det_{TM_p}(g^{-1/2}-g^{1/2})
}
}
if we take $c=(2\pi)^{2N(N_{\rm F}-N)}$.

It is trivial to generalize these results to a holomorphic vector bundle $E$, or equivalently the sheaf ${\cal O}_M(E)$ of holomorphic sections of $E$.
We assume that $E$ arises via \NaturalBundle\ from a vector space $V$ carrying a representation of $(G\times G_{\rm F})_{\Bbb C}$.
We find 
$$
\langle {\cal B}[{\cal O}_M(E)]|{\bf v}\rangle=
{\rm Tr}_V e^{-2\pi i(m^{\bf v}+m)}
\Big[\prod_{f\in {\bf v}}\prod_{g\in {\bf v}^\vee} {1 \over2 \sin\pi  m_{fg}}
{1 \over 2\sin\pi( m_{fg}+ m_{\rm ad})}\Big]^{1/2}
\,.
$$

Another class of natural D-branes are sheaves supported on the zero-section of $T^* {\rm Gr}(N,N_{\rm F})$.
Let us consider a vector bundle over $ {\rm Gr}(N,N_{\rm F})$ and call it $E$, abusing notation slightly.
We assume that $E$ is constructed from a representation $V$ of $(G\times G_{\rm F})_{\Bbb C}$.
We wish to compute 
the hemisphere partition function for the sheaf $\iota_* {\cal O}_{\bf Gr}(E)$, where $\iota$ is the inclusion.
Following the procedure for Case 2 in Section \BCcomplexes, we further pushforward $\iota_* {\cal O}_{\bf Gr}(E)$ by the inclusion $i:M\rightarrow X$, where 
\eqn\XTStar{
X=\{(Q,\tilde Q)|{\rm rk}\,Q={N} \}/GL(N)\,.
}
Since ${\bf Gr}$ is given in $X$ simply by the equations $\tilde Q^f=0$, we have a locally-free resolution of $i_* \iota_* {\cal O}_{\bf Gr}$,
\eqn\ResOGr{
\wedge^{\rm r} F^* 
\longrightarrow 
\ldots
\longrightarrow 
\wedge^2 F^* \longrightarrow 
F^*
\longrightarrow 
{\cal O}_X\,,
}
where ${\rm r}=NN_{\rm F}$ is the rank of the equivariant vector bundle $F$, of which $(\tilde Q^f)$ defines a section.
A resolution of $i_* \iota_* {\cal O}_{\bf Gr}(E)$ is obtained by tensoring each term in \ResOGr\ with the bundle $\hat E$ over $X$ that arises from $V$ via \NaturalBundle.
The complex \ResOGr\ can be translated into the boundary interaction by introducing oscillators satisfying $\{\eta^i{}_f,\bar\eta^g{}_j\}=\delta^i_j\delta^g_f$.
The Chan-Paton space ${\cal V}$ is obtained by tensoring with $V$ the Fock space built on the vacuum $|0\rangle$ annihilated by $\eta^f{}_j$, and the tachyon profile is given by
$
{\cal Q}=\tilde Q^f{}_i \eta^i{}_f {+\Phi^i{}_j Q^j{}_f \bar\eta^f{}_i }
$.
According to \RepVihat, we must assign the same abelian charges to $|0\rangle$ as in the ${\cal O}_M$ case.
Then $|0\rangle$ contributes the factor $e^{N^2 \pi i m_{\rm ad}}$.
We find the integral representation
\eqn\TStarZhemiOGrE{\eqalign{
Z_{\rm hem}(\iota_*{\cal O}_{\bf Gr}(E))
&=
\Big[{
 e^{\pi i m_{\rm ad}}
\over 2\pi i
}\Big]^{N^2}
\int{ d^N  \sigma \over (2\pi i)^N N!}
e^{t{\rm Tr}\,\sigma}
\prod_{j,f}
\big(1-e^{-2\pi i(\sigma_j-m_f+m_{\rm ad})}\big)
\cr
&\ \quad\times
\prod_{i<j}\sigma_{ij}{\sin\pi \sigma_{ji}\over\pi}
{\rm Tr}_V(e^{-2\pi i (\sigma+m)})
\prod_{i,j}\Gamma(\sigma_{ij}+ m_{\rm ad})
\cr
&\ \qquad
\times
\prod_{j,f}\Gamma(\sigma_j - m_f)\Gamma(1-\sigma_j + m_f - m_{\rm ad})\,.
}}
As we see by comparing with \ResultPreservedGeneralOneLoop\ an effect of the boundary interaction is to modify the boundary condition for $\tilde Q^f$ from the Neumann to the Dirichlet condition, as we expect for a brane supported on the zero-section.
Only the sequences of poles \UNPoles\ contribute, with other combinations of apparent poles canceled.%
\foot{%
Here $\vec r$ in \CI\ is given by $\vec r=(r,\ldots,r)$.
It is not possible to satisfy the conditions $r_{aw}>0$ in \CI\ if $I$ involves an anti-fundamental.
If $I$ involves the adjoint and fundamentals, the zeros from the product in the first line of \TStarZhemiOGrE\ cancel the poles.
}
We then find
\eqn\HemPFTGrOGr{\eqalign{
Z_{\rm hem}(\iota_*{\cal O}_{\bf Gr}(E))
&=
e^{N\pi i\sum_f m_f} \sum_{\bf v} {\rm Tr}_V \big(e^{-2\pi i (m^{\bf v}+m)}\big) e^{(t-N_{\rm F}\pi i){\rm Tr}\,m^{\bf v}} 
\cr
&\qquad\times
\Big(\prod_{f\in {\bf v}}\prod_{g\in{\bf v}^\vee}
{
2\pi i e^{-\pi i m_{\rm ad}}\Gamma(m_{fg})
\over
\Gamma(m_{fg} + m_{\rm ad})
}
\Big)
Z_{\rm vortex}(t;m)\,.
}}
By identifying this with $\sum_{\bf v}\langle {\cal B}[\iota_* {\cal O}_{\rm Gr}(E)]|{\bf v}\rangle \langle {\bf v}|{\tt 1}\rangle$ and using \vfTStarGr, we obtain
\eqn\TStarBOGrEv{\eqalign{
\langle {\cal B}[\iota_* {\cal O}_{\rm Gr}(E)]|{\bf v}\rangle
&=
e^{N_{\rm F}\pi i(\sum_f m_f-{\rm Tr}m^{\bf v})} e^{-N(N_{\rm F}-N)\pi i m_{\rm ad}}
 i^{N(N_{\rm F}-N)}
\cr
&\qquad\times
{\rm Tr}_V \big(e^{-2\pi i (m^{\bf v}+m)}\big)
\prod_{f\in {\bf v}}\prod_{g\in{\bf v}^\vee}
\left[
{
\sin \pi (m_{fg}+m_{\rm ad})
\over
\sin \pi m_{fg}
}
\right]^{1/2}\,.
}}
The matrix element $\langle{\bf v}| {\cal B}[\iota_* {\cal O}_{\rm Gr}(E)]\rangle$ is obtained by replacing $i$ with $-i$ in \TStarBOGrEv.

\newsec{Seiberg-like dualities}
\seclab\SecDualities

\subsec{Grassmannian model and the $(N,N_{\rm F})\leftrightarrow (N_{\rm F}-N,N_{\rm F})$ duality}
\subseclab\SecGrDuality

Recall from Section \SecProjGrass\ that the $U(N)$ theory with $N_{\rm F}\geq N $ fundamental chiral multiplets $Q_f$ with $r\gg 0$ is in the geometric phase with target space the Grassmannian ${\rm Gr}(N,N_{\rm F})$.
To simplify equations we can take the flavor symmetry group to be $SU(N_{\rm F})$ since the overall $U(1)$ is part of the gauge group.
Correspondingly, we require that the twisted masses $-m_f$ of $Q_f$ sum to zero:
 \eqn\EqMassSumZero{
 \sum_{f=1}^{N_{\rm F}}{m}_f=0\,.
 }

The hemisphere partition function was computed in \EqPFGrassmann.
Let us focus on the structure sheaf ${\cal O}$ and consider the map of parameters
\eqn\EqDualityMap{
(N,N_{\rm F}, t_{\rm ren},  m) \rightarrow (N_{\rm F}-N,N_{\rm F}, t_{\rm ren} , - m)\,.
}
The exponential factor in \EqPFGrassmann\ is invariant because of \EqMassSumZero.
The one-loop determinant is also manifestly invariant under \EqDualityMap\
and ${\bf v}\rightarrow {\bf v}^\vee$.
As shown in \BeniniUI\ the vortex partition function $Z^{\bf v}_{\rm vortex}$ is also invariant.
Thus we have the equality
$$
Z_{\rm hem}[{\rm Gr}(N,N_{\rm F});{\cal O};t_{\rm ren};  m]
=
Z_{\rm hem}[{\rm Gr}(N_{\rm F}-N,N_{\rm F});{\cal O};t_{\rm ren};- m]
$$
for the structure sheaf.
This equality extends to D-branes carrying vector bundles 
$$
Z_{\rm hem}[{\rm Gr}(N,N_{\rm F});E;t_{\rm ren};  m]
=
Z_{\rm hem}[{\rm Gr}(N_{\rm F}-N,N_{\rm F});E^\vee;t_{\rm ren};- m]
$$
if we define the map $E\mapsto E^\vee$, in a way compatible with tensor product, by the assignments
$$\eqalign{
{\rm tautological\ bundle} &\longmapsto 
({\cal O}^{N_{\rm F}}/{\rm tautological\ bundle})^*\,,
\cr
{\cal O}^{N_{\rm F}}/{\rm tautological\ bundle}&\longmapsto 
({\rm tautological\ bundle})^* \,.
}$$
We denoted by $*$ the dual bundle (in the usual sense), whose fiber is the dual of the fiber for the original bundle.
(Somewhat confusingly, the quotient, ${\cal O}^{N_{\rm F}}/{\rm tautological\ bundle}$, is sometimes called the dual tautological bundle.)
We also recall that the tautological bundle is constructed from the anti-fundamental representation of $GL(N)$ via \NaturalBundle.%
\foot{%
The assignment $V\mapsto {\rm Tr}_V[{\rm diag}(x_1^{-1},\ldots,x_N^{-1})\times {\rm diag}(x_{1}^{-1},\ldots,x_{N_{\rm F}}^{-1})]$ defines a map $D(X)\rightarrow K^{GL(N_{\rm F})}(X)\simeq \Bbb C[x_1^{\pm 1},\ldots,x_N^{\pm 1};x_{N+1}^{\pm 1},\ldots,x_{N_{\rm F}}^{\pm 1}]^{S_N\times S_{N_{\rm F}-N}}$ for $X={\rm Gr}(N,N_{\rm F})$ \NakajimaQAA.
}

\subsec{$T^*{\rm Gr}(N, N_F)$ model}
\subseclab\SecTStarDuality

The hemisphere partition function for ${\cal O}_{T^*{\rm Gr}(N,N_{\rm F})}$ was computed in \ZhemOMTStar.
We again impose the condition \EqMassSumZero\ on the fundamental masses.
Under the map
$$
N\to N_F-N\,,\ t\to t\,,\ {m}_f \to -{m}_f\,,\ \adm \to \adm\,,\ {\bf v}\to {\bf v}^\vee\,,
$$
the exponential factor and the one-loop determinant are invariant.
The vortex partition functions $Z_{\rm vortex}^{U(N), \bf v}(t;  m_f, \adm)\equiv Z_{\rm vortex}^{\bf v}(t;  m_f, \adm)$ are not invariant, but we found the relations
\eqn\TStartVortexDuality{
(1+(-1)^{N_F} e^{-t})^{(N_F-2N)(\adm-1)}Z_{\rm vortex}^{U(N), \bf v}(t;  m_f, \adm)
=
Z_{\rm vortex}^{U(N_F-N), \bf v^\vee}(t; - m_f, \adm)
}
by comparing the power series expansions in $e^{-t}$.%
\foot{%
Similar relations hold between instanton partition functions computed in different schemes for ALE spaces \ItoKPA.
}
Since the prefactor on the left hand side is independent of ${\bf v}$, we find a similar relation for the hemisphere partition functions.%
\foot{%
A similar relation also holds for the sphere partition functions.
}
In particular, in the limit ${\rm Re}\,t \gg 0$ the hemisphere partition function is invariant.
The same relation holds for the hemisphere partition functions of $\iota_*{\cal O}_{\bf Gr}$.
It can also be extended to include vector bundles as we did for Grassmannians in Section \SecGrDuality.

\subsec{$U(N)$ gauge group with fundamental and determinant matter fields}

Let us consider the Grassmannian model with an extra chiral multiplet in the $(-N_{\rm F})$-th power of the determinant representation with twisted mass ${m}_{\rm det}$.
For simplicity we impose the Dirichlet condition for the determinant matter and the Neumann condition for the fundamentals.
Then the hemisphere partition function is
$$
Z_{\rm hem}(N, N_{\rm F};t;  m_f,{m}_{\rm det})
=
\sum_{{\bf v}}
e^{t {\rm Tr}\, m^{\bf v}  }
Z^{\bf v}_{\rm 1\mathchar`-loop}( m_f,{m}_{\rm det})
Z_{\rm vortex}^{{\bf v}}(t;  m_f,{m}_{\rm det})
$$
with the one-loop determinant given by
$$
Z_{\rm 1\mathchar`-loop}^{\bf v}(  m_f,{m}_{\rm det})=
{- 2 \pi i e^{\pi i \left( -N_{\rm F}{\rm Tr}\, m^{\bf v} + {m}_{\rm det}\right)} \over \Gamma\left( 1 + N_{\rm F} {\rm Tr}\, m^{\bf v} -{m}_{\rm det}\right)}
\prod_{f\in{\bf v}} \prod_{g\in {\bf v}^\vee} \Gamma( m_{fg})
$$
and the vortex partition function defined in \DefVortex.
It was found in \GaddeDDA\ that the superconformal index of this model is invariant under
$$
N\to N_F-N,\ t\to t,\ {m}_f \to -{m}_f,\ {m}_{\rm det}\to {m}_{\rm det},\ {\bf v}\to {\bf v}^\vee\,.
$$
One can show that the vortex partition functions in this case are duality invariant, by noting that they are simply related to those of the Grassmannian model.
Thus the hemisphere partition function is also invariant under the duality map.

\subsec{$SU(N)$ gauge theories}

To study Seiberg-like dualities for $SU(N)$ theories, we use a trick introduced in \BeniniUI;
the hemisphere partition function of the $SU(N)$ gauge theory is related to that of the $U(N)$ gauge theory by
$$
Z^{SU(N)}_{\rm hem}(b)= \int^{\infty}_{-\infty} {dr\over 2\pi}e^{-r b}  Z^{U(N)}_{\rm hem}(r,\theta=0).
$$
Then the duality of the $U(N)$ hemisphere partition function implies a duality of the $SU(N)$ hemisphere partition function.

The $U(1)$ baryonic symmetry is defined by its action on the fundamentals $Q^i{}_f$ ($i=1,\ldots N,\, f=1,\ldots, N_{\rm F}$) and the anti-fundamentals $\tilde Q^{\tilde f}{}_i$ ($i=1,\ldots N,\, f=1,\ldots, N_{\rm A}$)
$$
Q^i{}_f \rightarrow e^{2 \pi i b/N} Q^i{}_f,\,\quad \tilde Q^{\tilde f}{}_i \rightarrow e^{-2 \pi i b/N} \tilde Q^{\tilde f}{}_i\,.
$$
It is the $U(1)$ part of the $U(N)$ gauge group that we ungauge.
The baryonic and the anti-baryonic operators
$$
B_{f_1,\ldots,f_N}=\varepsilon_{i_1 \ldots i_N} Q^{i_1}_{\ f_1} \cdots Q^{i_N}_{\ f_N},\,
\qquad
\tilde B^{\tilde f_1,\ldots,\tilde f_N}=\varepsilon^{i_1 \ldots i_N} \tilde Q^{\tilde f_1}_{\ i_1} \cdots \tilde Q^{\tilde f_N}_{\ i_N}
$$
in the $SU(N)$ theory are charged under this $U(1)$.
The pure-imaginary parameter $b$, which is dual to the FI parameter $r$, becomes the twisted mass for the baryonic symmetry. Indeed starting with the Coulomb branch representation \ResultPreservedGeneral\ of $Z^{U(N)}_{\rm hem}$, the delta function given by the $r$ integral
$$
\int^{\infty}_{-\infty} {dr\over 2\pi} e^{-r b} e^{r \Tr  \sigma}=\delta(ib-i\Tr \sigma)
$$
produces the hemisphere partition function for the $SU(N)$ theory.

\newsec{Monodromies and domain walls}
\seclab\SecWall

\subsec{Localization with domain walls}
\subseclab\SecLocWall

In this section we consider supersymmetric localization for theories with domain walls preserving B-type supersymmetries.
Let us assume that a domain wall is located along the circle $\vartheta=\pi/2$ of the sphere $\Bbb S^2$.
The domain wall connects theory ${\cal T}_1$ on the first hemisphere $0\leq\vartheta \leq \pi/2$ and another theory ${\cal T}_2$ on the second hemisphere $\pi/2 \leq\vartheta\leq \pi$.
As we review below, the theory ${\cal T}_2$ can be mapped to another theory ${\cal I}[ {\cal T}_2]$ on the first hemisphere.
A domain wall is then defined as a D-brane in the {\it folded theory} ${\cal T}_1\times {\cal I}[{\cal T}_2]$ on the first hemisphere $0\leq\vartheta \leq \pi/2$.
When both ${\cal T}_1$ and ${\cal T}_2$ are in geometric phases, the BPS domain walls, or line operators, are in a one-to-one correspondence with objects in the derived category of equivariant coherent sheaves in the product of the target spaces.

Let us consider an involution%
\foot{%
If we regard 2d ${\cal N}=(2,2)$ supermultiplets as 4d ${\cal N}=1$ multiplets independent of two coordinates $(x^3,x^4)$,
the involution ${\cal I}_0$ acts as a reflection $(\vartheta,\varphi, x^3,x^4) \mapsto (\pi-\vartheta,\varphi, x^3,-x^4)$ followed by a $U(1)_R$ transformation.
The SUSY parameters transform as ${\cal I}_0\cdot \epsilon(\vartheta,\varphi)=\gamma_{\hat 1}\epsilon(\pi-\vartheta,\varphi)$, ${\cal I}_0\cdot\bar \epsilon(\vartheta,\varphi)=\gamma_{\hat 1}\bar\epsilon(\pi-\vartheta,\varphi)$.
Invariant parameters give the supercharges that commute with ${\cal I}_0$.
}
${\cal I}_0$ that acts on a chiral multiplet $(\phi,\psi,F)$ as
$$\eqalign{
{\cal I}_0\cdot\phi(\vartheta,\varphi)=\phi(\pi-\vartheta,\varphi)\,,
\quad
{\cal I}_0\cdot\psi(\vartheta,\varphi)=
-
\gamma_{\hat 1}\psi(\pi-\vartheta,\varphi)\,,
\cr
{\cal I}_0\cdot \bar\psi(\vartheta,\varphi)=
-
\gamma_{\hat 1}\bar\psi(\pi-\vartheta,\varphi)\,,
\quad
{\cal I}_0\cdot F(\vartheta,\varphi)= -F(\pi-\vartheta,\varphi)\,.
}$$
On a vector multiplet $(A_\mu,\sigma_{1,2},\lambda,{\rm D})$, we define
$$\eqalign{
&{\cal I}_0\cdot A_\vartheta(\vartheta,\varphi)=-A_\vartheta(\pi-\vartheta,\varphi)\,,
\quad
{\cal I}_0\cdot A_\varphi(\vartheta,\varphi)=
A_\varphi(\pi-\vartheta,\varphi)\,,
\cr
&{\cal I}_0\cdot \sigma_1(\vartheta,\varphi)=
-\sigma_1(\pi-\vartheta,\varphi)\,,
\quad
{\cal I}_0\cdot \sigma_2(\vartheta,\varphi)= \sigma_2(\pi-\vartheta,\varphi)\,,
\cr
&{\cal I}_0\cdot\lambda(\vartheta,\varphi)=\gamma_{\hat 1}\lambda (\pi-\vartheta,\varphi)\,,
\quad
{\cal I}_0\cdot\bar\lambda(\vartheta,\varphi)=\gamma_{\hat 1}\bar\lambda (\pi-\vartheta,\varphi)\,,
\cr
&\hskip 2.5cm
{\cal I}_0\cdot {\rm D}(\vartheta,\varphi)= {\rm D}(\pi-\vartheta,\varphi)\,.
}$$
One can define a more general involution ${\cal I}\equiv {\cal I}_{1}\circ {\cal I}_0$ by composing ${\cal I}_0$ with a discrete flavor$+$gauge  symmetry transformation ${\cal I}_1$ that acts on each chiral multiplet as multiplication by $+1$ or $-1$.
If the theory has superpotential $W$, the signs need to be chosen so that $W({\cal I}\cdot\phi)=+W(\phi)$. 
Then ${\cal L}_W$ in \LW\ changes sign and the action is invariant under ${\cal I}$.
The theory ${\cal I}[{\cal T}]$ is obtained from the original theory ${\cal T}$ by mapping the fields using ${\cal I}$., and by replacing the sign of the theta angle $\theta$.

The trivial domain wall, which we will call the identity domain wall $\Bbb W[{\bf 1}]$, corresponds to a single theory ${\cal T}$ with gauge group $G$ on the full sphere $0\leq \vartheta\leq \pi$.
If we apply ${\cal I}$ to the part of the theory on $\pi/2 \leq \vartheta\leq \pi$, then we get the product theory ${\cal T}\times {\cal I}[{\cal T}]$ with gauge group $G\times G$ on the hemisphere $0\leq \vartheta\leq \pi/2$.
If ${\cal T}$ has gauge group $G$, the product theory has gauge group $ G\times  G$.
Thus the identity domain wall provides an example of a supersymmetric boundary condition that reduces gauge symmetry; along the boundary the unbroken gauge group is the diagonal subgroup $(G\times G)_{\rm diag}\simeq G$.

\bigskip
\centerline{
\epsfxsize 2.3 in \epsfbox{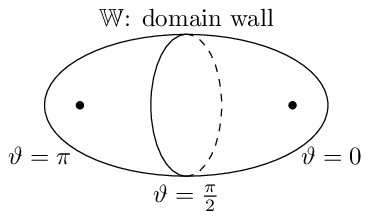}
\hskip 5 mm
\epsfxsize 2 in \epsfbox{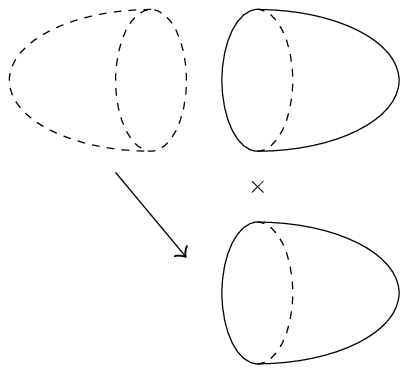}
\hskip 12mm
}
\centerline{\sl \hskip 0.2cm (a) \hskip 6.3cm (b)}
 \noindent{\ninepoint\sl
\centerline{ {\bf Figure 2}
 {{(a) A sphere with a domain wall $\Bbb W$.  (b) Folding the full sphere. 
 }}
}
}\bigskip
\noindent

If ${\cal T}$ is in a geometric phase with low-energy target space $X$ and if we take ${\cal I}={\cal I}_0$, the identity domain wall is realized by the boundary condition corresponding to the diagonal $\Delta X$ of $X\times X$:
$$
{\cal B}[\Bbb W({\bf 1})]={\cal B}[{\cal O}_{\Delta X}]\,.
$$
The general pairing \Spheregf\ between the (twisted) chiral and anti-chiral operators can be written as
$$
\langle {\tt g}|{\tt f}\rangle
=
\langle {\tt g}|\Bbb W({\bf 1})|{\tt f}\rangle
=
\langle {\cal B}[{\cal O}_{\Delta X}]|\cdot
|{\tt f}\rangle_1\otimes |{\tt g}\rangle_2\,.
$$

In the rest of the section, we will be studying the expectation values of more general domain walls $\Bbb W$ on $\Bbb S^2$
\eqn\WallVev{
\langle {\Bbb W}\rangle_{\Bbb S^2}=
\langle {\tt 1}| {\Bbb W}| {\tt 1}\rangle
=\langle {\cal B}[{\Bbb W}]|\cdot |{\tt 1}\rangle_1\otimes |{\tt 1}\rangle_2
}
or more generally the matrix elements (see Figure 2)
$$
\langle {\tt g}| {\Bbb W}| {\tt f}\rangle
=\langle {\cal B}[{\Bbb W}]|\cdot |{\tt f}\rangle_1\otimes |{\tt g}\rangle_2\,.
$$

\subsec{Monodromy domain walls, 4d line operators, and Toda theories}
\subseclab\SecToda

We now apply the machinery we have developed to find a 2d gauge theory realization of certain 4d line operators bound to a surface operator \refs{\GukovJK,\AldayFS,\GaiottoTF}.
To avoid clutter, details of calculations are relegated to Appendix \AppTodaDetails.

The relevant 4d theory is the ${\cal N}=2$ theory with gauge group $U(N_{\rm F})$ with $2N_{\rm F}$ fundamental hypermultiplets.
Some of its physical observables are captured by two-dimensional $A_{N_{\rm F}-1}$ Toda conformal field theories on a sphere with four punctures of specific types \refs{\AldayAQ,\WyllardHG}, via the AGT relation.
In particular the basic surface operator of the 4d theory corresponds to a fully degenerate field of the Toda theory \refs{\AldayFS,\DrukkerJP}.
It was argued in \refs{\AldayFS} that 4d line operators bound to a surface operator correspond to monodromies of the conformal blocks, with the insertion point of the degenerate field varied along closed paths.
In the limit where the four-dimensional gauge coupling becomes weak, the correlation function of the Toda theory with the degenerate insertion coincides with the $\Bbb S^2$ partition function of an ${\cal N}=(2,2)$ gauge theory described below  \DoroudXW. 
In this limit, the 4d line operator becomes a 2d line operator, or equivalently a domain wall.
Our aim is to find its intrinsic description within the 2d gauge theory.

The 2d theory in question has gauge group $G=U(1)$, $N_{\rm F}$ chirals $\phi_f$ of charge $+1$, and $N_{\rm F}$ chirals $\tilde \phi_{ f}$ of charge $-1$, with no superpotential.
We denote the twisted masses of the chirals  by $m=({m_f},\tilde m_{ f})_{f=1}^{N_{\rm F}}$.
Correspondingly  the flavor symmetry group is  $G_{\rm F}=U(N_{\rm F})_1\times U(N_{\rm F})_2$, under which $(\phi_f)$ and $(\tilde\phi_f)$ are in $({\bf N_{\rm F}},{\bf 1})$ and  $({\bf 1},{\bf N_{\rm F}})$, respectively.
For $r\gg0$, the IR theory has as the target space a toric Calabi-Yau that we denote by $X$.
There are $N_{\rm F}$ classical vacua $\sigma=-m_v$ labeled by $v=1,\ldots, N_{\rm F}$.

As we show in Appendix \AppTodaDetails\  the $\Bbb S^2$ partition function takes the form
$\langle {\tt 1}|{\tt 1}\rangle = \sum_v \langle {\tt 1}|v\rangle\langle v|{\tt 1}\rangle$, where
$$
\langle v|{\tt 1}\rangle = (2\pi i)^{N_{\rm F}-1/2} e^{-t m_v} \bigg[ \prod_{f\neq v}{\Gamma(m_{fv})\over \Gamma(1-m_{fv})} \prod_f {\Gamma(m_v+\tilde m_f)\over \Gamma(1-m_v-\tilde m_f)}\bigg]^{1/2}
Z^{ v}_{\rm vortex}(t, m)\,,
$$
and $\langle{\tt 1}| v\rangle=\langle v|{\tt 1}\rangle|_{t\rightarrow \bar t}$.
The vortex partition functions as defined in \DefVortex\ are given in \TodaVortex.
Their explicit expressions imply that the matrix elements $\langle v|{\tt 1}\rangle$ as functions of $e^{-t}$ obey the differential equation
\eqn\DiffEqt{
\left[
e^{-t}\prod_f (\partial_t -\tilde m_f) +(-1)^{N_{\rm F}-1} \prod_f (\partial_t + m_f)
\right]
\langle v|{\tt 1}\rangle
=0\,,
}
which has regular singularities at $e^{-t}=0, (-1)^{N_{\rm F}}, \infty$.%
\foot{%
These are the singularities in the quantum K\"ahler moduli space ${\cal M}_K$
 of the non-compact Calabi-Yau $X$, and the equation \GHyperEq\ with $m\rightarrow 0$ can be identified with the Picard-Fuchs equation for the periods of the mirror Calabi-Yau manifold, and can be easily obtained from the period integrals of the mirror Langdau-Ginzburg model \HoriKT.
}
The monodromy along a path $\gamma$ on ${\cal M}_K=\Bbb P^1 \backslash \{0,(-1)^{N_{\rm F}},\infty\}$ is given in the form
\eqn\MonodvOne{
\langle v|{\tt 1}\rangle
\rightarrow
\sum_{w=1}^{N_{\rm F}}  M(\gamma)_{vw} \langle w|{\tt 1}\rangle\,.
}
When $z$ moves along $\gamma$ and then along $\gamma'$, the corresponding modnoromy matrix is $M(\gamma') M(\gamma)$.

\bigskip
\centerline{
\epsfxsize 2.4 in \epsfbox{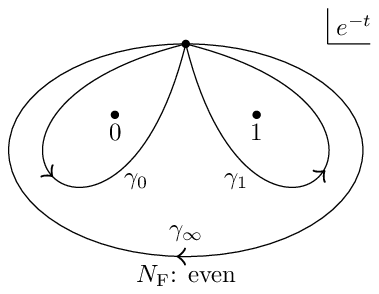}
\hskip 5mm
\epsfxsize 2.4 in \epsfbox{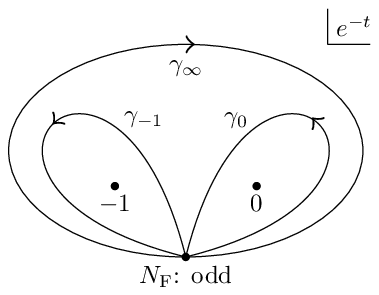}
\hskip 22mm
\
}
  \noindent{\ninepoint\sl
\centerline{ {\bf Figure 3}
 {{Paths for monodromies.
 }}
}
}\bigskip
\noindent

Let us consider the three paths $(\gamma_0, \gamma_{\pm 1}, \gamma_\infty)$ depicted in Figure 3, where we have $\gamma_{1}$ for $N_{\rm F}$ even and $\gamma_{-1}$ for $N_{\rm F}$ odd.
In Appendix \AppTodaDetails\ we derive the monodromy matrices
\eqn\MonodMatToda{\eqalign{
&
M(\gamma_0)_{vw}= \delta_{vw} e^{2  \pi i m_v},
\cr
&
M(\gamma_{\pm 1})_{vw}=
\delta_{vw} 
{-}  e^{{-}\pi i \sum_f (m_f + \tilde{m}_f) }  S_{vw}\,,
\cr
&
M(\gamma_\infty)_{vw}=
\delta_{vw} e^{- 2  \pi i m_v }
+
 e^{  \pi i  \sum_f (m_f + \tilde{m}_f)}e^{ -2\pi im_{ w}}  S_{vw}\,,
}}
 where
$$\eqalign{
 S_{vw}
&=
\left[
{
\prod_f 2i \sin\pi (m_v+\tilde m_f) 2i \sin\pi (m_w+\tilde m_f)
\over
\prod_{f\neq v}2i \sin\pi m_{fv}
\prod_{f\neq w}2i \sin\pi m_{fw}
}
\right]^{1/2}
\times
\left\{
\matrix{
(-1) & {\rm for}\ N_{\rm F}& {\rm even}\,,
\cr
e^{\pi i m_{wv}} &  {\rm for}\ N_{\rm F}& {\rm odd}\,.
}
\right.
}$$
Because of the relation $M(\gamma_0) M(\gamma_{\pm 1})M(\gamma_\infty)=1$, only $M(\gamma_0)$ and $M(\gamma_{\pm 1})$ are independent.
In view of  \WallVev\ and $\langle {\tt g}|v\rangle=\langle v|{\tt g}\rangle|_{t\rightarrow \bar t}$, the monodromy for each path $\gamma$ should be realized as a domain wall $\Bbb W(\gamma)$ such that
$$
\langle {\cal B}[\Bbb W(\gamma)]|\cdot |w\rangle_1\otimes |v\rangle_2
=\langle v|\Bbb W(\gamma)|w\rangle = M(\gamma)_{vw}\,.
$$

It is clear from \MonodMatToda\ that the domain wall $\Bbb W (\gamma_0)$ is simply the gauge Wilson loop with charge $+1$.
Geometrically it corresponds to a sheaf supported on the diagonal $\Delta X$.

Denote by $L$ and $\tilde L$ the topologically trivial equivariant line bundles 
constructed from the representations $({\bf det},{\bf 1})$ and $({\bf 1}, {\bf det})$ of $G_{\rm F}=U(N_{\rm F})_1\times U(N_{\rm F})_2$, respectively.
By comparing \MonodMatToda\ with \BOYnv\ and \vBOYn, we find for $\gamma_{\pm 1}$%
\foot{%
By the tensor product ($\otimes$) of two sheaves, we mean the tensor product of the complexes corresponding to the sheaves.
}
$$\eqalign{
&\quad\
\langle {\tt 1}|\Bbb W(\gamma_{\pm 1})|{\tt 1}\rangle
= \sum_{v,w} \langle {\tt 1}|v\rangle M(\gamma_{\pm 1})_{vw} \langle w|{\tt 1}\rangle
 \cr
&=\langle {\tt 1}|{\tt 1}\rangle 
+
(-1)^{N_{\rm F}-1}\langle{\cal B}(L^{-1/2}\otimes \tilde L^{1/2}\otimes {\cal O}_Y(\lfloor -N_{\rm F}/2\rfloor))|{\tt 1}\rangle
\langle {\cal B}({\cal O}_Y(-\lfloor N_{\rm F}/2\rfloor))|1\rangle_{t\rightarrow\bar t}
\,,
}$$
where $\lfloor x\rfloor$ denotes the largest integer not more than $x$.
Thus
\eqn\BWToda{
\eqalign{
&\ \quad
\big \langle {\cal B}[\Bbb W(\gamma_{\pm 1})]\big|
\cr
&=
\big \langle {\cal B}[{\cal O}_{\Delta}]\big|
+(-1)^{N_{\rm F}-1}
\big\langle {\cal B}\big[
 {\cal O}_Y\left(\lfloor N_{\rm F}/2\rfloor\right)\otimes (L^{-1} \otimes  \tilde L)^{1\over 2}
\boxtimes
{\cal O}_Y\left(
-\lfloor N_{\rm F}/2\rfloor\right)
\big]\big|\,.
}}
Here $\boxtimes$ denotes the external tensor product \CG.%
\foot{%
If $p_i: X_1\times X_2\rightarrow X_i$ are the projections and ${\cal E}_i$ are complexes of holomorphic vector bundles  ($i=1,2$),
${\cal E}_1\boxtimes {\cal E}_2$ is the complex $p_1^* {\cal E}_1\otimes p_2^* {\cal E}_2$ over $X_1\times X_2$, where $p_i^*$ are the pullbacks by $p_i$.
}

We expect that a monodromy in the K\"ahler moduli space acts on the derived category as a Fourier-Mukai transform.
It would be interesting to compare \BWToda\ with the kernel of the corresponding Fourier-Mukai transform.

We computed the monodromies by first decomposing the hemisphere partition function into the vortex partition functions, and then by computing their monodromies.
It is also possible to compute monodromies, or more generally perform analytic continuation from one region to another, using the integral representation \ResultPreservedGeneral.
We give an example of such analytic continuation in Appendix \AppFlop.

\subsec{Monodromy domain walls and the affine Hecke algebra}
\subseclab\SecAHA

Next let us consider the theory realizing $M=T^* \Bbb P^1=T^*{\rm Gr}(1,2)$, a special case of the model studied in Section \SecTstarGr.
This is almost identical to the model with $N_{\rm F}=2$ considered in Section \SecToda, but it includes a neutral chiral multiplet $\Phi$ with twisted mass $m_{\rm ad}$, interacting via the superpotential ${\cal W}=\tilde Q^f \Phi Q_f$.
Since the superpotential affects the hemisphere partition function only by constraining the twisted masses, we can recycle the computations there.
The difference in the conventions in Sections \SecTstarGr\ (and here) and \SecToda\ (there) requires  a replacement $m_f^{\rm there}= -m_f^{\rm here}$, $\tilde m_f^{\rm there}= 1+m_f^{\rm here}-m_{\rm ad}^{\rm here}$.
We also demand that $m_1+m_2=0$.

We are interested in the monodromy of the matrix element $\langle v|{\tt 1}\rangle$ in the $T^* \Bbb P^1$ model, computed in \vfTStarGr.
Thus the monodromy matrices are identical to \MonodMatToda\ with the replacement above:
\eqn\MonodMatTStarPOne{\eqalign{
M(\gamma_0)_{vw}&= \delta_{vw} e^{{-}2 \pi i  m_v}\,,
\cr
M(\gamma_1)_{vw}&=
\delta_{vw} 
{-} e^{{2\pi i m_{\rm ad}}} S_{vw}\,,
\cr
M(\gamma_\infty)_{vw}&=
\delta_{vw} e^{ {2\pi i m_w}}
+ e^{-2\pi i m_{\rm ad}} {e^{ 2\pi i m_w}} S_{vw} \,,
}}
with
\eqn\SvwToda{
S_{vw}=-
\bigg[
{
\prod_f
2i\sin\pi (m_{vf} +  m_{\rm ad})
2i \sin\pi (m_{wf} + m_{\rm ad}) 
\over
\prod_{f\neq v}
 2i  \sin\pi m_{vf}
\prod_{f\neq w}
2i\sin\pi m_{wf}
}
\bigg]^{1/2}\,.
}
Let us set
$$
q=e^{2\pi i m_{\rm ad}}\,,
\quad
X=M(\gamma_0)^{-1}\,,
\quad
T=-1+{q\over 1-q} S\,. 
$$
The relation  $M(\gamma_0) M(\gamma_{1})M(\gamma_\infty)=1$ implies that
\eqn\AHAOne{
(T+1)(T-q)=0\,.
}
The explicit expression \SvwToda\ can be used to show another relation
\eqn\AHATwo{
T X^{-1} - XT = (1-q)X\,.
}

The two relations \AHAOne\ and \AHATwo\ define the so-called $sl_2$ affine Hecke algebra, and we have followed the notation in \CG.
We used the monodromies to motivate and derive the relations, but we can study the domain wall realization of the algebra on its own right.
The generator $X$ is simply the gauge charge $-1$ Wilson loop, and corresponds geometrically to the sheaf $\pi_\Delta^* {\cal O}(-1)$, where $\pi_\Delta$ is the projection from the diagonal of $T^* \Bbb P^1 \times T^* \Bbb P^1$ to the diagonal of the base $\Bbb P^1\times \Bbb P^1$:
$$
X_{vw}=\langle {\cal B}(\pi_\Delta^* {\cal O}(-1))| \cdot |w\rangle_1\otimes |v\rangle_2\,.
$$
For $T$, or a related operator $c=-T-1=-{q\over 1-q} S$, we find from \TStarBOGrEv\ and \SvwToda
\eqn\cvw{\eqalign{
c_{vw}&
= -q^{1/2} 
\langle v|{\cal B}(\iota_* {\cal O}_{\Bbb P^1}(-1)\rangle 
\langle {\cal B}(\iota_* {\cal O}_{\Bbb P^1}(-1) |w\rangle 
\cr
&=q^{-1/2}\langle {\cal B}(\iota_* {\cal O}_{\Bbb P^1}(-1)\boxtimes \iota_* {\cal O}_{\Bbb P^1}(-1))|
 \cdot |w\rangle_1\otimes |v\rangle_2\,.
}}

The $sl_2$ affine Hecke algebra is a basic example of an algebra that can be constructed geometrically as a convolution algebra \CG.
The sheaf we found for $X$ is precisely what appears in the construction.
On the other hand, our sheaf for $c=-1-T$ is slightly different from the one in the convolution algebra, though their supports coincide.
It is desirable to understand in more generality the relation between the algebras realized by domain walls and convolution.

\vskip 1.0cm

\centerline{\bf Acknowledgments}

We would like to thank F.~Benini, C.~Cordova, A.~Gerasimov, J.~Gomis, J.~Halverson, K.~Hori, S.~Komatsu, S.~Lee, D.~Morrison, N.~Nekrasov, S.~Shatashvili, M.~Romo, Y.~Tachikawa, P.~Yi, and P.~Zhao for helpful discussions.
We also thank K.~Hori, M.~Romo, S.~Sugishita, and S.~Terashima for coordinating the arXiv submission.
This work was completed at the Simons Summer Workshop, and we thank the organizers for hospitality.
The research of D.H. is supported in part by a JSPS Research Fellowship for Young Scientists.
The research of T.O. is supported in part by the Grant-in-Aid for Young Scientists (B) No. 23740168, and by the Grant-in-Aid for Scientific Research (B) No. 20340048.

\appendix{A}{Spinor conventions and supersymmetry transformations}
\applab\SUSYTrans

By default we think of a spinor $\psi=(\psi_\alpha)_{\alpha=1,2}$ as a column vector.
The indices are raised and lowered by the charge conjugation matrix
$$
C= (C^{\alpha\beta})=\pmatrix{0 & 1 \cr -1 & 0}\,,\quad
C^{-1}=(C_{\alpha\beta})=\pmatrix{0 & -1 \cr 1 & 0}
$$
as $\psi^\alpha= C^{\alpha \beta} \psi_\beta$, $\psi_\alpha=C_{\alpha\beta}\psi^\beta$.
When the upper index of $\psi$ is contracted with the lower index of $\lambda$, we write
$$
\psi\lambda=\psi^\alpha \lambda_\alpha=\psi^T C^T \lambda\,,
$$
where $T$ indicates the transpose.
The gamma matrices $\gamma_m$ ($m=1,2,3$) have the index structure $\gamma_m=(\gamma_m{}_\alpha{}^\beta)$.
A spinor bilinear is defined as
$$
\psi\gamma_{m_1}\ldots\gamma_{m_n}\lambda=\psi^T C^T \gamma_{m_1}\ldots\gamma_{m_n}\lambda\,.
$$

We always take the SUSY parameters $\epsilon$ and $\bar\epsilon$ to be bosonic.
We assume that they are conformal Killing spinors satisfying \CKSEq.
In this convention fields in a vector multiplet transform under SUSY as%
\eqn\DeformedVec{\eqalign{
&
\delta \lambda=(i\CV_m \gamma^m -\D)\epsilon\,,
\qquad
\delta \bar\lambda=(i\bar\CV_m \gamma^m +\D)\bar\epsilon,
\cr
&
\delta A_\mu =-{i\over 2}\left(\bar\epsilon \gamma_\mu \lambda + \bar\lambda \gamma_\mu \epsilon\right)\,,
\quad
\delta \sigma_1={1 \over 2}\left(\bar\epsilon \lambda + \bar\lambda \epsilon \right)\,,
\quad
\delta \sigma_2=-{i \over 2}\left(\bar\epsilon \gamma^3 \lambda + \bar\lambda \gamma^3 \epsilon \right)\,,
\cr
&
\delta \D=-{i \over 2}\bar\epsilon \Dirac \lambda -{i \over 2}[\sigma_1,\bar\epsilon\lambda] -{1 \over 2}[\sigma_2,\bar\epsilon \gamma^3 \lambda]
+{i \over 2}\epsilon\Dirac\bar\lambda +{i \over 2}[\sigma_1,\bar\lambda\epsilon]+{1\over 2}[\sigma_2,\bar\lambda \gamma^3 \epsilon],
}}
where
$$\eqalign{
&
\CV_m = \left(D_1 \sigma_1+{f(\vartheta) \over \ell \sin \vartheta}D_2\sigma_2\,,\
D_2\sigma_1 - {\ell \sin \vartheta \over f(\vartheta)}D_1 \sigma_2\,,\
F_{\1\2}+i[\sigma_1,\sigma_2]+{1 \over f(\vartheta)} \sigma_1\right)\,,
\cr
&
\bar\CV_m = \left(-D_1 \sigma_1+{f(\vartheta) \over \ell \sin \vartheta}D_2\sigma_2\,,\ 
-D_2\sigma_1 - {\ell \sin \vartheta \over f(\vartheta)}D_1 \sigma_2\,,\ 
F_{\1\2}-i[\sigma_1,\sigma_2]+{1 \over f(\vartheta)} \sigma_1\right)\,.
}$$
For a chiral multiplet of R-charge $q$, the SUSY transformation laws are given by
\eqn\DeformedChi{\eqalign{
  \delta \phi = & \bar \epsilon \psi \,,
\qquad  \delta \bar \phi =  \epsilon \bar \psi\,,  \cr
  \delta \psi = & + i \gamma^\mu \epsilon D_\mu \phi + i\epsilon \sigma_1 \phi + \gamma^3 \epsilon \sigma_2 \phi - i {q\over 2f(\vartheta)}
  \gamma_3 \epsilon \phi + \bar \epsilon {\rm F}\   \cr
  \delta \bar \psi = & - i \bar \epsilon \gamma^\mu D_\mu \bar \phi + i \bar \epsilon \bar \phi \sigma_1
  + \bar \epsilon \gamma^3 \bar \phi \sigma_2 - i {q\over 2f(\vartheta)} \bar \epsilon \gamma_3 \bar \phi + \epsilon \bar {\rm F}\
   \cr 
  \delta {\rm F}  = & \epsilon \Big( i \gamma^\mu D_\mu \psi  - i \sigma_1 \psi + \gamma^3 \sigma_2 \psi - i \lambda\phi \Big)
  - i {q\over 2} \psi \gamma^\mu D_\mu \epsilon \    \cr
  \delta \bar {\rm F}  = & \bar \epsilon \Big( i \gamma^\mu D_\mu \bar\psi - i\bar\psi \sigma_1 - \gamma^3 \bar \psi \sigma_2
  + i \bar \phi \lambda\Big) - i {q\over 2} \bar \psi \gamma^\mu D_\mu \bar \epsilon\,.
}}
The twisted mass ${\rm m}$ can be introduced by replacing $\sigma_2 \rightarrow \sigma_2 +{\rm m}$.

\appendix{B}{Spherical harmonics}

\applab\AppMonoHarm

We will first review the Jacobi polynomials that appear in the scalar monopole harmonics.
Although  we only deal with the situations with vanishing fluxes, a special case of monopole harmonics will appear in the construction of spinor spherical harmonics.
We will also review the vector spherical harmonics.
In this appendix, we take the metric to be that of the round unit sphere
\eqn\MetricRoundUnit{
ds^2=d\vartheta^2+\sin^2\vartheta d\varphi^2\,.
}
The symbol $q\in (1/2)\Bbb Z$ denotes the monopole charge and should not be confused with the R-charge of a chiral multiplet.

\subsec{Jacobi polynomials and scalar monopole harmonics}

Jacobi polynomials are defined as \AbraHand
$$
P^{\alpha\beta}_n(x):={(\alpha+1)_n \over n!}\ _2 F_1\left(-n,1+\alpha+\beta+n;\alpha+1;{1-x \over 2}\right),
$$
where $_2 F_1$ is the hypergeometric function and $(x)_n$ is the Pochhammer symbol
$$
(a)_n:=a(a+1)(a+2)\cdots (a+n-1)={\Gamma(a+n) \over \Gamma(a)}.
$$
The variable $x$ takes values in $[-1,1]$.
An alternative definition is known as Rodrigues' formula:
$$
P^{\alpha\beta}_n(x)={(-1)^n \over 2^n n!}(1-x)^{-\alpha}(1+x)^{-\beta} {d^n \over dx^n}\{(1-x)^{\alpha +n}(1+x)^{\beta+n}\},
$$
where $n, n+\alpha, n+\beta, n+\alpha+\beta \in\NZ$.
When $n, n+\alpha, n+\beta, n+\alpha+\beta \in \NZ$ and $x \in \R$, we can also write
$$
P^{\alpha\beta}_n(x)=\sum_{s=\max\{0,-\beta\}}^{\min\{n,n+\alpha\}} {(n+\alpha)!(n+\beta)! \over s! (n+\alpha-s)! (\beta+s)! (n-s)!}\left({x-1 \over 2}\right)^{n-s}\left(x+1 \over 2\right)^s\,.
$$
For $\alpha, \beta >-1$, they satisfy the orthogonality relations
$$
\int^1_{-1} (1-x)^\alpha (1+x)^\beta P^{\alpha\beta}_n(x) P^{\alpha\beta}_m(x) dx ={2^{\alpha+\beta+1} \over 2n+\alpha+\beta+1}
{\Gamma(n+\alpha+1) \Gamma(n+\beta+1) \over n! \Gamma(n+\alpha+\beta+1)}\delta_{nm}\,.
$$
The polynomials $\{P^{\alpha,\beta}_n(x)\}^\infty_{n=0}$ form a complete orthogonal system in $L^2_{\alpha,\beta}([-1,1])$, {\it i.e.}, the space of functions which are square integrable with weight $(1-x)^\alpha (1+x)^\beta$.

Let us review the basic properties of the monopole scalar harmonics \WuDira. 
When the monopole charge $q$ is non-zero, the scalar harmonics consist of sections of a topologically non-trivial line bundle ${\cal O}(2q)$.
Since we are most interested in the boundary of a hemisphere, we work in the patch $0<\vartheta<\pi$.

We define
$$\eqalign{
&
Y_{qjm}(\vartheta,\varphi):=M_{qjm}(1-x)^{\alpha/2}(1+x)^{\beta/2}P^{\alpha\beta}_n(x) e^{im\varphi},
\cr
&
M_{qjm}:=2^m \sqrt{{2j+1 \over 4\pi}{(j-m)!(j+m)! \over (j-q)!(j+q)!}},
\cr
&
x:=\cos\vartheta,\ \alpha:=-q-m,\ \beta:=q-m,\ n:=j+m.
}$$
For $q=0$, $Y_{jm}:=Y_{0jm}$ give the usual spherical harmonics.
For given $q \in \Z/2$, $j$ and $m$ take values
$$
j=|q|, |q|+1, |q|+2, \ldots\,,\qquad m=-j, -j+1, \ldots, j\,.
$$
$\{Y_{qjm}\}_{j,m}$ form a complete orthonormal system in the space of square integrable sections of the line bundle ${\cal O}(2q)$.

The covariant derivative for the sections of ${\cal O}(2q)$ is given by $D_\mu=\partial_\mu -iq\omega_\mu$, where $\omega_\mu=(0,-\cos\vartheta)$ is the spin connection. The monopole scalar harmonics are the eigenfunctions of the Laplacian:
$$\eqalign{
-D^\mu D_\mu Y_{qjm}
&\equiv \left[-{1 \over \sin\vartheta} {\partial \over \partial\vartheta} \sin\vartheta {\partial \over \partial\vartheta} -{1 \over \sin^2\vartheta} \left( {\partial^2 \over \partial \varphi^2} +2iq\cos\vartheta{\partial \over \partial \varphi} -q^2\cos^2\vartheta \right)\right]Y_{qjm}
\cr
&=[j(j+1)-q^2] Y_{qjm}.
}$$

The monopole harmonics provide an orthonormal basis with respect to the natural inner product:
\eqn\OrthonormalSphereScalar{
\int_{\Bbb S^2} Y_{qjm}(\vartheta,\varphi)^*\, Y_{q j'm'}(\vartheta,\varphi)=\delta_{jj'} \delta_{mm'}\,,
}
where the measure is $ d\vartheta d\varphi \sin\vartheta$ and the complex conjugate is related to the original harmonics as
\eqn\YConj{
Y_{qjm}^*=(-1)^{q+m} Y_{-q,j,-m}\,.
}

Under $\vartheta\rightarrow \pi-\vartheta$, $Y_{jm}$ is even for $j+m$ even, and is odd for $j+m$ odd.
In particular
$$\eqalign{
\partial_\vartheta Y_{jm}|_{\vartheta=\pi/2}=0 &{\rm \quad if \quad } j+m {\rm \ is\ even}\,,
\cr
Y_{jm}|_{\vartheta=\pi/2}=0 &{\rm \quad if \quad } j+m {\rm \ is\ odd}\,.
}$$
The orthogonality relations on the hemisphere can be obtained from \OrthonormalSphereScalar\ by doubling the integration region to the full sphere.

\subsec{Spinor and vector spherical harmonics}

We write $\Dirac \equiv \gamma^\mu D_\mu$.
Let us consider the spectral problem with respect to the modified Dirac operator $$
\gamma^3\Dirac
=
\pmatrix{
 &
\displaystyle
 \partial_\vartheta - {i\over\sin \vartheta} \partial_\varphi + {1\over 2} \cot\vartheta 
\cr 
\displaystyle
- \partial_\vartheta - {i\over\sin \vartheta} \partial_\varphi -{1\over 2} \cot\vartheta
 &}
=:\pmatrix{& {\underline D}^\dagger \cr {\underline D} &}
$$ on $\Bbb S^2$. 
One can check that the eigenspinors are given by
\eqn\chiDef{
\chi^\pm_{jm}(\vartheta,\varphi):={1 \over 2}\pmatrix{
(1 \mp i)Y_{-1/2,jm}(\vartheta,\varphi)
\cr
(j+1/2)^{-1}(-i \pm 1){\underline D}Y_{-1/2,jm}(\vartheta,\varphi)
}\,,
}
which satisfy
$$
\gamma^3\Dirac \chi^\pm_{jm} = \pm (j+1/2)\chi^\pm_{jm}\,.
$$
The range of the quantum numbers is given by
$$
j={1\over 2}\,,\,{3\over 2}\,,\ldots\,,\qquad
m=-j,-j+1,\ldots,j\,.
$$
The eigenspinors form an orthonormal basis on $\Bbb S^2$:
$$
\int_{\Bbb S^2}(\chi^s_{jm})^\dagger \chi^{s'}_{j'm'}=\delta_{ss'}\delta_{jj'}\delta_{mm'}\,.
$$

Next let us review the vector spherical harmonics described {\it e.g.}, in \BarrVect.
We define the one-forms
\eqn\DefVecC{\eqalign{
&
(C^1_{jm})_\mu(\vartheta, \varphi):={1 \over \sqrt{j(j+1)}}
\pmatrix{
\partial_\vartheta Y_{jm}(\vartheta, \varphi)
\cr
i m Y_{jm}(\vartheta, \varphi)
}\,,
\cr
&
(C^2_{jm})_\mu(\vartheta, \varphi):={1 \over \sqrt{j(j+1)}}
\pmatrix{
-(i m/ \sin \vartheta)Y_{jm}(\vartheta, \varphi)
\cr
\sin\vartheta \partial_\vartheta Y_{jm}(\vartheta, \varphi)
}.
}}
With the quantum numbers taking values
$$
j=1,2,3,\ldots\,,\qquad
m=-j,-j+1,\ldots,j\,,
$$
the whole sequence $\{C^\lambda_{jm}\}_{\lambda,j,m}$ forms an orthonormal basis of one-forms on $\Bbb S^2$.
Moreover they are eigenvectors of the vector Laplacian:
$$
-D^\mu D_\mu C^{1(2)}_{jm} =\left[ j(j+1)-1\right] C^{1(2)}_{jm}\,.
$$
They also have the properties
$$\eqalign{
&
D_\mu (C^1_{jm})^\mu = -\sqrt{j(j+1)} Y_{jm}\,,\ D_\mu (C^2_{jm})^\mu = 0\,,
\cr
&
\varepsilon^{\mu\nu} D_\mu (C^1_{jm})_\nu = 0\,,\ \varepsilon^{\mu\nu} D_\mu (C^2_{jm})_\nu = -\sqrt{j(j+1)}Y_{jm}\,.
}$$

\appendix{C}{Eigenvalue problems on a round hemisphere}
\applab\EigenProb

In this Appendix we study the eigenvalue problems and their solutions, which we use in Section \SecOneLoop\ to compute the one-loop determinants.

We are interested in the Neumann and the Dirichlet boundary conditions at $\vartheta=\pi/2$:
$$
\partial_\vartheta\Phi|_{\vartheta=\pi/2}=0\quad ({\rm Neumann})
\quad{\rm and} \quad \Phi|_{\vartheta=\pi/2}=0
\quad ({\rm Dirichlet})\,.
$$
One can check that the Laplacian $-D^\mu D_\mu$ is self-adjoint on the hemisphere $0\leq\vartheta\leq\pi/2$ with these boundary conditions.
For the harmonics $Y_{jm}$, the conditions respectively reduce to
$$
P^{-m,-m}_{j+m}(0)=0\,,\quad {\rm and}\quad \partial_x P^{-m,-m}_{j+m}(x)|_{x=0}=0.
$$
The property $P^{\alpha,\beta}_n(-x)=(-1)^nP^{\beta,\alpha}_n(x)$ implies that the eigenmodes that survive the boundary conditions are given by
$$\eqalign{
&
Y_{jm},\ j-m={\rm even},\ {\rm eigenvalue}= j(j+1)\quad ({\rm Neumann})\,,
\cr
&
Y_{jm},\ j-m={\rm odd},\ \ {\rm eigenvalue}= j(j+1) \quad ({\rm Dirichlet})
\,.
}$$
We have indicated the eigenvalues of the Laplacian $-D^\mu D_\mu$.
Since $-D^\mu D_\mu$ is self-adjoint on the hemisphere when either boundary condition is imposed, the surviving modes form an orthogonal system.
The precise normalizations can be inferred from the relations among such modes
\eqn\OrthonormalHemiScalar{
\int_{0\leq \vartheta\leq\pi/2} Y_{jm}(\vartheta,\varphi)^* \, Y_{j'm'}(\vartheta,\varphi)={1\over 2} \delta_{jj'} \delta_{mm'}\,,
}
which can be obtained from \OrthonormalSphereScalar\ by doubling the integration region to $0\leq\vartheta\leq \pi$.

Let us consider two types of boundary conditions for a spinor $\psi=(\psi_1,\psi_2)^T$:
$$
(\psi_1 + \psi_2)|_{\vartheta=\pi/2}=0\quad ({\rm A}) \qquad {\rm and} \quad
(\psi_1 - \psi_2)|_{\vartheta=\pi/2}=0\quad ({\rm B})\,.
$$
Suppose that another spinor $\lambda$ obeys the same boundary condition as $\psi$.
Then
$$
\langle \psi , \gamma^3 \Dirac \lambda \rangle 
\equiv  \int_{\vartheta\leq \pi/2}  \psi^\dagger \gamma^3 \Dirac \lambda
=
\langle \gamma^3 \Dirac\psi , \lambda \rangle
- \int d\varphi\, \psi^\dagger \gamma^1 \gamma^3 \lambda|_{\vartheta=\pi/2}\,.
$$
For both (A) and (B),
$$
\psi^\dagger \gamma^1 \gamma^3 \lambda|_{\vartheta=\pi/2} \propto \left[(\psi^\dagger)^1\lambda_2 - (\psi^\dagger)^2 \lambda_1\right]|_{\vartheta=\pi/2}=0.
$$
Thus the Dirac operator $\gamma^3 \Dirac$, together with the boundary condition either (A) or (B),
is self-adjoint on the hemisphere.

For $\chi^\pm_{jm}$ the condition (A) reduces to
$$
[(2j+1)\mp (1-2m)] P^{1/2-m,-1/2-m}_{j+m}(0) \pm (j-m+1	)P^{3/2-m,1/2-m}_{j+m-1}(0)=0.
$$
The modes that survive the condition are
$$\eqalign{
&
\chi^+_{jm}\,,\ j-m={\rm odd}\,,\ {\rm eigenvalue}= j+1/2\,,
\cr
&
\chi^-_{jm}\,,\ j-m={\rm even}\,,\ {\rm eigenvalue}\ =-(j+1/2)\,.
}$$
Similarly (B) reduces to
$$
[(2j+1)\pm (1-2m)] P^{1/2-m,-1/2-m}_{j+m}(0) \mp (j-m+1	)P^{3/2-m,1/2-m}_{j+m-1}(0)=0\,,
$$
and the surviving modes are
$$\eqalign{
&
\chi^+_{jm}\,,\ j-m={\rm even}\,,\ {\rm eigenvalue}= j+1/2\,,
\cr
&
\chi^-_{jm}\,,\ j-m={\rm odd}\,,\ {\rm eigenvalue}\ =-(j+1/2)\,.
}$$
Among the surviving modes we have
\eqn\OrthonormalHemiSpinor{
\int_{\vartheta\leq \pi/2}\chi^s_{jm}(\vartheta,\varphi)^\dagger \chi^{s'}_{j'm'}(\vartheta,\varphi)
=
{1\over2}
\delta_{ss'}\delta_{jj'}\delta_{mm'}\,,
}
\eqn\OrthonormalHemiSpinorTwo{
\int_{\vartheta\leq \pi/2}
\chi^s_{jm}(\vartheta,\varphi)\gamma_3 \chi^{s'}_{j'm'}(\vartheta,\varphi)
=
{s'(-1)^{m-1/2}\over2}\delta_{s,-s'}\delta_{jj'}\delta_{m,-m'}
\,.
}

Finally we consider the boundary condition 
$$
A_\vartheta |_{\vartheta=\pi/2}=\partial_\vartheta A_\varphi |_{\vartheta=\pi/2}=0\,.
$$
for vector harmonics \DefVecC.
The modes that survive are
$$\eqalign{
&
C^1_{jm},\ j-m={\rm even},\ {\rm spectrum}\ j(j+1), {\rm degeneracy}\ j+1,
\cr
&
C^2_{jm},\ j-m={\rm odd},\ {\rm spectrum}\ j(j+1), {\rm degeneracy}\ j.
}$$

\appendix{D}{Hemisphere partition functions for exact complexes}
\applab\AppExact

The aim of this appendix is to argue that the map \HPFmap\ is well-defined.
Namely we argue that the hemisphere partition function for each object of the derived category $D(X\ {\rm or}\ M)$ does not depend on the choice of a complex of vector bundles used in the construction.

As an example in Case 1, let us consider the resolved conifold.
The gauge group is $G=U(1)$, and there are four chiral fields $\phi=(\phi^1,\phi^2,\phi^3,\phi^4)$ with gauge charges $w_a=(+1,+1,-1,-1)$.
The flavor group is $G_{\rm F}=U(1)^4=\prod_{a=1}^r U(1)_a$, where $\phi^a$ has charge $+1$ for $U(1)_a$ and charge zero for $U(1)_{b\neq a}$.

Let $m=(m_a)$ be the complexified twisted masses for $\phi^a$.
For $r\gg 0$, the model is in the geometric phase and flows to the non-linear sigma model with target space the resolved conifold $X$.
We want to show that for an exact equivariant complex $({\cal E},d)$ of vector bundles given by 
$$
0\longrightarrow {\cal E}^1\longrightarrow \ldots
\longrightarrow {\cal E}^n \longrightarrow 0\,,
$$ 
the partition function $Z_{\rm hem}({\cal E})$ vanishes.
Following the definition of \MapEToBForX
, we let $V^i$ be the representation of $G\times G_{\rm F}$ from which the vector bundle ${\cal E}^i$ arises via \NaturalBundle.
We assume that the values of $m_a$ are generic.
Under this assumption, the integral
$$
Z_{\rm hem}({\cal E})
=\int^{i\infty}_{-i\infty} {d\sigma\over 2\pi i}
{\rm Str}_V[ e^{-2\pi i \rho(\sigma,m)}]
e^{t\sigma} \Gamma(\sigma+m_1) \Gamma(\sigma+m_2) \Gamma(-\sigma+m_3) \Gamma(-\sigma+m_4)\,,
$$
where we wrote explicitly the representation $\rho_*(\sigma,m)$ of Lie$(G\times G_{\rm F})$,
is evaluated by residues to give
$$
Z_{\rm hem}({\cal E})
=\sum_{v=1}^2 {\rm Str}_V[ e^{-2\pi i \rho_*(-m_v,m)}]
e^{-t m_v} 
\sum_{k=0}^\infty {(-1)^k \over k!}
\prod_{a\neq v} \Gamma(w_a(- m_v-k) +m_a)\,.
$$
This involves two sequences of poles at $\sigma=-m_v,-m_v-1,\ldots$ ($v=1,2$).
As noted in \refs{\BeniniUI,\DoroudXW}, the beginning of each sequence corresponds to a solution of 
the condition
$$
(w_a\sigma +m_a)\phi^a=0
$$
with $\phi^a$ satisfying the D-term equation
$$
\sum_a w_a |\phi^a|^2={r \over 2\pi}\,.
$$
Such values of $(\sigma,\phi)$ describe a fixed point in $X$ under
the action of the flavor group $G_{\rm F}$.%
\foot{%
For a more general $X$ for which $G_{\rm F}$ is non-abelian, we should consider a fixed point with respect to the maximal torus of $G_{\rm F}$.}
We now recall that the tachyon profile ${\cal Q}$ has to satisfy the condition that
$
\rho(g){\cal Q}(g^{-1}\cdot \phi) \rho(g)^{-1}={\cal Q}(\phi)
$
for any $g\in G\times G_{\rm F}$.
For $g=(e^{-2\pi i \sigma},e^{-2\pi i m})\in G\times G_{\rm F}$ and $\phi$ under consideration then, 
$$
\rho(g){\cal Q}(\phi) ={\cal Q}(\phi) \rho(g)\,.
$$
This relation together with Hodge decomposition shows that there are complete cancellations between ${\rm Im}\,d^i$ and ${\rm Ker}\,d^{i+1}$ so that ${\rm Str}_V[ e^{-2\pi i \rho_*(\sigma,m)}]$ vanishes at all poles, and hence $Z_{\rm hem}=0$ for an exact complex ${\cal E}$.

For more general $X$, if a given exact complex can be made equivariant with twisted masses generic enough so that the poles become simple, the same argument can be applied to show that $Z_{\rm hem}$ vanishes.

Next let us consider the Fermat quintic $M$ as an example of Case 2.
The chiral fields are $(P,x_a)$.
The fields $x^a$, $a=1,\ldots,5$, parametrize $X$.
The superpotential $W=P (x_1^5+\ldots+x_5^5)$ does not allow us to introduce real twisted masses.
Given an object in $D(M)$, we push it forward to $D(X)$, where $X=\Bbb P^4$ and resolve it there.

In order to argue that the map $D(M)\rightarrow \Bbb C$ is well-defined,
suppose that we have two resolutions in $X$ of the same object of $D(M)$.
For the resolutions, which are quasi-isomorphic in $X$, we construct the boundary interactions according to \MapEToB.
The difference of their hemisphere partition functions is clearly the hemisphere partition function of their mapping cone, which is exact.
Thus if $Z_{\rm hem}$ vanishes for any exact complex in $X$, then the map $Z_{\rm hem}: D(M)\rightarrow \Bbb C$ is well-defined.

We have not found such a proof yet.
As an alternative, we offer an example of exact complex for which $Z_{\rm hem}$ indeed vanishes.
Consider the following complex ${\cal E}$ of vector bundles over $X=\Bbb P^4$:
$$
0
\rightarrow 
{\cal O}(n)
\rightarrow
{\cal O}(n+1)^5
\rightarrow
{\cal O}(n+2)^{10}
\rightarrow
{\cal O}(n+3)^{10}
\rightarrow
{\cal O}(n+4)^5
\rightarrow 
{\cal O}(n+5) 
\rightarrow
0\,.
$$
In terms of fermionic oscillators $\{\eta_a,\bar\eta_b\}=\delta_{ab}$, this complex is realized as the Fock space ${\cal V}$ built on the vacuum $|0\rangle$ satisfying $\eta_a|0\rangle=0$.
The differential is ${\cal Q}_0=x_a \eta_a$, and the tachyon profile is ${\cal Q}={\cal Q}_0+\sum_a P x_a^4\bar\eta_a$.
This is exact since $\{{\cal Q},\bar {\cal Q}\}$ is everywhere positive.
The boundary interaction $({\cal V},{\cal Q})$ then contributes
$$
{\rm Str}_{\cal V}(e^{-2\pi i\sigma})\propto \sin^5\pi\sigma\,,
$$
which has order 5 zeros at $\sigma\in \Bbb Z$.
It then follows that the hemisphere partition function vanishes,
$$
Z_{\rm hem}({\cal E})=
\int^{i\infty}_{-i\infty}{\rm Str}_{\cal V}(e^{-2\pi i\sigma}) e^{t\sigma}
\Gamma(\sigma)^5 \Gamma(1-5\sigma)
=0\,,
$$
when the integral is evaluated by closing the contour to the left.

Finally, let us consider another example of Case 2, $M=T^* {\rm Gr}(N,N_{\rm F})$ considered in Section \SecTstarGr.
As in the previous example, we want to show that $Z_{\rm hem}$ vanishes for an exact complex on the ambient space $X$ given as in \XTStar.
The general result \ResultResidue\ with the definition \CI\ of $C(I)$ implies that we need to find decompositions of the vector $\vec r=(r,\ldots,r)$ by the weights of fundamental, anti-fundamental, and adjoint representations, with positive coefficients.
One can show that anti-fundamental weights can never appear in such decompositions.
The poles are associated with fixed points on $T^*{\rm Gr}(N,N_{\rm F})$ with respect to the $U(1)^{N_{\rm F}}$($\subset G_{\rm F}$) action.
Indeed the decomposition $\vec r=\sum_{(a,w)\in I} r_{aw} \vec w$ implies that the D-term equations can be solved by setting $\phi_a^w =(r_{aw}/2\pi)^{1/2}$ for $(a,w)\in I$ (with other $\phi_a^w=0$), and the poles $\sigma$ satisfy $e^{-2\pi i (w\cdot \sigma+m_a)}=1$ for $(a,w)\in I$.
Thus at the poles $\rho(g)$ and ${\cal Q}_0(\phi)$ commute with each other, and ${\rm Str}_V[ e^{-2\pi i \rho_*(\sigma,m)}]$ vanishes, as in the case of the resolved conifold.
Since the poles are simple for generic twisted mass parameters, the hemisphere partition function vanishes.

\appendix{E}{Complete intersection CYs in a product of projective spaces}
\applab\AppCICY

In this appendix we generalize the result for the quintic obtained in Section \SecQuintic.
Let us consider a direct product of projective spaces $X=\prod_{r=1}^m \Bbb P^{N_r -1}$.
We take sections $s_a$ of the line bundles $ {\cal O} (l^1_a,\ldots,l^m_a)$ for $a=1,\ldots,k$ and assume that the intersection $M$ of their zero-loci $s_a^{-1}(0)$ is a smooth manifold.
For $M$ to be Calabi-Yau, $l_a^r$ must satisfy
$$
\sum_a l_a^r=N_r\,.
$$
This geometry is realized by a gauge theory with gauge group $G=U(1)^m=\prod_{r=1}^m U(1)_r$ and the following matter content: the chiral multiplet fields 
$$
\phi_{r, 1}, \ldots, \phi_{r, N_m}
$$ 
charged only under $U(1)_r$ with charge 1, and 
$$
P_a,\, a=1,\ldots, k
$$ 
that have $U(1)^m$ charges $(-l^1_a, \ldots, -l^m_a)$ and R-charge $-2$. 
We also include a superpotential $W=\sum^k_{a=1} P_a {\rm G}_a(\phi)$, where ${\rm G}_a(\phi)$ are the polynomials that define the sections $s_a$.
For $r\gg 0$ the gauge theory flows to the nonlinear sigma model whose target space $M$.

Let us take as the Chan-Paton space ${\cal V}$ the fermionic Fock space generated by the Clifford algebra $\{\eta_a, \bar\eta_b\}=\delta_{ab},\, a,b=1,\ldots, k$ and the Clifford vacuum $| 0 \rangle$ satisfying $\eta_a |0\rangle=0$. The tachyon profile is given by
$
{\cal Q}=
{\rm G}_a \eta_a + P_a \bar\eta_a
$
and is a matrix factorization, ${\cal Q}^2= W$.
Via \MapEToB\ this corresponds to the Koszul resolution
$$
\wedge^k E
\mathop{\longrightarrow}^{i_s}
\cdots
\mathop{\longrightarrow}^{i_s}
\wedge^2 E
\mathop{\longrightarrow}^{i_s}
E
\mathop{\longrightarrow}^{i_s}
{\cal O}_X(n_1, \ldots, n_m)\,,
$$
of the sheaf ${\cal O}_M(n_1, \ldots, n_m)$,
where
$$
E=\bigoplus_{a=1}^k {\cal O}_X(n_1-l_a^1, \ldots, n_m-l_a^m)
$$
and $i_s$ is the contraction by the section $s=(s_a)$ of the vector bundle
$\bigoplus_{a=1}^k {\cal O}_X(l_a^1, \ldots,l_a^m)$.
Following the rule \RepVihat\  we assign gauge charges 
$$
(n_1  +\sum_a l_a^1/2,\ldots, n_m +\sum_a l_a^m/2)
=(n_1  + N_1/2,\ldots, n_m + N_m/2)
$$ 
to $| 0 \rangle$.
Thus
\eqn\ZhemCICY{\eqalign{
&\quad
Z_{\rm hem}[{\cal O}_M(n_1, \ldots, n_m)]
\cr
&=\int_{i \R^m} {d\sigma^m \over (2\pi i)^m}
e^{-2 \pi i  n_r \sigma_r }
\Big[ 
\prod_{a=1}^k {2\over i} \sin (\pi l_a^r \sigma_r )
\Big]
e^{ t_r \sigma_r} 
\Big[ 
\prod_{r=1}^m \Gamma(\sigma_r)^{N_r} 
\Big]
\prod_{a=1}^k \Gamma\left(1- l_a^r \sigma_r\right)
\cr
&=
(- 2 \pi i)^k
\int_{i \R^m} {d\sigma^m \over (2\pi i)^m} e^{ (t_r-2 \pi i n_r)\sigma_r}{\prod_r \Gamma(\sigma_r)^{N_r} \over \prod_a \Gamma ( l_a^r \sigma_r ) }\, .
}}
This integral can be evaluated by residues, and is given by the coefficient of $\prod_r \sigma_r^{-1}$ in the Laurent expansion of the integrand, up to exponentially suppressed terms for ${\rm Re}\,t\gg 0$.

We wish to compare this with the large volume formula
\eqn\ChargeAroot{
\int_M 
{\rm ch}(
{\cal E}
)
 e^{B+i\omega} \sqrt{\hat A(TM)}
}
for the central charge of ${\cal E}\in D(M)$.
The complexified K\"ahler form $B+i\omega$ depends linearly on the complexified FI parameters $t=(t_r)$ in the large volume limit.
Note the relation
$$
\prod_j \sqrt{x_j \over e^{x_j/2} - e^{-x_j/2}} - \prod_j \Gamma \left(1+{i x_j \over 2 \pi}\right)={\cal O}(x_j^{ 3})\,,
$$
which is valid when $\sum_j x_j=0$.
This implies that the polynomial terms in $t$, appearing in \ChargeAroot\ with the first three highest orders, also appear in the integral
\eqn\ChargeGamma{
\int_M 
{\rm ch}(
{\cal E}
)
 e^{B+i\omega}
\hat \Gamma(TM)\,.
}
Here $\hat \Gamma$ is the multiplicative characteristic class%
\foot{%
We learned of the relevance of the Gamma class $\hat\Gamma$ to the hemisphere partition function in talks by D.~Morrison and K.~Hori.
Our use of the Gamma class was motivated by their talks.
}
 defined via the splitting principle as
\eqn\GammaClass{
\hat\Gamma(E)=\prod_j\Gamma \left(1+{i x_j \over 2 \pi}\right)\,,
}
where $x_j$ are the Chern roots of a vector bundle $E$.
Using the exact sequence
$$
0 \longrightarrow TM \longrightarrow TX|_M \longrightarrow \bigoplus_{a=1}^k {\cal O}(l_a^1,\ldots,l_a^m)|_M \longrightarrow 0
$$
and the Euler sequence
$$
0 \longrightarrow {\cal O} \longrightarrow {\cal O}(1)^{\oplus N_r} \longrightarrow T{\Bbb P}^{N_r-1}\rightarrow 0
$$
for each $r$, we can write
$$
\hat \Gamma(TM)
={i^* \hat \Gamma (TX) \over i^* \hat \Gamma(\bigoplus_a {\cal O}(l_a^1,\ldots,l_a^m))}
=
\prod_{r=1}^m 
\Gamma\left(1+ { i{\bf e}_r \over 2\pi}\right)^{N_r}
\bigg/ \prod_{a=1}^k  \Gamma\left(1+ {\sum_r l_a^r{\bf e}_r\over 2\pi i}\right)
\,,
$$
where ${\bf e}_r= i^* h_r$, and the hyperplane classes $h_r\in H^2(\Bbb P^{N_r-1})$ satisfy $\int_X \prod_r  h_r^{N_r-1}=1$.
Thus we can rewrite the large volume formula for the central charge as
\eqn\ChargeCICY{\eqalign{
&\int_M {\rm ch}({\cal O}_M(n_1,\ldots,n_m)) e^{B+i\omega} \sqrt{\hat A(TM)}
\cr
& \quad
\sim
\int_M e^{{i\over 2\pi}\sum_r (t_r-2\pi i n_r) {\bf e}_r}{\prod_r \Gamma \left(1+{i \over 2\pi} {\bf e}_r\right)^{N_r} \over \prod_a \Gamma \left(1+{i \over 2\pi}\sum_r l_a^r {\bf e}_r\right) }
\cr
&\quad =
(-2\pi i)^k \int_X \prod_{r=1}^m\left( i h_r \over 2 \pi\right)^{N_r} e^{{i\over 2\pi}\sum_r (t_r-2\pi i n_r) h_r} 
{\prod_r \Gamma \left({i \over 2\pi} h_r\right)^{N_r} \over \prod_a \Gamma \left({i \over 2\pi} \sum_r l_a^r h_r\right) }\,.
}}
In the last line we used the fact that the Poincar\'e dual of the homology class $[s^{-1}_a(0)]$ is $c_1( {\cal O}(l_a^1,\ldots,l_a^m))=\sum_r l_a^r h_r$. 
Comparing \ChargeCICY\ with \ZhemCICY, we see that the hemisphere partition function agrees with the central charge in the large volume limit, up to an overall numerical factor,
for the polynomial terms in $t$ with the first three highest orders.

\appendix{F}{Vortex partition functions}

Basic building blocks of the hemisphere partition function for theories with gauge group $G=U(N)$ and $N_{\rm F}\geq N$ fundamental chiral multiplets are the vortex partition functions \ShadchinYZ.
Here we give certain expressions that arise in the sphere and the hemisphere partition functions.
We take them as {\it definitions} of the vortex partition functions in the presence of other matter fields in various representations.
Conceptually the vortex partition functions are equivariant integrals on the moduli space of vortex solitons with appropriate integrands, but the first principle derivations have been given only for some of the representations.
One may regard the definitions here as predictions.

Let $-m_f$ be the twisted masses of the fundamentals.
We define the vortex partition function specified by  ${\bf v}\equiv \{f_1 <\ldots <f_N\} \subseteq \{1,\ldots, N_F\}$ as
\eqn\DefVortex{
Z_{\rm vortex}^{\bf v}(t_{\rm ren} ,m)\equiv
\sum_{k_1,\ldots,k_N=0}^\infty
\prod_{j<l}
(-1)^{k_{jl}}
\Big(1 -{ k_{jl}\over m_{f_j f_l}}\Big)
\prod_{a\notin {\bf v}} Z^{\bf v}_{R_a}(\vec k;m_a;\vec \beta)
 e^{-|\vec k|t_{\rm ren}}\,.
}
In the product, $a$ runs over all chiral multiplets in irreducible representations $R_a$ of $U(N)$, except the fundamentals corresponding to $f\in {\bf v}$.
Let $(x)_k=x(x+1)\ldots(x+k-1)$ be the Pochhammer symbol.
For the fundamental representation $Z_{\rm fund}^{\bf v}$ appears in the form
$$
Z_{\rm fund}^{\bf v}(\vec k;-m_f)={ (-1)^{\sum_j k_j}\over
\prod_{j=1}^N (1+m_f-m_{f_j})_{k_j}}\,.
$$
For anti-fundamental,  adjoint, and $\det^n$ representations, the $Z^{\bf v}_{R}$ is given by
$$\eqalign{
&Z_{\rm antifund}^{\bf v}(\vec k;m)=
\prod_{j=1}^N (m-m_{f_j} )_{k_j}\,,
\qquad
Z_{\rm adj}^{\bf v}(\vec k;m)=
\prod_{i,j=1}^N {(m_{f_i f_j}-k_i+m)_{k_j}\over (m_{f_i f_j}-k_i+m)_{k_i}}\,,
\cr
&\qquad\qquad\qquad\qquad\qquad
Z_{{\rm det}^n}^{\bf v}(\vec k;m)={1\over (1+m{ +n}\sum_j m_{f_j})_{|\vec k|}}\,.
}$$
More generally, each infinite sum specified by $I$ in \ResultResidue, normalized so that the series starts with $1$, defines an analog of the vortex partition function.

We study several Seiberg-like dualities in Section \SecDualities.
The vortex partition functions for the $T^* {\rm Gr}$ models are not duality invariant; rather, they satisfy a non-trivial relation \TStartVortexDuality.
We found numerically that similar relations%
\foot{%
For $N_{\rm A}\leq N_{\rm F}-2$, the vortex partition functions are invariant under the duality map
$
N\to N_F-N,\ t_{\rm ren} \to t_{\rm ren}-N_{\rm A}\pi i,\ {m}_f \to -{m}_f-{1 / 2},\ \tilde m_a \to - \tilde m_a+{1/ 2},\ {\bf v}\to {\bf v}^\vee
$.
}
hold for $U(N)$ theories with $N_{\rm F}$ fundamental and $N_{\rm A}$ anti-fundamental matter fields with $N_{\rm A}=N_{\rm F},N_{\rm F}-1$.
By denoting the vortex partition function as $Z_{\rm vortex}^{(N,N_{\rm F},N_{\rm A}), \bf v}(t_{\rm ren}; m_f, \tilde m_a)$, for $N_{\rm A}=N_{\rm F}$ we have
$$\eqalign{
&(1+(-1)^{N_F-N+1} e^{-t_{\rm ren} })^{-(N_F-N)+\sum_{f=1}^{N_{\rm F}} {m}_f +\sum_{a=1}^{N_{\rm A}} \tilde m_{a}}Z_{\rm vortex}^{(N,N_{\rm F},N_{\rm A}), \bf v}(t_{\rm ren};  m_f,\tilde m_a)
\cr
& \qquad =
Z_{\rm vortex}^{(N_F-N,N_{\rm F},N_{\rm A}), \bf v^\vee}(t_{\rm ren}- N_{\rm A} \pi i; - m_f-1/2,- \tilde m_a+1/2)\,,
}$$
and for $N_{\rm A}=N_{\rm F}-1$,
$$\eqalign{
&\quad\exp((-1)^{N_{\rm F}-N+1} e^{-t_{\rm ren}})Z_{\rm vortex}^{U(N), \bf v}
\cr
&
\qquad \qquad \qquad
=
Z_{\rm vortex}^{U(N_F-N), \bf v^\vee}(t_{\rm ren}- N_{\rm A} \pi i; - m_f-1/2,- \tilde m_a+1/2)\,.
}
$$

\appendix{G}{Detailed calculations for a $U(1)$ theory}
\applab\AppTodaDetails

Let us consider the 2d gauge theory in Section \SecToda.
The $\Bbb S^2$ partition function is
$$\eqalign{
Z_{\Bbb S^2}(X)&=c
\sum_{\bf v}
e^{-(t+\bar t) m_v}
\prod_{f\neq v}{\Gamma(m_{fv})\over \Gamma(1-m_{fv})}
\prod_f {\Gamma(m_v+\tilde m_f)\over \Gamma(1-m_v-\tilde m_f)}
Z^{ v}_{\rm vortex}(t,  m)
Z^{ v}_{\rm vortex}(\bar t , m)\,,
}$$
where we chose $w_0=0$ for the ambiguity $w_0$ in \Spheregf, and $c$ is a normalization constant to be determined.
The vortex partition function is as defined in \DefVortex:
\eqn\TodaVortex{
Z^{v}_{\rm vortex}(t, m)=\sum^\infty_{k=0}e^{-kt} (-1)^{k N_F}\prod^{N_{\rm F}}_{f=1} {(\widetilde{m}_f + m_v)_k \over (1-m_{fv})_k}.
}
We can write $Z_{\Bbb S^2}=\sum_{ v} \langle {\tt 1}| v\rangle \langle  v|{\tt 1}\rangle$
if we set
$$
\langle v|{\tt 1}\rangle
=c^{1/2}
e^{-t m_v}  
\bigg[
\prod_{f\neq v}{\Gamma(m_{fv})\over \Gamma(1-m_{fv})}
\prod_f {\Gamma(m_v+\tilde m_f)\over \Gamma(1-m_v-\tilde m_f)}
\bigg]^{1/2}
Z^{ v}_{\rm vortex}(t, m)
$$
and
$$
\langle{\tt 1}| v\rangle
=c^{1/2} 
e^{-\bar t m_v}  
\bigg[
\prod_{f\neq v}{\Gamma(m_{fv})\over \Gamma(1-m_{fv})}
\prod_f {\Gamma(m_v+\tilde m_f)\over \Gamma(1-m_v-\tilde m_f)}
\bigg]^{1/2}
Z^{ v}_{\rm vortex}(\bar t, m)\,.
$$

We can compute the cylinder partition function $\langle{\cal B}({\cal O}_X(n_2))|{\cal B}({\cal O}_X(n_1))\rangle$ by a generalization of \ZcylAB,
\eqn\ZcylABTwo{
{\rm ind}_{F\otimes E^*}(\displaystyle{\not} D)
=
\sum_{p:\, {\rm fixed \, points}}
{
1
\over
\det_{TX_p}(g^{-1/2}-g^{1/2})
}
{\rm Tr}_{F_p}(g) {\rm Tr}_{E_p}(g^{-1})\,.
}
We find
$$\eqalign{
\langle
{\cal B}({\cal O}_X(n_2))|{\cal B}({\cal O}_X(n_1))\rangle
&=
\sum_v e^{2\pi i n_{21} m_v}
\bigg[\prod_{f \neq v}2i\sin\pi m_{fv} \prod_f 2i\sin\pi (m_v+\tilde m_f)\bigg]^{-1}\,,
}$$
where $n_{ab}:=n_a-n_b$.
This can be written as $\sum_v \langle {\cal B}({\cal O}_X(n_2))|v\rangle
\langle v|{\cal B}({\cal O}_X(n_1))\rangle$
by setting
$$
\langle {\cal B}({\cal O}_X(n))|v\rangle
=
 e^{2\pi i n m_v}
\bigg[\prod_{f \neq v}2i\sin\pi m_{fv} \prod_f 2i \sin\pi (m_v+\tilde m_f)\bigg]^{-1/2}
$$
and
$$
\langle v|{\cal B}({\cal O}_X(n))\rangle
=
 e^{-2\pi i n m_v}
\bigg[\prod_{f \neq v}2i\sin\pi m_{fv} \prod_f 2i \sin\pi (m_v+\tilde m_f)\bigg]^{-1/2}\,.
$$

The hemisphere partition function for $ {\cal B}({\cal O}_X(n))$ is
$$\eqalign{
Z_{\rm hem}( {\cal B}({\cal O}_X(n)))
&=\int {d\sigma \over 2\pi i} 
e^{-2\pi i n \sigma}
e^{t \sigma} \prod^{N_{\rm F}}_{f=1} \Gamma(\sigma+ m_f) \Gamma(-\sigma+\widetilde{m}_f)
\cr
&=
 \sum^{N_F}_{v=1}
e^{2\pi i n m_v}
 Z^v_{\rm cl}(t,  m) Z^v_{\rm 1\mathchar`-loop}(m) Z^{v}_{\rm vortex}(t,m)\,.
}$$
where
$$
 Z^v_{\rm cl}(t, m)=e^{-t  m_v}\,,
\qquad
Z^v_{\rm 1\mathchar`-loop}(m)=\prod_{ f\neq v} \Gamma( m_{fv})\prod_{f} \Gamma( {\widetilde{m}}_f + m_v)\,.
$$
We can write
$$
Z_{\rm hem}( {\cal B}({\cal O}_X(n)))=
\sum_{v=1}^{N_{\rm F}} \langle  {\cal B}({\cal O}_X(n))| v\rangle \langle  v|{\tt 1}\rangle
=
 \langle  {\cal B}({\cal O}_X(n))|{\tt 1}\rangle\,.
$$
if we set $c=(2\pi i)^{2N_{\rm F}-1}$.

We will also be interested in the brane for the structure sheaf of $Y$, the submanifold defined by setting to zero the chiral fields $\tilde\phi_f$.
This corresponds to Case 1 of Section \BCcomplexes.
Let us introduce fermionic oscillators satisfying $\{\eta_f,\bar\eta_g\}=\delta_{fg}$, $\eta_f|0\rangle=0$.
A locally free resolution of ${\cal O}_Y$ is given by a complex of equivariant vector bundles which corresponds to
$$
\Bbb C \bar\eta_1\ldots\bar\eta_{N_{\rm F}}|0\rangle
\rightarrow
\ldots
\rightarrow
\bigoplus_{f<g} \Bbb C \bar\eta_f\bar\eta_g|0\rangle
\rightarrow
\bigoplus_f \Bbb C \bar\eta_f|0\rangle
\rightarrow
\underline{ \Bbb C|0\rangle}
$$
with the differential ${\cal Q}=\tilde \phi_f\eta^f$.
The underline indicates the degree-zero location.
Including the twist by ${\cal O}_X(n)$, we find
\eqn\BOYnv{\eqalign{
\langle {\cal B}({\cal O}_Y(n))|v\rangle
&=
\prod_f (1-e^{{+}2\pi i(m_v+\tilde m_f)})
\times \langle {\cal B}({\cal O}_X(n))|v\rangle
\cr
&=
{
(-1)^{N_{\rm F}}
e^{2\pi i n m_v}
e^{N_{\rm F}\pi im_v}
e^{\pi i \sum_f \tilde m_f}
}
\bigg[
{\prod_f 2i \sin\pi (m_v+\tilde m_f)
\over
\prod_{f \neq v}2i \sin\pi m_{fv} 
}
\bigg]^{1/2}
}}
and
\eqn\vBOYn{\eqalign{
\langle v
|
 {\cal B}({\cal O}_Y(n))
\rangle
&=
{
e^{-2\pi i n m_v}
e^{-N_{\rm F}\pi im_v}
e^{-\pi i \sum_f \tilde m_f}
}
\bigg[
{\prod_f 2i \sin\pi (m_v+\tilde m_f)
\over
\prod_{f \neq v}2i \sin\pi m_{fv} 
}
\bigg]^{1/2}\,.
}}
We wish to derive the monodromies of $\langle  v|{\tt 1}\rangle$ along paths on the $(e^{-t})$-plane.
To simplify the computations let us set $z=(-1)^{N_{\rm F}}e^{-t}$.
The differential equation \DiffEqt\ becomes
\eqn\GHyperEq{
\bigg[z\prod^{N_{\rm F}}_{f=1}\left(z {d \over dz} +\widetilde{m}_f\right) - \prod^{N_{\rm F}}_{f=1}\left(z {d \over dz} - m_f\right) \bigg]G(z)=0\,,
}
which has $N_{\rm F}$ basic solutions 
\eqn\Gvz{
 G_v(z)=z^{m_v}{}_{N_{\rm F}}F_{N_{\rm F}-1}\bigg( \matrix{{\scriptstyle\{\tilde{m}_f+m_v\}^{N_{\rm F}}_{f=1}} \cr{\scriptstyle \{1-m_f+m_v\}^{N_{\rm F}}_{f\neq v}}}\bigg| z \bigg)
}
 analytic on the complex $z$-plane minus the branch cuts $(-\infty, 0] \cup [1,\infty)$.
In terms of the functions $G_v$ and  the coefficients
$$
A_{v}=
(2\pi i)^{N_{\rm F}-1/2}
\bigg[
\prod_{f\neq v}{\Gamma(m_{fv})\over \Gamma(1-m_{fv})}
\prod_f {\Gamma(m_v+\tilde m_w)\over \Gamma(1-m_v-\tilde m_w)}
\bigg]^{1/2}
\times
\left\{
\matrix{
1 & {\rm for}\ N_{\rm F} & {\rm even}\,,
\cr
e^{-\pi i m_v} & {\rm for}\ N_{\rm F} & {\rm odd}\,,
}
\right.
$$
we can write
\eqn\vOneToda{
\langle  v|{\tt 1}\rangle=
A_v G_v(z)\,.
}
On $G_v$, the monodromy along a path $\tilde \gamma$ acts as
$$
G_v(z) \rightarrow \sum_w {\bf M}(\tilde \gamma)_{vw}G_w(z)
$$
for some matrix $ {\bf M}(\tilde \gamma)_{vw}$.
If a path $\tilde \gamma$ on the $z$-plane corresponds to the path $\gamma$ on the $(e^{-t})$-plane, the matrix ${\bf M}(\tilde \gamma)$ is related to $M(\gamma)$ in \MonodvOne\ by a diagonal similarity transformation
\eqn\Similar{
M(\gamma)_{vw}
= A_{v}
{\bf M}(\tilde \gamma)_{vw}A^{-1}_{w}\,.
}

For the small loop $\tilde \gamma_0$ going around $z=0$ counterclockwise, the monodromy acts as $
 G_v (z) \rightarrow  e^{ 2  \pi i m_v} G_v(z)$.
Thus
$
{\bf M}(\tilde \gamma_0)_{vw} = e^{ 2 \pi i m_v} \delta_{vw}
$.

In order to obtain monodromies along other paths, let us consider independent solutions of \GHyperEq\ around $z=\infty$ \GomisKV
$$
\tilde{G}_v(z):=z^{-\tilde{m}_v}{}_{N_{\rm F}}F_{N_{\rm F}-1}\bigg( \matrix{{\scriptstyle\{m_f+\tilde{m}_v\}^{N_{\rm F}}_{f=1}} \cr{\scriptstyle \{1+\tilde{m}_{vf}\}^{N_{\rm F}}_{f\neq v}}}\bigg| {1 \over z} \bigg),\ v=1,\ldots , N_F.
$$
They are analytic on $\Bbb C\backslash (-\infty, 1]$.
We can relate $G_v(z)$ defined near $z=0$ and $\tilde G_v(z)$ defined near $z=\infty$ by analytic continuation upon choosing a path that connects the two regions.
The relation, the connection formula, depends on whether the path goes above ($\epsilon=+1$) or below ($\epsilon=-1$) the singularity at $z=1$:
$$
G_v(z)=\sum^{N_{\rm F}}_{w=1} e^{i \pi \epsilon (m_v+\tilde{m}_w)}
\prod^{N_{\rm F}}_{f\neq v}{\Gamma(1+m_{vf}) \over \Gamma(1-\tilde{m}_w-m_f)}
\prod^{N_{\rm F}}_{f\neq w}{\Gamma(\tilde{m}_{fw}) \over \Gamma(\tilde{m}_f+m_v)}
\tilde{G}_w(z)\,.
$$
By exchanging $z\leftrightarrow z^{-1}$ and  $m \leftrightarrow \tilde{m}$ we obtain the inverse formula
$$
\widetilde{G}_v(z)=\sum^{N_{\rm F}}_{w=1} e^{i \pi \epsilon (\tilde{m}_v+m_w)}
\prod^{N_{\rm F}}_{f\neq v}{\Gamma(1+\tilde{m}_{vf}) \over \Gamma(1-m_w-\widetilde{m}_f)}
\prod^{N_{\rm F}}_{f\neq w}{\Gamma(m_{fw}) \over \Gamma(m_f+\widetilde{m}_v)}
G_w(z)\,,
$$
where the two regions are connected along a path below ($\epsilon=+1$) or above ($\epsilon=-1$) $z=1$.

Let us define a path $\tilde \gamma_{\epsilon_1 \epsilon_2 \epsilon_3}$ as follows.
It first goes from $z=0$ to $+\infty$ above or below $z=1$ for $\epsilon_1=+1$ or $\epsilon_1=-1$, respectively.
Then for $\epsilon_2=+1(-1)$, it moves along a very large circle clockwise(counterclockwise), and does not move for $\epsilon_2=0$.
Finally $\epsilon_3=1$ or $\epsilon_3=-1$ if the path goes from $z=+\infty$ back to 0 below or above $z=1$.
The monodromy along $\tilde\gamma_{\epsilon_1 \epsilon_2 \epsilon_3}$ is%
\foot{%
The expressions of the form $\prod_{f\neq v,w} C_f$ mean $(\prod_{f} C_f)/C_vC_w$ in this appendix.
}
$$\eqalign{
 G_v(z) &
\rightarrow
\sum_{w} \sum_g e^{ {\pi i \epsilon_1 } (m_v+\tilde{m}_g)}
\prod_{f\neq v}{\Gamma(1+m_{vf}) \over \Gamma(1-\tilde{m}_g-m_f)}
\prod_{f\neq g}{\Gamma(\tilde{m}_{fg}) \over \Gamma(\tilde{m}_f+m_v)}
\cr
& \qquad \times
e^{2\pi i  \epsilon_2 \tilde{m}_g}
e^{{\pi i \epsilon_3} (\tilde{m}_g+m_w)}
\prod_{f\neq g}{\Gamma(1+\tilde{m}_{gf}) \over \Gamma(1-m_w-\widetilde{m}_f)}
\prod_{f\neq w}{\Gamma(m_{fw}) \over \Gamma(m_f+\widetilde{m}_g)}
G_w(z)
\cr
&
=\sum_w { e^{\pi i (\epsilon_1 m_v + \epsilon_3 m_w)}}
{\prod_{f\neq v}\Gamma(1+m_{vf}) \prod_{f\neq w}\Gamma(m_{fw}) \over
\prod_{f}\Gamma(\widetilde{m}_f+m_v) \Gamma(1-m_w-\widetilde{m}_f)}
\cr
& \qquad \times
  \pi \sum_g
 e^{i \pi ({+}\epsilon_1+2\epsilon_2 {+}\epsilon_3) \tilde{m}_g}
{ \prod_{f\neq v, w} \sin\pi(m_f+\widetilde{m}_g) \over
\prod_{f\neq g} \sin\pi\tilde{m}_{fg} }
G_w(z)
\,.
}$$
If $n=\epsilon_2{+}(\epsilon_1+\epsilon_3)/2$ satisfies $|n|\leq 1$,%
\foot{%
For such $n$ we have the identity \GomisKV\
$$\eqalign{
&\sum_g e^{2 \pi i n \tilde{m}_g}
{ \prod_{f\neq v, w} \sin\pi(m_f+\tilde{m}_g) \over
\prod_{f\neq g} \sin\pi\tilde{m}_{fg} }
\cr
&=
\delta_{vw}e^{-2 \pi i n m_v}
{ \prod_{f\neq v} \sin\pi m_{fv} \over
\prod_f \sin\pi(\tilde{m}_f+m_v) }
+
(-1)^{N_{\rm F}-1} 2 n i e^{n i \pi\left[\sum_f \tilde{m}_f + \sum_{f \neq v, w} m_f \right]}\,.
}$$
\vskip-5mm
}
 we can rewrite the monodromy in the form
$$
G_v(z) \rightarrow
 \sum_w \CM^{\epsilon_1\epsilon_2 \epsilon_3}_{vw} G_w(z)\,,
$$
where
$$\eqalign{
\CM^{\epsilon_1 \epsilon_2 \epsilon_3}_{vw}&=
\pi
{
e^{ \pi i (\epsilon_1 m_v + \epsilon_3 m_w)}
}
{\prod_{f\neq v}\Gamma(1+m_{vf}) \prod_{f\neq w}\Gamma(m_{fw}) \over
\prod_{f}\Gamma(\tilde{m}_f+m_v) \Gamma(1-m_w-\tilde{m}_f)}
\cr
& \quad   \times
\left[
\delta_{vw} e^{-2\pi i n  m_v}
{ \prod_{f\neq v} \sin\pi m_{fv} \over
\prod_f \sin\pi(\widetilde{m}_f+m_v) }
+
(-1)^{N_{\rm F}-1} 2 n  i 
e^{n  \pi i \left(\sum_f \tilde{m}_f + \sum_{f \neq v,w} m_f \right)}
\right]
\cr
&=
\delta_{vw} e^{ -2 \epsilon_2  \pi i m_v }
+
2 n  \pi i e^{ i \pi [n\sum_f (m_f + \tilde{m}_f) {+(\epsilon_1-n)m_v} {+ (\epsilon_3-n)m_w}]} {\bf S}_{vw}\,.
}$$
The matrix
$$
{\bf S}_{vw}\equiv (-1)^{N_{\rm F}-1}{\prod_{f\neq v}\Gamma(1+m_{vf}) \prod_{f\neq w}\Gamma(m_{fw}) \over
\prod_{f}\Gamma(\tilde{m}_f+m_v) \Gamma(1-m_w-\tilde{m}_f)}.
$$
satisfies the equations%
\foot{%
The second equation can be proved by using the identity
$$\sum_g{ \prod_{f} \sin(\tilde{m}_f+m_g) \over \prod_{f\neq g} \sin(m_f-m_g)}
 ={(-1)^{N_{\rm F}-1} \over 2i}\left( e^{i \pi \sum_f (m_f+\tilde{m}_f)}- e^{-i \pi \sum_f (m_f+\tilde{m}_f)}\right)\,.
$$
\vskip -6mm
}
$$\eqalign{
&{\bf S}_{vv}={(-1)^{N_{\rm F}-1} \over \pi} {\prod_f \sin\pi(\tilde{m}_f + m_v) \over \prod_{f\neq v} \sin\pi m_{fv}},
\cr
&
\sum^{N_{\rm F}}_{g=1}{\bf S}_{vg}{\bf S}_{gw}={1\over 2i\pi}\left( e^{i \pi \sum_f (m_f+\tilde{m}_f )}- e^{-i \pi \sum_f (m_f + \tilde{m}_f )}\right){\bf S}_{vw}\,.
}$$

In particular the monodromy matrices for the basic paths in Figure 3 are
$$\eqalign{
&
{\bf M}(\tilde\gamma_0)_{vw}
= \delta_{vw} e^{2  \pi i m_v},
\cr
&
{\bf M}(\tilde\gamma_1)_{vw}
=
\CM^{-1, 0, -1}_{vw}=
\delta_{vw} 
{-}2\pi i e^{{-}\pi i \sum_f (m_f + \tilde{m}_f) } {\bf S}_{vw}\,,
\cr
&
{\bf M}(\tilde \gamma_\infty)_{vw}
=\CM^{1, 1, -1}_{vw}=
\delta_{vw} e^{- 2  \pi i m_v }
+
2  \pi i e^{ \pi i \sum_f (m_f + \tilde{m}_f)}e^{ -2\pi im_{w}} 
{\bf S}_{vw}.
}$$
One can check that ${{\bf M}(\tilde\gamma_0){\bf
M}(\tilde\gamma_1){\bf M}(\tilde\gamma_\infty)=1}$ as expected.%
\foot{%
We defined ${\bf M}(\tilde\gamma)$ for all $\tilde\gamma$ using a base point on a common Riemann sheet.
For a discussion on the choice of base point and relations satisfied by monodromy matrices, see \FerrariSV.
}
After the similarity transformation \Similar, we obtain the monodromy matrices \MonodMatToda.

\appendix{H}{Grade restriction rule and analytic continuation}
\applab\AppFlop

In this appendix we explain how to use the integral representation \ResultPreservedGeneral\ to analytically continue a hemisphere partition function from one region to another in the K\"ahler moduli space.
This involves choosing a complex of bundles representing a given object in the derived category so that each bundle satisfies the so-called grade restriction rule \HerbstJQ.
We will use a D2-brane on the resolved conifold as an example.

We first review a derivation of the grade restriction rule from the integral representation of $Z_{\rm hem}$, as explained in a talk by K.~Hori.
Let us consider a general $U(1)$ gauge theory with $N_F$ chiral multiplets with gauge charges $Q_f$ and twisted masses $m_f$, $f=1,\ldots,N_F$, satisfying $\sum_f Q_f=0$.
We impose the Neumann boundary condition on all chiral fields and include a Wilson loop with gauge charge $n$.
The hemisphere partition function is then
$$
\int^{i\infty}_{-i\infty} {d\sigma \over 2\pi i} e^{t\sigma}e^{- 2\pi i n \sigma}
\prod^{N_F}_{f=1} \Gamma(Q_f \sigma + m_f),
$$
where $t=r -i\theta$.
In the limit $\sigma\rightarrow \pm i\infty$, the absolute value of the integrand behaves as 
$
\exp\Big(\big(-\pi{\cal S}\pm (2\pi n+\theta)\big)|\sigma|\Big)
$,
where ${\cal S}=\sum_{Q_f>0} Q_f$.
When the {\it grade restriction rule}%
\foot{%
The energy for large $|\sigma_1-i\sigma_2|$ is bounded from below only if \GRR\ is satisfied \HerbstJQ.
}
\eqn\GRR{
-{{\cal S} \over 2}<n+{\theta \over 2\pi}<{{\cal S} \over 2}
}
is obeyed, the $\sigma$-integral along the imaginary axis is absolutely convergent, and the hemisphere partition function can be analytically continued from $r\gg 0$ to $r\ll 0$.

Let us consider a $U(1)$ gauge theory with chiral multiplet fields $(\phi_1, \phi_2)$ with charge $+1$, and $(\tilde\phi_1, \tilde\phi_2)$ with charge $-1$.
We denote their twisted masses as $(m_1, m_2)$ and $(\tilde{m}_1, \tilde{m}_2)$ respectively.
The theory flows to the nonlinear sigma model whose target space $X$ is defined by the equation $|\phi_1|^2+|\phi_2|^2-|\tilde\phi_1|^2-|\tilde\phi_2|^2=r/2\pi$ and the $U(1)$ quotient.
In the phase $r\gg 0$, this is the resolved conifold, the total space of ${\cal O}_{\Bbb P^1}(-1)^{\oplus 2} \to \Bbb P^1$, where $(\phi_1, \phi_2)$ parametrize the base $\Bbb P^1$ and $(\tilde \phi_1, \tilde\phi_2)$ are the fiber coordinates.
In the flopped phase $r \ll 0$ the roles of $(\phi_1, \phi_2)$ and $(\tilde \phi_1, \tilde\phi_2)$ are exchanged.
Let $i^\pm :\Bbb P^1\rightarrow X$ be the embeddings in the $\pm r\gg 0$ phases respectively. 

We are interested in transporting the sheaf $i^+_{*}{\cal O }_{\Bbb P^1}$ from $r\gg 0$ to $r\ll 0$, through the window $-2\pi <\theta<0$, for which the grade restriction rule is obeyed only by $n=0,1$.
In particular, we will perform an analytic continuation of its hemisphere partition function.

To study this problem, let us introduce fermionic oscillators satisfying $\{\eta_f,\bar\eta_g\}=\{\tilde\eta_f,\bar{\tilde\eta}_g\}=\delta_{fg}$ ($f,g=1,2$), with the corresponding Clifford vacua such that $\eta_f|0\rangle=\tilde\eta_g|\tilde 0\rangle=0$.
We assume that $|\tilde 0\rangle$ is neutral under gauge and flavor symmetries, and identify $|\tilde 0\rangle=\tilde\eta_2\tilde\eta_1|0\rangle$.
Consider the following two complexes of vector spaces
\eqn\CompConi{
0 \longrightarrow \Bbb C \bar\eta_1\bar\eta_2|0\rangle \longrightarrow \Bbb C \bar\eta_1|0\rangle \oplus \Bbb C \bar\eta_2|0\rangle \longrightarrow \underline{\Bbb C |0\rangle} \longrightarrow 0\,,
}
\eqn\CompTConi{
0 \longrightarrow \Bbb C \bar{\tilde \eta}_1\bar{\tilde \eta}_2|\tilde 0\rangle \longrightarrow \Bbb C \bar{\tilde\eta}_1|\tilde 0\rangle \oplus \Bbb C \bar{\tilde \eta}_2|\tilde 0\rangle \longrightarrow \underline{\Bbb C |\tilde 0\rangle}  \longrightarrow 0\,,
}
with the underline indicating degree zero.
The differentials are ${\cal Q}=\sum_{f=1,2}\phi_f\eta_f$, $\tilde {\cal Q}=\sum_{f=1,2}\tilde\phi_f\tilde \eta_f$ respectively.
These represent complexes of equivariant vector bundles.
In the phase $r\gg 0$, $\{{\cal Q},\bar{\cal Q}\}$ is positive definite, implying that \CompConi\ is  exact and represents the zero object in the derived category.
On the other hand, in the same phase,  \CompTConi\ is the Koszul resolution \CG\ of $i^+_*{\cal O }_{\Bbb P^1}$ supported on $\{\tilde\phi_1=\tilde\phi_2=0\}$, which is the D-brane we are interested in.
Again the roles of \CompConi\ and \CompTConi\ are swapped for $r\ll 0$.

The gauge charges of $|\tilde 0\rangle$,  $\bar{\tilde\eta}_f|\tilde 0\rangle$, $\bar{\tilde\eta}_1 \bar{\tilde\eta}_2|\tilde 0\rangle$ are $0, 1, 2$ respectively.
The last one is outside the range \GRR.
As a consequence, the hemisphere partition function for \CompTConi
\eqn\ZhemiOPOne{
(- 2 \pi i)^2 e^{ \pi i (\tilde m_1+ \tilde m_2)} \int^{i\infty}_{-i\infty} {d\sigma \over 2\pi i} 
 e^{ (t - 2 \pi i )\sigma} { \Gamma(\sigma + m_1) \Gamma(\sigma + m_2) \over \Gamma ( 1 + \sigma - \tilde m_1) \Gamma ( 1+ \sigma - \tilde m_2)}
}
does not converge absolutely along the imaginary axis.
For $r\gg 0$, convergence requires us to choose the $\sigma$ contour so that asymptotically $\sigma\rightarrow \pm i (1\pm\epsilon)\infty$, and this gives 
\eqn\ZhemIPlusOP{\eqalign{
Z_{\rm hem}(i^+_*{\cal O}_{\Bbb P^1})
&=
(- 2 \pi i)^2 e^{\pi i (\tilde m_1+ \tilde m_2)} \sum_{v=1}^2 e^{-m_v(t-2 \pi i)}
{\prod^2_{f \neq v} \Gamma(m_f-m_v) \over \prod^2_{f=1} \Gamma(1-m_v - \tilde m_f)}
\cr
& \qquad \qquad\times 
{}_2 F_1 \bigg( \matrix{{\scriptstyle\{\tilde m_f+m_v\}^{2}_{f=1}} \cr{\scriptstyle \{1-m_f+m_v\}^{2}_{f\neq v}}}\bigg| e^{-t}\bigg)\,.
}}
For $r\ll 0$ we need  $\sigma\rightarrow \pm i (1\mp\epsilon)\infty$, and \ZhemiOPOne\ vanishes, as it should for the zero object.
The two functions are not related by analytic continuation.

In order to analytically continue $Z_{\rm hem}(i_{*}{\cal O }_{\Bbb P^1})$ from $r\gg 0$ to $r\ll 0$, we may evaluate \ZhemiOPOne\ by residues and apply the connection formula, as we did in Appendix \AppTodaDetails.
Here we explain an alternative method found in \HerbstJQ.

The problematic term $\Bbb C \bar{\tilde\eta}_1 \bar{\tilde\eta}_2|\tilde 0\rangle$ can be eliminated from the complex \CompTConi by binding the D-brane \CompTConi\ with the other D-brane \CompConi, which is empty for $r\gg 0$.
Let $f$ be the unique cochain map from \CompTConi\ to \CompConi, with degrees shifted for the latter, such that $\Bbb C \bar{\tilde\eta}_1 \bar{\tilde\eta}_2|\tilde 0\rangle$ in  \CompTConi\ is mapped to $\Bbb C|0\rangle$ in \CompConi\ by the identity map.
The bound state of the two D-branes is the mapping cone $C(f)$:
$$
\xymatrix{
&
\Bbb C \bar{\tilde\eta}_1 \bar{\tilde\eta}_2|\tilde 0\rangle
\ar[r] \ar[rd]^{\bf 1} \ar@{}[d]|{\bigoplus}
&
\Bbb C \bar{\tilde\eta}_1|\tilde 0\rangle \oplus \Bbb C \bar{\tilde \eta}_2|\tilde 0\rangle 
\ar[r]  \ar@{}[d]|{\bigoplus}
&
 \Bbb C |\tilde 0\rangle 
\\
 \Bbb C \bar \eta_1\bar \eta_2| 0\rangle 
\ar[r]
& 
 \Bbb C \bar \eta_1| 0\rangle \oplus  \Bbb C \bar \eta_2| 0\rangle
\ar[r]
&
\Bbb C| 0\rangle
&
\\}
$$
The pair, which carries the gauge charge $2$ and is connected by the identity map, can be neglected in computing $Z_{\rm hem}$ for $C(f)$.%
\foot{%
As in \HerbstJQ\ one can change the basis to show that $C(f)$ decomposes into a complex
$
{\cal V}^{-3}
\rightarrow 
{\cal V}^{-2}
\rightarrow 
{\cal V}^{-1}
\rightarrow 
{\cal V}^{0}
$
and a trivial pair $\tilde {\cal V}^{-2}\rightarrow \tilde{\cal V}^{-1}$,
where $({\cal V}^{-3},{\cal V}^{-2},{\cal V}^{-1},{\cal V}^{0};$ $\tilde {\cal V}^{-2}, \tilde{\cal V}^{-1})$ carry the same quantum numbers as $(
\Bbb C \bar \eta_1\bar\eta_2 | 0\rangle, \Bbb C \bar \eta_1| 0\rangle \oplus  \Bbb C \bar \eta_2| 0\rangle, \Bbb C \bar{\tilde\eta}_1|\tilde 0\rangle \oplus \Bbb C \bar{\tilde \eta}_2|\tilde 0\rangle 
, \Bbb C |\tilde 0\rangle;$ $\Bbb C \bar{\tilde\eta}_1 \bar{\tilde\eta}_2|\tilde 0\rangle,\Bbb C | 0\rangle)$.
}
The other terms carry gauge charges $0$ or $1$.
The hemisphere partition function can be written as
$$\eqalign{
Z_{\rm hem}(
C(f)
) &=\int  {d\sigma \over 2\pi i} 
 e^{ t \sigma}  
 \bigg[ 
 1
 -
e^{-2\pi i \sigma}\left(e^{2\pi i \tilde m_1}+e^{2\pi i \tilde m_2}\right)
 +
(e^{2\pi i m_1}+e^{2\pi i m_2})e^{2\pi i(\tilde m_1 +\tilde m_2-\sigma)}
  \cr
 & 
\qquad
-e^{2\pi i(m_1+m_2 +\tilde m_1 +\tilde m_2)} 
\bigg]
\prod_{f=1}^2 \Gamma(\sigma + m_f) \Gamma ( - \sigma + \tilde m_f)
 \, .
}
$$
This integral along the imaginary axis is now absolutely convergent for $-2\pi <\theta<0$, and interpolates the hemisphere partition functions in the two phases.

In the phase $r\gg 0$, the contribution from \CompConi\ is trivial, and $Z_{\rm hem}(
C(f) )$ coincides with $Z_{\rm hem}(i^+_*{\cal O}_{\Bbb P^1})$ in \ZhemIPlusOP.
In the phase $r \ll 0$, the contribution from \CompTConi\ becomes trivial and $Z_{\rm hem}(C(f))$ coincides with the hemisphere partition function for \CompConi
$$\eqalign{
Z_{\rm hem}(i^-_*{\cal O}_{\Bbb P^1}(2)[1])
&=-(- 2 \pi i)^2 e^{\pi i (m_1+m_2+2\tilde m_1+ 2\tilde m_2)}
\cr
& \quad \times
 \int {d\sigma \over 2\pi i} 
 e^{ (t - 2 \pi i )\sigma} { \Gamma(-\sigma + \tilde m_1) \Gamma(-\sigma + \tilde m_2) \over \Gamma ( 1- \sigma - m_1) \Gamma ( 1- \sigma - m_2)}
\cr
&=-(- 2 \pi i)^2 e^{\pi i (m_1+m_2+2 \tilde m_1+ 2 \tilde m_2)}
\cr
& \quad \times 
\sum_{v=1}^2 e^{\tilde m_v(t-2 \pi i)}
{\prod^2_{f \neq v} \Gamma(\tilde m_f- \tilde m_v) \over \prod^2_{f=1} \Gamma(1- \tilde m_v - m_f)}
{}_2 F_1 \bigg( \matrix{{\scriptstyle\{ m_f + \tilde m_v\}^{2}_{f=1}} \cr{\scriptstyle \{1- \tilde m_f + \tilde m_v\}^{2}_{f\neq v}}}\bigg| e^{t}\bigg).
}$$
One can check that the relation between $Z_{\rm hem}(i^+_*{\cal O}_{\Bbb P^1})$ and $Z_{\rm hem}(i^-_*{\cal O}_{\Bbb P^1}(2)[1]) $ is consistent with the connection formulas in Appendix \AppTodaDetails.

\listrefs

\bye